\documentclass [12pt]{article}
\usepackage{graphicx,amssymb,amsfonts,latexsym,amsmath,amsthm,times}
\usepackage{epsfig}
\usepackage{fancyhdr}
\usepackage{color}
\usepackage[english]{babel}
\usepackage{soul}
\numberwithin{equation}{section}

\begin{document}


\thispagestyle{plain}

\vspace*{2cm} \normalsize \centerline{\Large \bf Modeling of Sensory
Characteristics} \centerline{\Large \bf Based on the Growth of Food
Spoilage Bacteria}

\vspace*{1cm}

\centerline{\bf D. Valenti$^a$\footnote{Corresponding author.
E-mail: davide.valenti@unipa.it}, G. Denaro$^b$, F. Giarratana$^c$,
A. Giuffrida$^c$, S. Mazzola$^b$}\centerline{\bf G. Basilone$^b$, S.
Aronica$^b$, A. Bonanno$^b$, B. Spagnolo$^{a,d,e}$}

\vspace*{0.5cm}

\centerline{$^a$ Dipartimento di Fisica e Chimica, Universit\`a di
Palermo,}\centerline{Group of Interdisciplinary Theoretical Physics
and CNISM, Unit\`a di Palermo,}\centerline{Viale delle Scienze,
Ed.~18, I-90128~Palermo, Italy}
\medskip
\centerline{$^b$ Istituto per l'Ambiente Marino Costiero, CNR,
U.O.S. di Capo Granitola}\centerline{Via del Faro 3, I-91020
Campobello di Mazara (TP), Italy}
\medskip
\centerline{$^c$ Dipartimento di Scienze Veterinarie, Universit\`a
di Messina,}\centerline{Polo Universitario dell'Annunziata, 98168
Messina, Italy}
\medskip
\centerline{$^d$ Radiophysics Department, Lobachevsky State
University,}\centerline{Nizhniy Novgorod, Russia}
\medskip
\centerline{$^e$ Istituto Nazionale di Fisica Nucleare, Sezione di
Catania, Italy}


\vspace*{1cm}

\noindent {\bf Abstract}

During last years theoretical works shed new light and proposed new
hypothesis on the mechanisms which regulate the time behaviour of
biological populations in different natural systems. Despite of
this, a relevant physical and biological issue such as the role of
environmental variables in ecological systems is still an open
question. Filling this gap of knowledge is a crucial task for a
deeper comprehension of the dynamics of biological populations in
real ecosystems.

The aim of this work is to study how dynamics of food spoilage
bacteria influences the sensory characteristics of fresh fish
specimens. This topic is worth of investigation in view of a better
understanding of the role played by the bacterial growth on the
organoleptic properties, and becomes crucial in the context of
quality evaluation and risk assessment of food products. We
therefore analyze and reproduce the time behaviour, in fresh fish
specimens, of sensory characteristics starting from the growth
curves of two spoilage bacterial communities.

The theoretical study, initially based on a deterministic model, is
performed by using the temperature profiles obtained during the
experimental analysis. As a first step, a model of predictive
microbiology is used to reproduce the experimental behaviour of the
two bacterial populations. Afterwards, the theoretical bacterial
growths are converted, through suitable differential equations, into
"sensory" scores, based on the Quality Index Method (QIM), a scoring
system for freshness and quality sensory estimation of fishery
products. As a third step, the theoretical curves of QIM scores are
compared with the experimental data obtained by sensory analysis.
Finally, the differential equations for QIM scores are modified by
adding terms of multiplicative white noise, which mimics the effects
of uncertainty and variability in sensory analysis. A better
agreement between experimental and theoretical QIM scores is
observed, in some cases, in the presence of suitable values of noise
intensity respect to the deterministic analysis.

\vspace*{0.5cm}

\noindent {\bf Key words:} population dynamics, predictive
microbiology, stochastic ordinary differential equations

\noindent {\bf AMS subject classification:} 82C05, 92D25, 92D40,
60H10


\vspace*{1cm}

\setcounter{equation}{0}

\section{Introduction}\label{S:1}

In this work we review a recent result~\cite{Giu13} obtained in the
context of predictive microbiology~\cite{Whi93}-\cite{Bar95} (a
theoretical approach to describe microbial dynamics in food
products) and sensory analysis (a tool which permits to evaluate the
food quality through sensory indicators). The previous study, based
on a deterministic model, which allowed to reproduce experimental
findings for fresh fish specimens, is deepened by improving fitting
procedures and statistical analysis.\\
We recall here that the general idea of predictive microbiology is
to model the dynamics of microbial populations, in particular
bacteria, responsible for food spoilage, taking into account the
changes of environmental variables such as temperature (T), pH, and
water activy (aw). Mathematical models can be therefore very useful
for practical applications, since some food products, such as
ripened meats and cheeses, are obtained under a continuous
modification of T, pH, and relative humidity (RH). For example, a
class of models, devised to reproduce the bacterial dynamics in food
products, exploits generalized Lotka-Volterra equations, which
describe the time evolution of two different microbial populations
both in deterministic and stochastic
regime~\cite{Den99}-\cite{Giu09b}.\\
Predictive models are classified as primary, secondary and
tertiary~\cite{Whi93}. Primary models provide microbial dynamics.
Secondary models describe the dynamics of environmental variables
which appear in primary models. Finally, tertiary models combine
primary and secondary models, allowing to take into account the
effects of environmental variables on the microbial
growth~\cite{Dal02}. We note that the real bacterial growth can be
overestimated if the competitive natural microflora is not taken
into account by these
models~\cite{Bar95}.\\
It is clear that the growth of spoilage bacteria in a fish product
determines a loss of quality, which appears through a worsening of
the sensory characteristics. These are "measured" by a scoring
system for freshness and quality sensory estimation of fishery
products, known as Quality Index Method (QIM) and initially
developed by the Tasmanian Food Research
Unit~\cite{Bre85}.\\
The QIM scoring allows to assign demerit points to each sensory
parameter considered, providing by a summation of the partial scores
an overall sensory score, named Quality Index (QI).\\
A crucial point is how to relate the sensory characteristics of a
fresh fish products to bacterial populations responsible for fish
spoilage. This subject has been widely debated, since the sensory
modifications in fish specimens is due to the growth of spoilage
bacteria on skin, gills, and flesh~\cite{Kou00}-\cite{Lou03}. These
sites, in fact, are those taken into account by the sensory
evaluation schemes such as QIM. However, the bacterial penetration
through the skin can occur very slowly~\cite{Giu05}. As a
consequence the sensory modifications in a fresh fish specimens is
expected to depend mainly on the spoilage bacteria located on skin
and gills.\\
Because of this, to predict correctly the whole-fish
freshness, one should model separately the dynamics of the specific
spoilage
bacteria (SSB) on skin, gills, and flesh~\cite{Giu13}.\\
Here we analyze the connection between sensory characteristics of
fresh fish specimens and two bacterial populations responsible for
fish spoilage. The sensory characteristics are "measured" by the QIM
scoring system~\cite{Bre85}.\\
The analysis consists in a theoretical study, initially based on a
set of deterministic differential equations for bacterial dynamics
and QIM scores, which allows: i) to model the experimental behaviour
of two SSB populations, obtained separately on skin, gills and flesh
under different storage conditions, i.e. varying the temperature;
ii) to reproduce the QI scores, obtained for Gilthead seabream
(\emph{Sparus aurata}) specimens, starting from the theoretical
bacterial curves obtained at the previous step.\\
It is important to recall that environmental perturbations can
affect significantly the dynamics of real physical and biological
systems~\cite{Lac10b}-\cite{Pan05}. In particular, the interplay
between fluctuations and nonlinearity in physical, biological, and
social systems as well as in financial markets can give rise to
counterintuitive phenomena, such as stochastic resonance, noise
enhanced stability, resonant activation, noise delayed extinction,
enhanced stochastic temporal and spatio-temporal oscillations,
effects of intrinsic noise, induced chaotic transitions from
periodic attractors, and pattern
formation~\cite{Gam98}-\cite{Roj14}. Variability of environmental
parameters, such as temperature and food availability, influence the
dynamics of real biological systems, favouring the survivance of
some populations and contributing to the extinction of other ones.\\
Natural systems are open systems, because of their continuous
interaction with the environment, which influences their dynamics by
deterministic and random perturbations. Moreover, natural systems
are governed by nonlinear dynamics. These two characteristics, i.e.
nonlinearity and external random perturbations, make them complex
systems, so that population dynamics, such as spatio-temporal
dynamics of phytoplankton species in a marine ecosystem, has to be
studied by stochastic approaches, which take into account the
presence of random fluctuations~\cite{Den13a}-\cite{Val15}.\\
As a final step, we take therefore into account the effects of
uncertainty and variability in sensory analysis, and modify the
differential equation for QI scores by adding terms of
multiplicative white noise. The stochastic model is solved for
different values of noise intensities. The QI time behaviours
obtained are compared with the experimental ones by calculating the
root mean square error (RMSE). As a result, we find that some curves
of theoretical QI scores, obtained in the presence of suitable noise
intensities, are in a better agreement with those observed
experimentally respect to those calculated by the deterministic
model.

\section{Experimental data}\label{S:2}

\subsection{Fish specimens and storage conditions}\label{SS:2.1}

Bacterial concentrations were obtained from Gilthead seabream (300
-–500 g) raised in three Italian fish farms (farm~1,~2,~and~3).
After death the fish were subdivided in four groups and stored at
different temperatures. Group 1 consisted of 147 fish subdivided in
seven batches, each containing twenty-one specimens. This group was
used to carry out seven replicated trials (three trials from farm~1,
two from farm~2, and two from farm~3) in such a way to characterize
the variability of the fish shelf life and to obtain a better
parametrization for Eqs.~(\ref{QI_S})-(\ref{QI_F}) with particular
regard to coefficients $\beta_1$ and $\beta_2$ (see
Section~\ref{S:3}).\\
Group~2 consisted of one batch of twenty-eight fish from farm~1,
while Group~3, as well as Group~4, consisted of one batch of
twenty-one fish, always coming from farm~1.\\
The storing temperature were monitored for all groups. Measures of
bacterial concentrations and sensory evaluations were carried out
after 0, 72, 168, 216, 336, 408 and 504 h from the beginning of
storage, by sampling three fish for each time interval.
Microbiological assays were performed by sampling, with sterile
instruments, 10 g of dorsal skin, 5 g of gills, and 20 g of dorsal
flesh; this last kind of sample was obtained from the opposite side
where skin was sampled, rinsing the skin with 70\% ethanol and
removing the flesh aseptically.

\subsection{Microbiological analyses and sensory evaluation}\label{SS:2.1}

Microbiological analyses revealed the presence of two different
bacterial populations, i.e. Pseudomonas spp. and Shewanella spp.,
recorded as sulphide non-producers (white colonies) and
sulphide-producers (black colonies), respectively.\\
Sensory evaluation was carried out
by using the QIM scheme developed for raw whole Gilthead
seabream~\cite{Hui00}, considering variables connected with surface
and eyes appearance, odour, elasticity of the muscle and gills,
taking into account a maximum of 15 demerit points. The sensory
evaluation was performed by an expert panel of three persons.

\section{Deterministic model}\label{S:3}

The time behaviour of QIM parameters were obtained by modeling
separately the dynamics of: i) $QI_S$, related to scores for
surface/eyes appearance and odour (0–-10 demerit points); ii) $QI_G$
connected with gills scores (0-–4 demerit points); iii) $QI_F$
related to scores assigned to the flesh evaluation (0-–1 demerit
points). Our main hypothesis, based on previous studies (see
Section~\ref{S:1}), is that $QI_S$, $QI_G$ and $QI_F$ depend on the
bacterial counts performed on skin, gills and flesh, respectively,
according to the following differential equations~\cite{Giu13}
\begin{eqnarray}
\frac{dQI_S}{dt}&=&\frac{dN_{wS}}{dt}\beta_{1S}+\frac{dN_{bS}}{dt}\beta_{2S}
\label{QI_S}\\
\frac{dQI_G}{dt}&=&\frac{dN_{wG}}{dt}\beta_{1G}+\frac{dN_{bG}}{dt}\beta_{2G}
\label{QI_G}\\
\frac{dQI_F}{dt}&=&\frac{dN_{wF}}{dt}\beta_{1F}+\frac{dN_{bF}}{dt}\beta_{2F},
\label{QI_F}
\end{eqnarray}
where $N_{wi}(t)$ and $N_{bi}(t)$ ($i=S,G,F$) are the
concentrations, expressed in Log CFU $g^{-1}$, of sulphide
non-producers (white colonies) and sulphide-producers (black
colonies) bacteria, respectively, at time $t$, on skin (S), gills
(G) and flesh (F). Here CFU is an acronym for "colony forming
units", whose value provides a measure of bacterial concentration,
while $\beta_1$ and $\beta_2$ are two coefficients that convert the
bacterial concentrations into demerit points.
The bacterial concentrations $N_{wi}(t)$ and $N_{bi}(t)$ are modeled
by the following differential equations~\cite{Bar94}
\begin{eqnarray}
\frac{dN_{wi}(t)}{dt}&=&\mu^{max}_w\thinspace N_{wi}(t)
\frac{Q_{wi}}{1+Q_{wi}}\left(1-\frac{N_{wi}(t)}{N^{max}_{wi}(t)}\right)\label{Nwi}\\
\frac{dQ_{wi}}{dt}&=&\mu^{max}_w\thinspace Q_{wi}\label{Qwi}\\
\frac{dN_{bi}(t)}{dt}&=&\mu^{max}_b\thinspace N_{bi}(t)
\frac{Q_{bi}}{1+Q_{bi}}\left(1-\frac{N_{bi}(t)}{N^{max}_{bi}}\right)\label{Nbi}\\
\frac{dQ_{bi}}{dt}&=&\mu^{max}_b(t)\thinspace Q_{bi}\label{Qbi},
\end{eqnarray}
where $\mu^{max}_w$ and $\mu^{max}_b$ are the maximum specific
growth rates of the white and black population, respectively.
$N^{max}_{wi}$ and $N^{max}_{bi}$ ($i=S,G,F$) are the theoretical
maximum population densities of the white and black population,
respectively, on skin, gills and flesh under monospecific growth
conditions, that is in the absence of interspecific interaction.
Finally, $Q_{wi}$ and $Q_{bi}$ ($i=S,G,F$) are the physiological
states of the white and black population, respectively, on skin,
gills and flesh. We note that the physiological state, which
represents the state of bacterial life functions, plays a crucial
role in the whole microbial dynamics, since it measures how
efficient the bacterial metabolism is and, as a consequence,
determines the values of growth rate. Specifically, maximum growth
rates of Pseudomonas~spp. and Shewanella~spp. were calculated
according to Refs.~\cite{Ross03,Rat83,Neu97}, modified as described
below
\begin{eqnarray}
\mu^{max}_w&=&\exp[b_0+b_1\thinspace T+b_2\thinspace
pH+b_3\thinspace T\thinspace pH + b_4\thinspace T^2 + b_5\thinspace
pH^2] \label{mu_max_w}\\
\mu^{max}_b&=&[c_0\thinspace(T+c_1)]^2.\label{mu_max_b}
\end{eqnarray}
Here, the values of the parameters in
Eqs.~(\ref{mu_max_w}),~(\ref{mu_max_b}) are $b_0= -12.4$,
$b_1=0.03318$, $b_2=2.948013$, $b_3=0.011715$, $b_4=0.004123$,
$b_5=-0.25717$, and $c_0=0.027$, $c_1=2.08$, respectively. These
parameters were obtained by fitting data for specific growth under
different constant values of temperature and pH.

\section{Results and discussion}\label{S:4}

\subsection{Bacterial dynamics}\label{SS:4.1}

\begin{figure}[h!]
\begin{center}
\includegraphics[width=4.0cm]{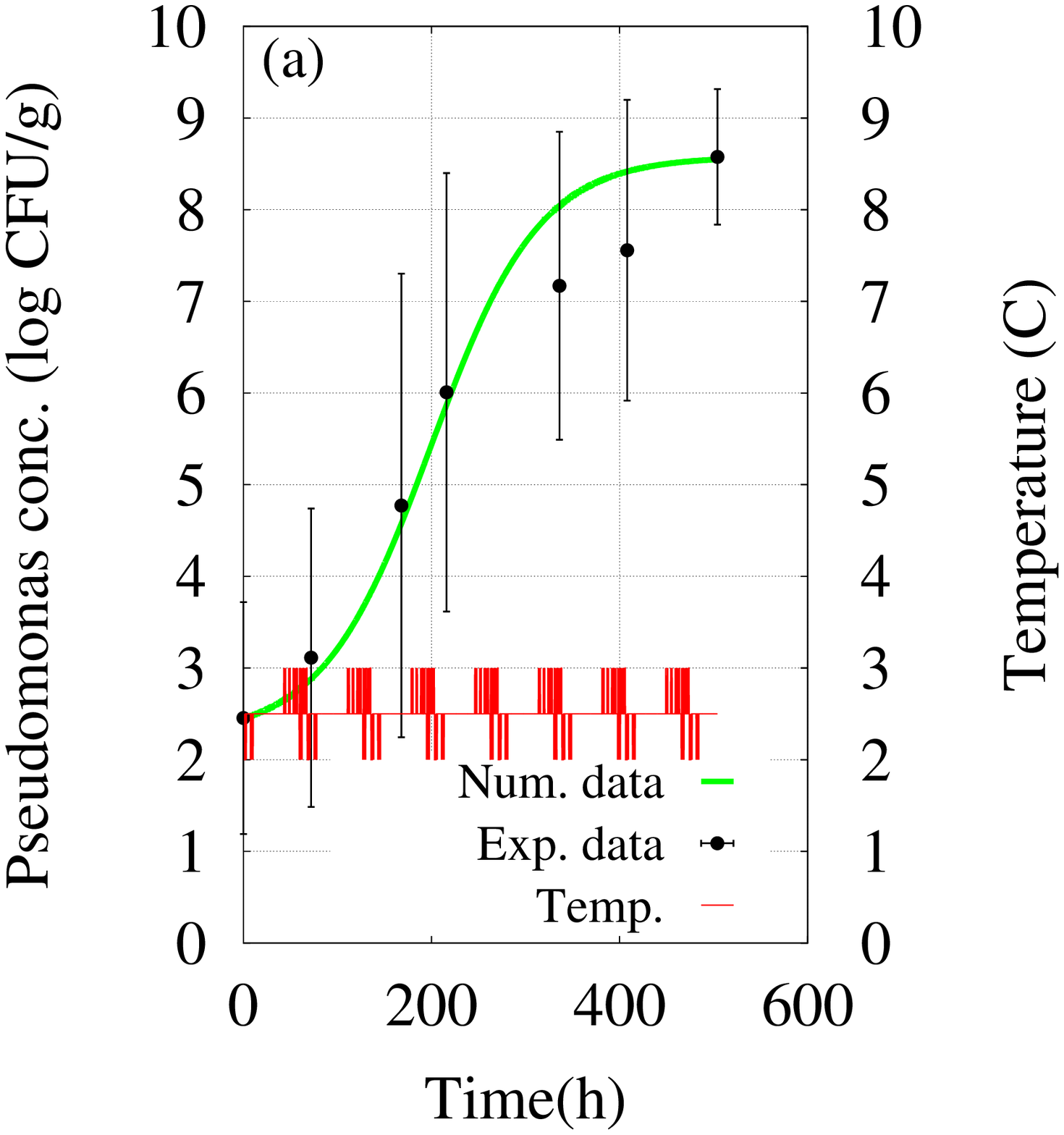}
\includegraphics[width=4.0cm]{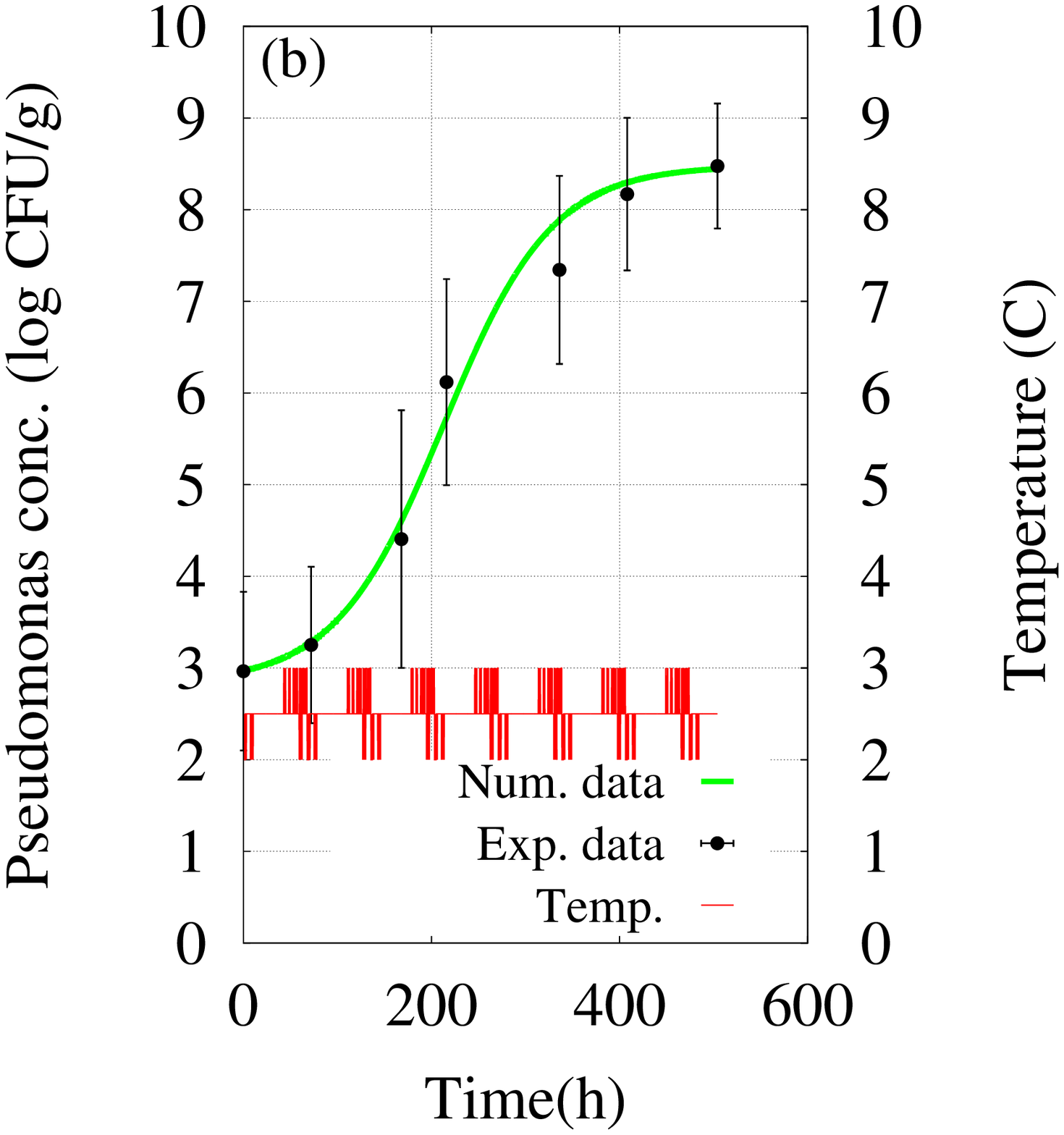}
\includegraphics[width=4.0cm]{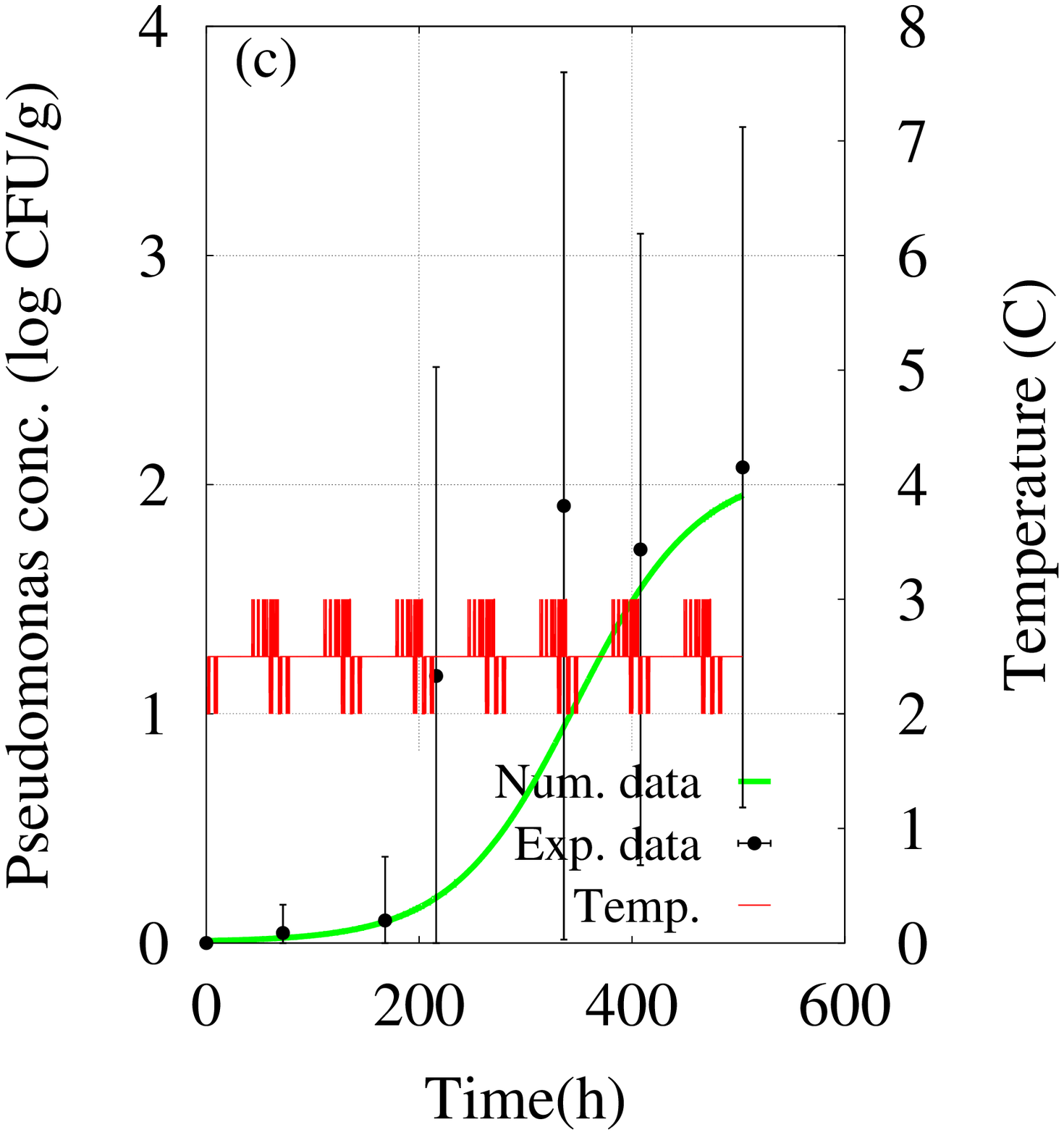}\\
\includegraphics[width=4.0cm]{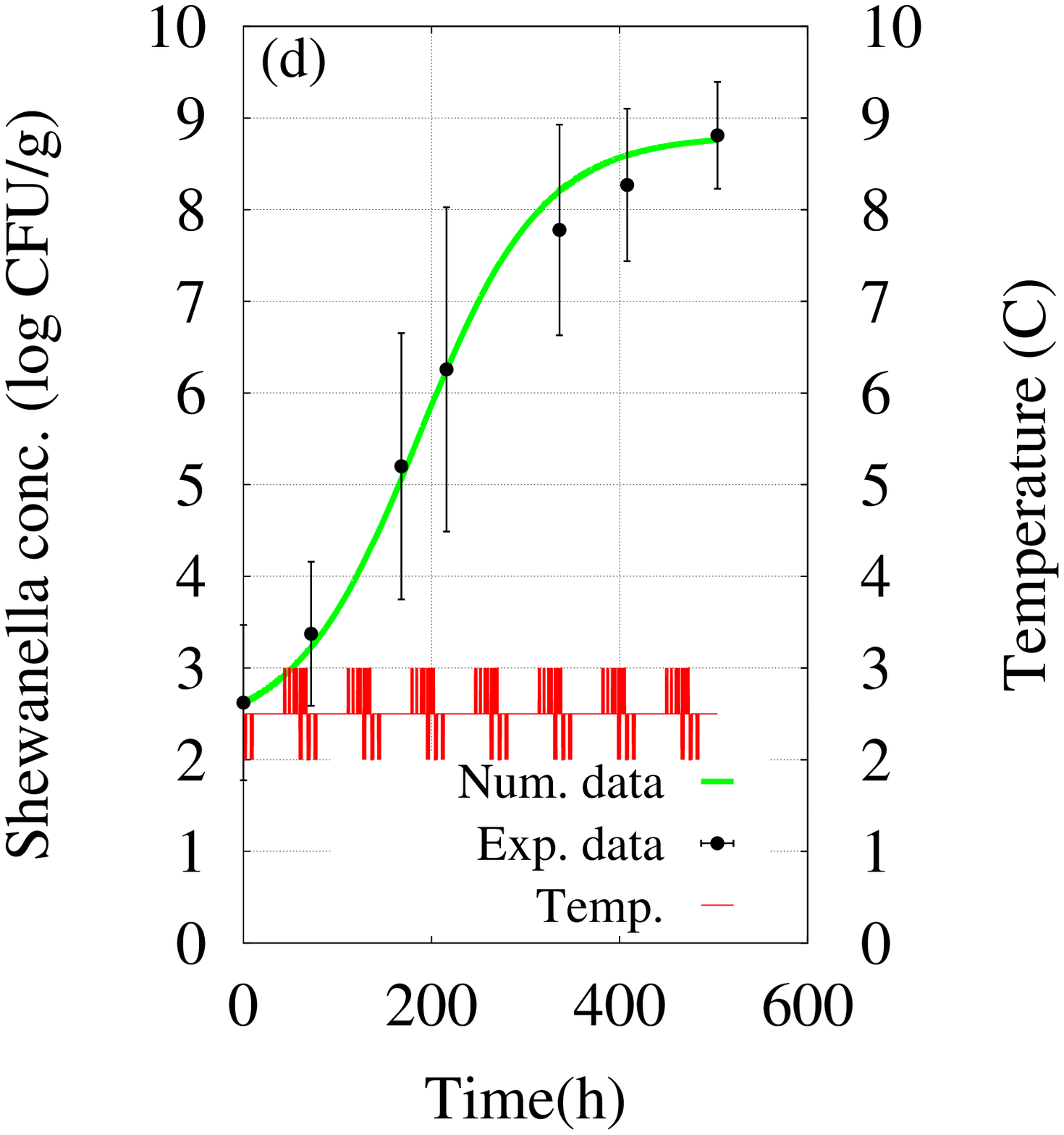}
\includegraphics[width=4.0cm]{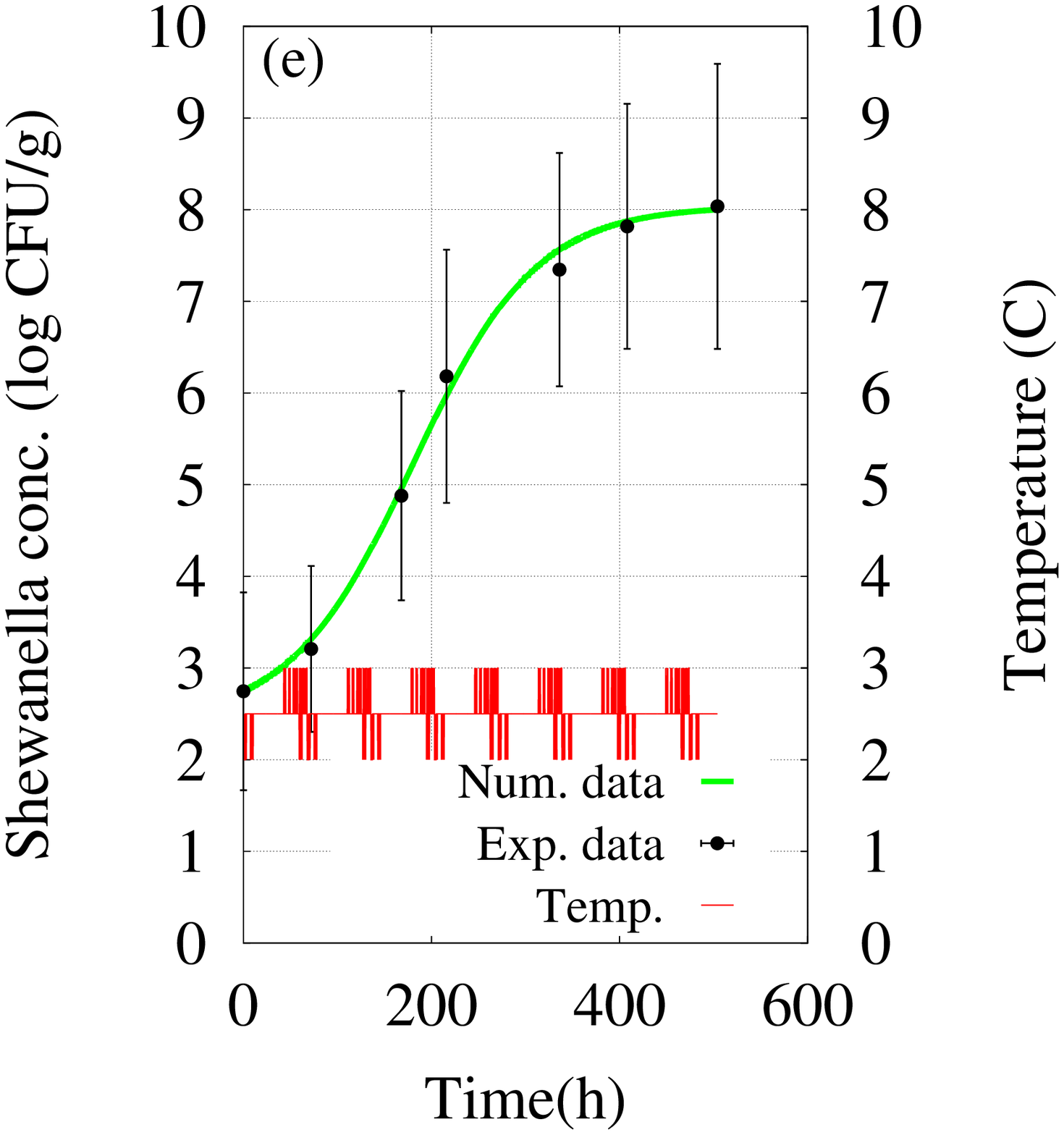}
\includegraphics[width=4.0cm]{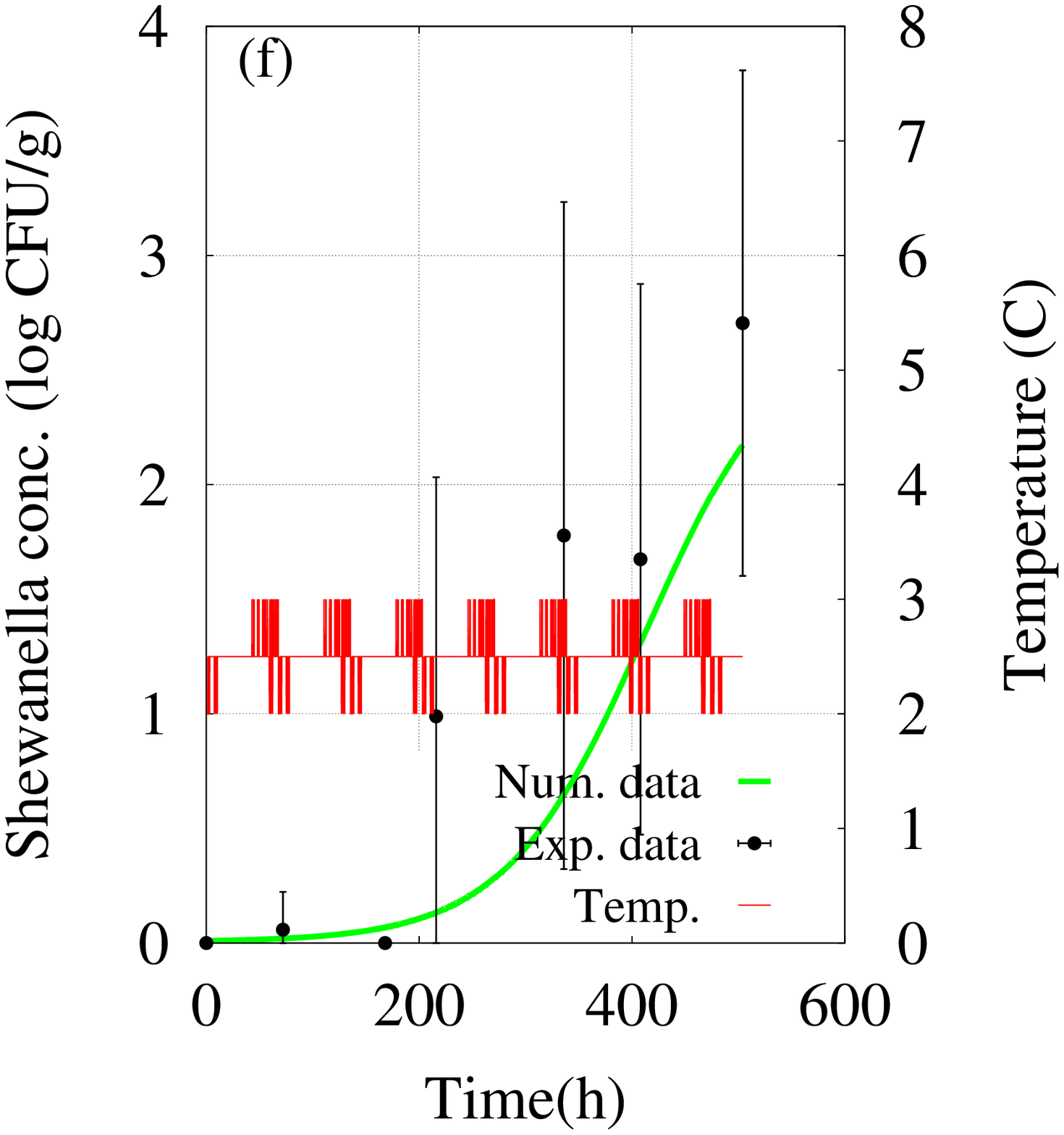}
\end{center}
\vspace{-0.8cm}\caption{\small \emph{Group~1. Comparison between
experimental data (black dots) and theoretical curves (green line)
of bacterial growth: Pseudomonas (white colonies) in panels a
(skin), b (gills), c (flesh); Shewanella (black colonies) in panels
d (skin), e (gills), f (flesh). Vertical bars indicate experimental
errors. Red curves represent the temperature profiles.}}
\vspace{-0.2cm} \label{Bacteria_Group1}
\end{figure}

\begin{figure}[h!]
\begin{center}
\includegraphics[width=4.0cm]{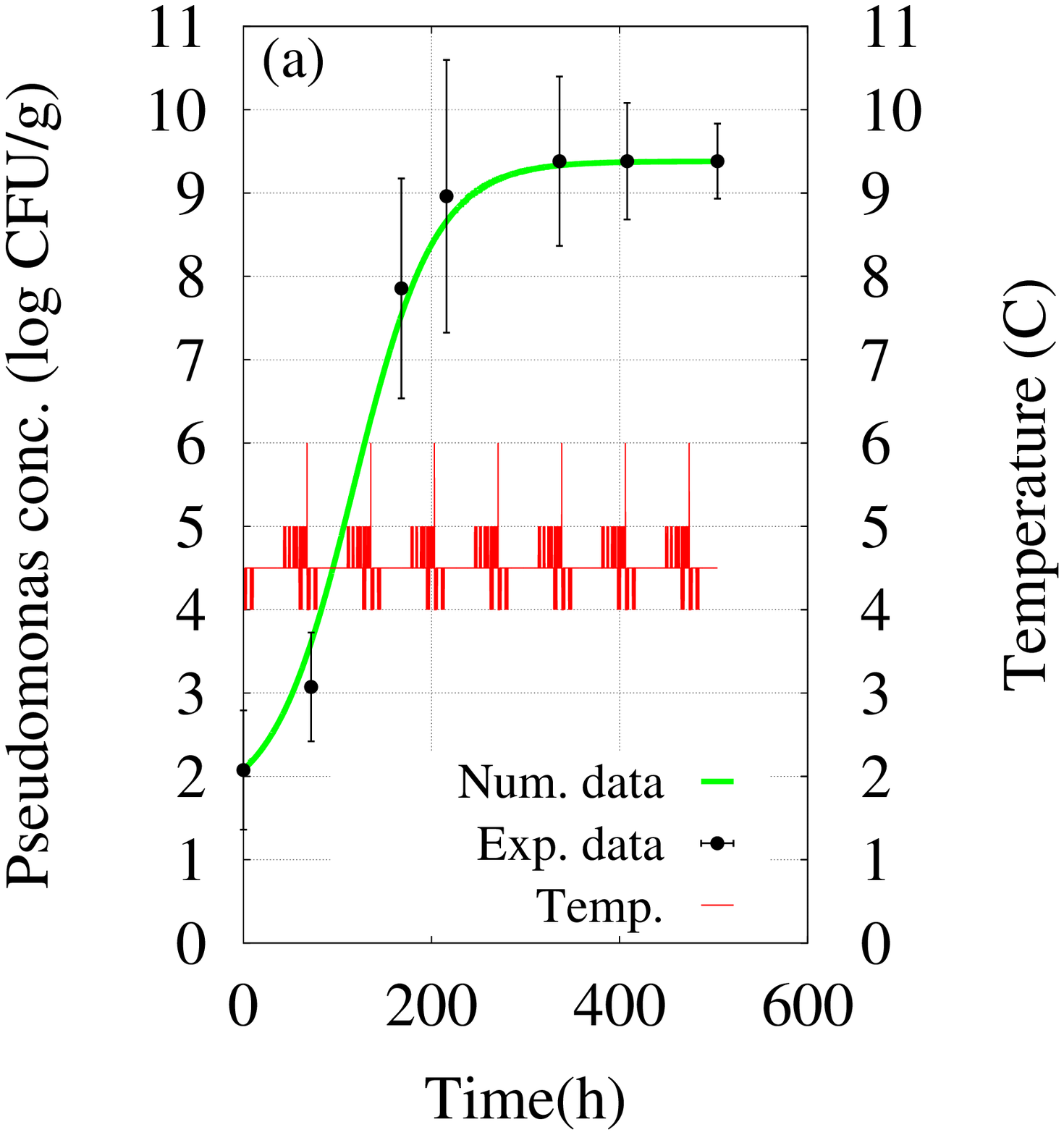}
\includegraphics[width=4.0cm]{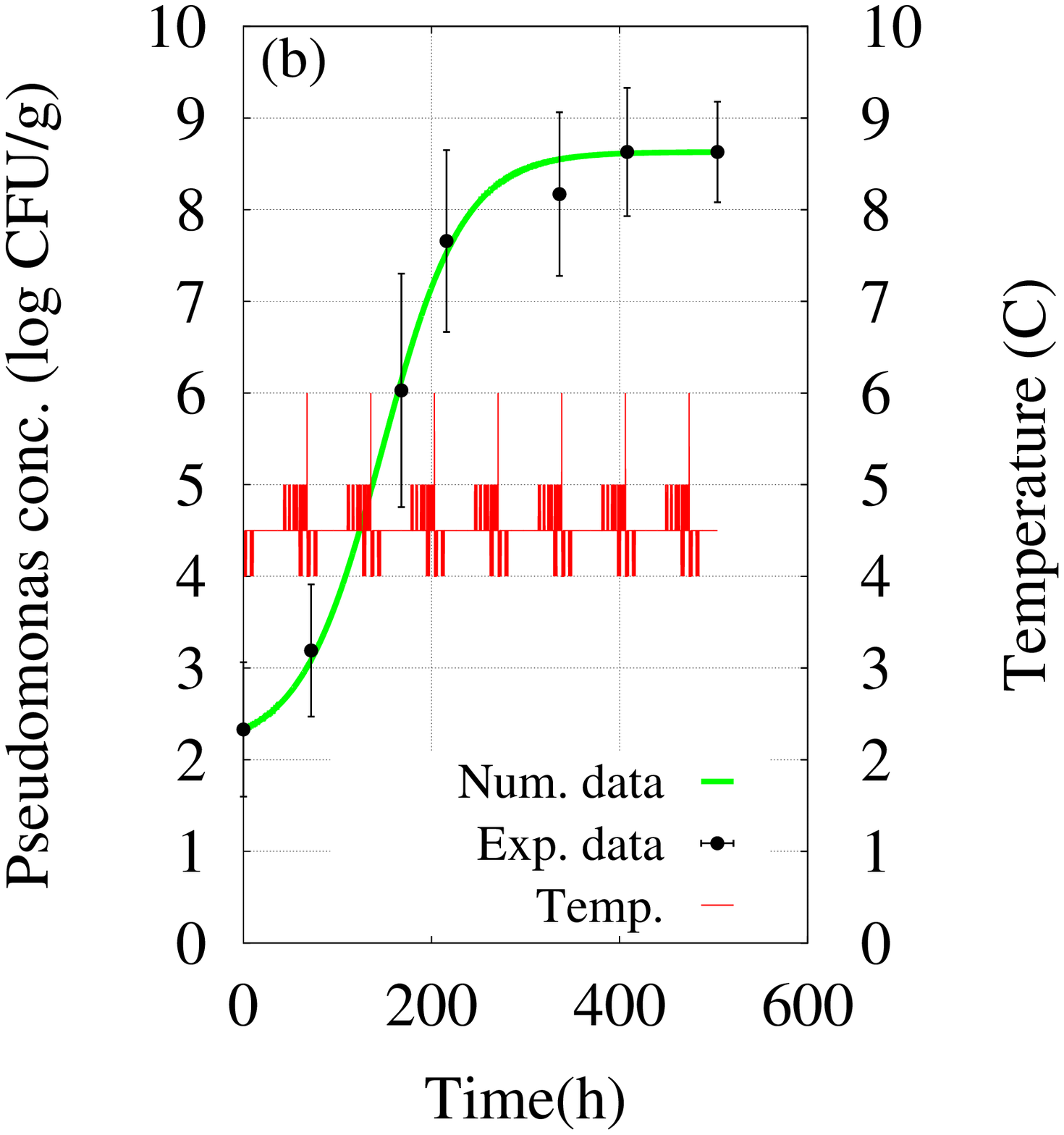}
\includegraphics[width=4.0cm]{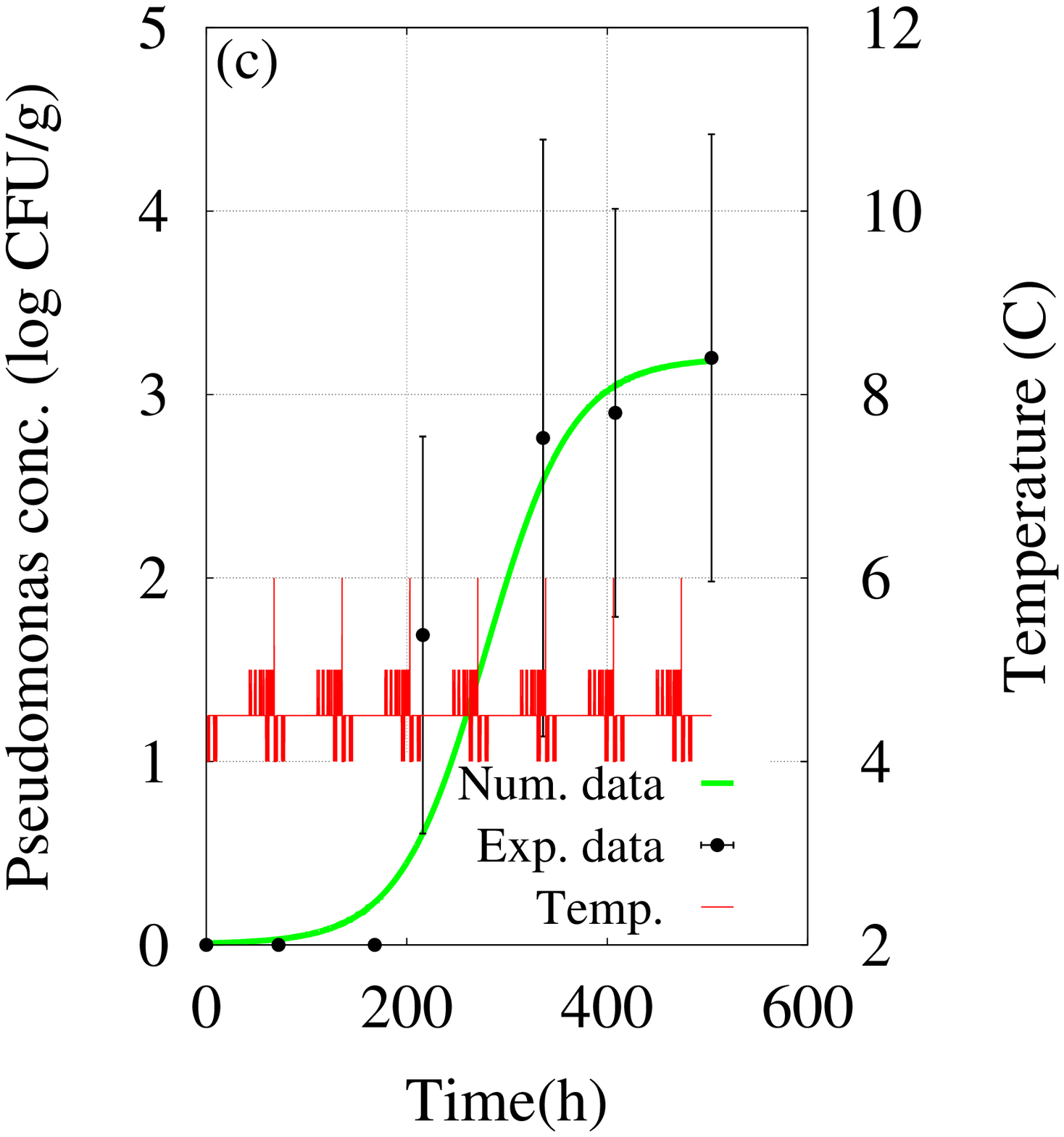}\\
\includegraphics[width=4.0cm]{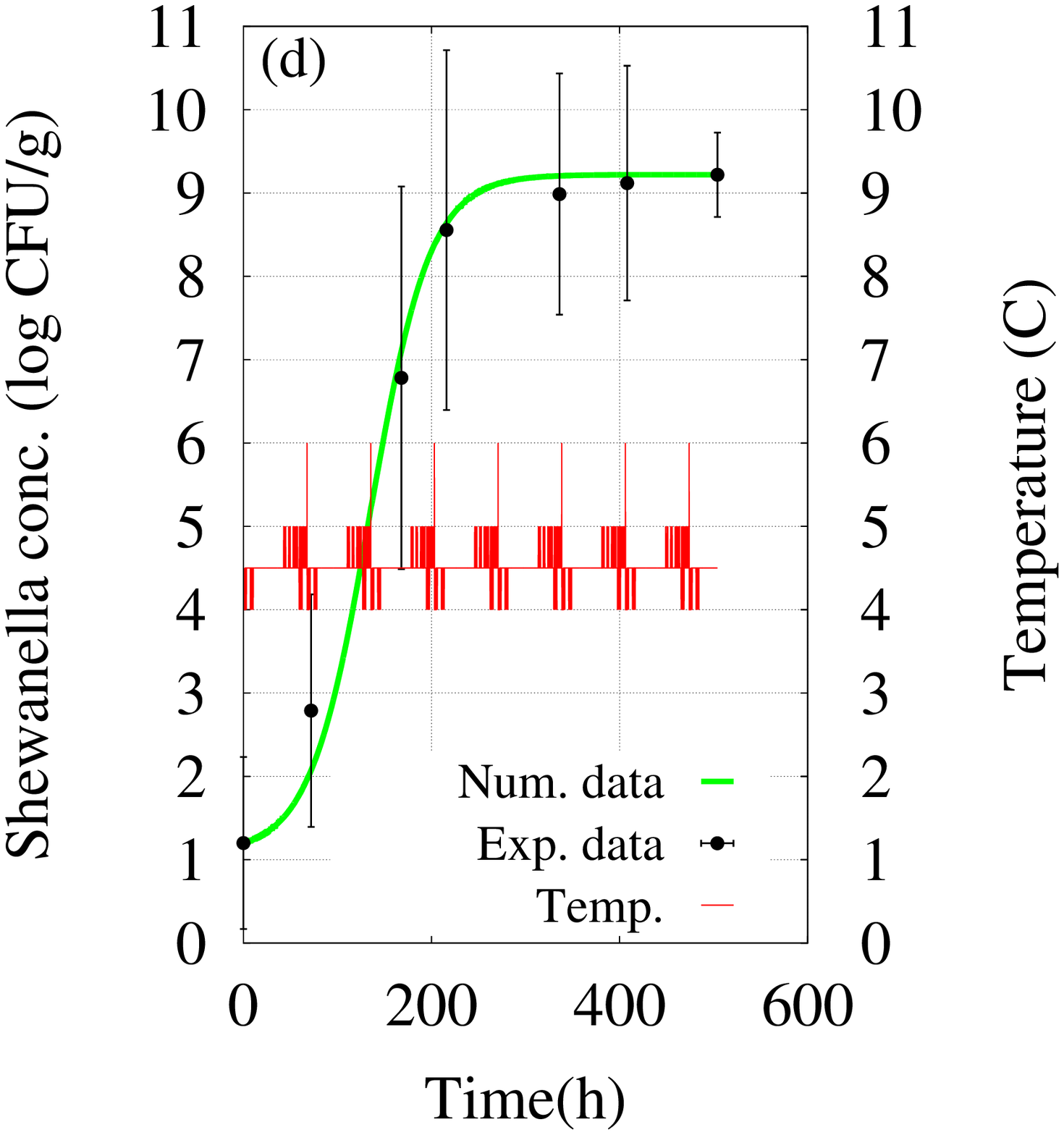}
\includegraphics[width=4.0cm]{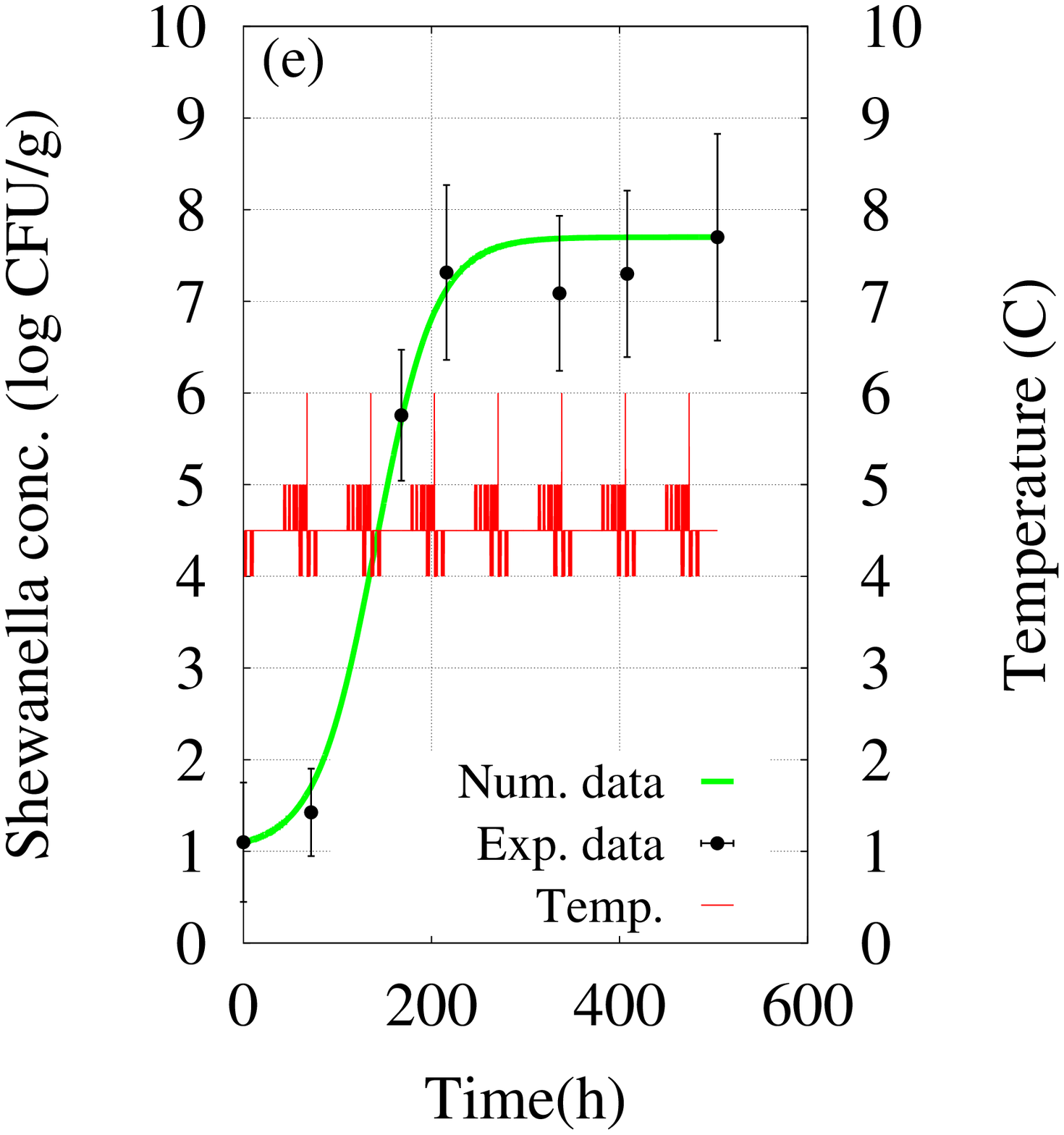}
\includegraphics[width=4.0cm]{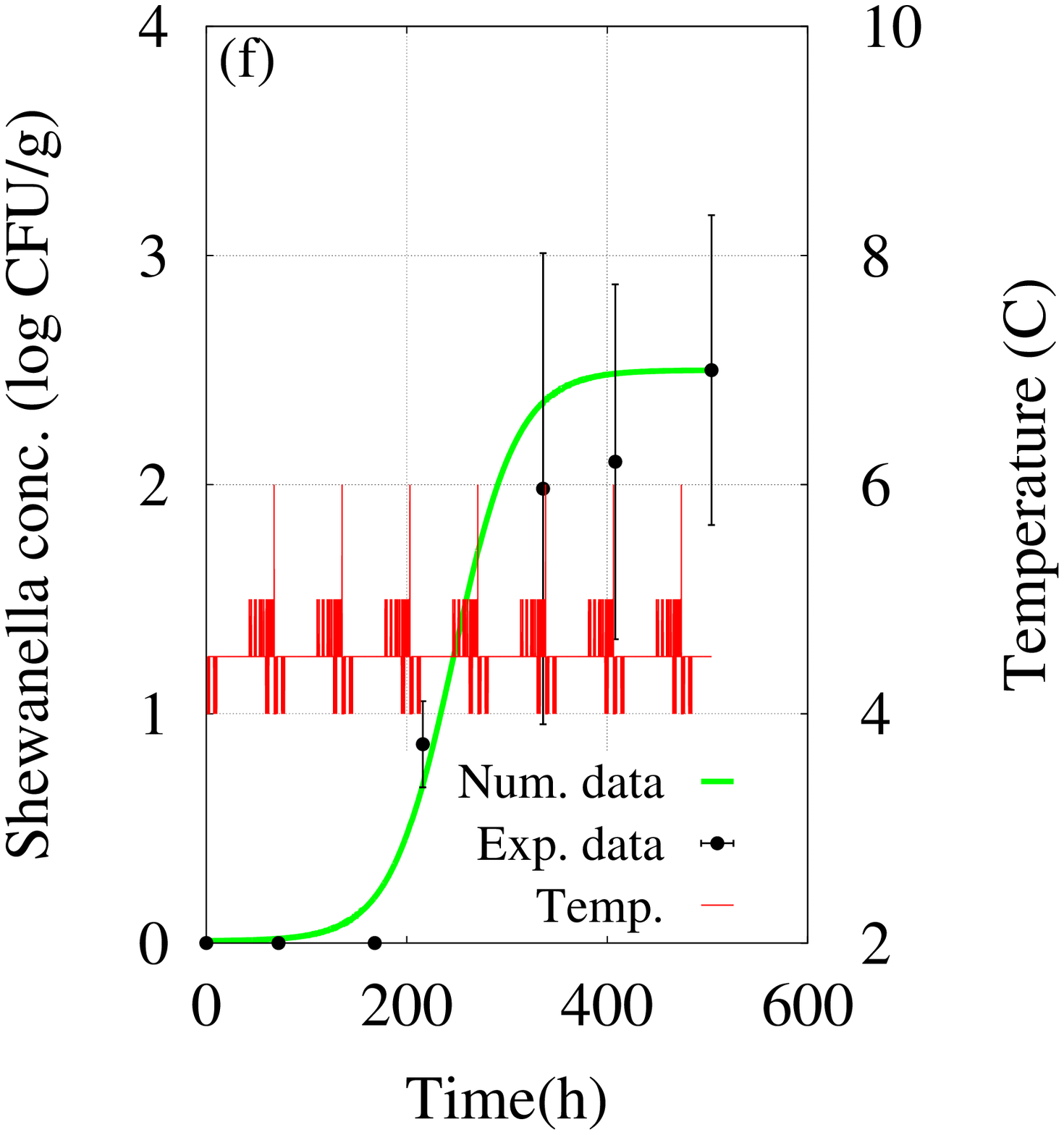}
\end{center}
\vspace{-0.8cm}\caption{\small \emph{Group~2. Comparison between
experimental data (black dots) and theoretical curves (green line)
of bacterial growth: Pseudomonas (white colonies) in panels a
(skin), b (gills), c (flesh); Shewanella (black colonies) in panels
d (skin), e (gills), f (flesh). Vertical bars indicate experimental
errors. Red curves represent the temperature profiles.}}
\vspace{-0.2cm} \label{Bacteria_Group2}
\end{figure}

\begin{figure}[h!]
\begin{center}
\includegraphics[width=4.0cm]{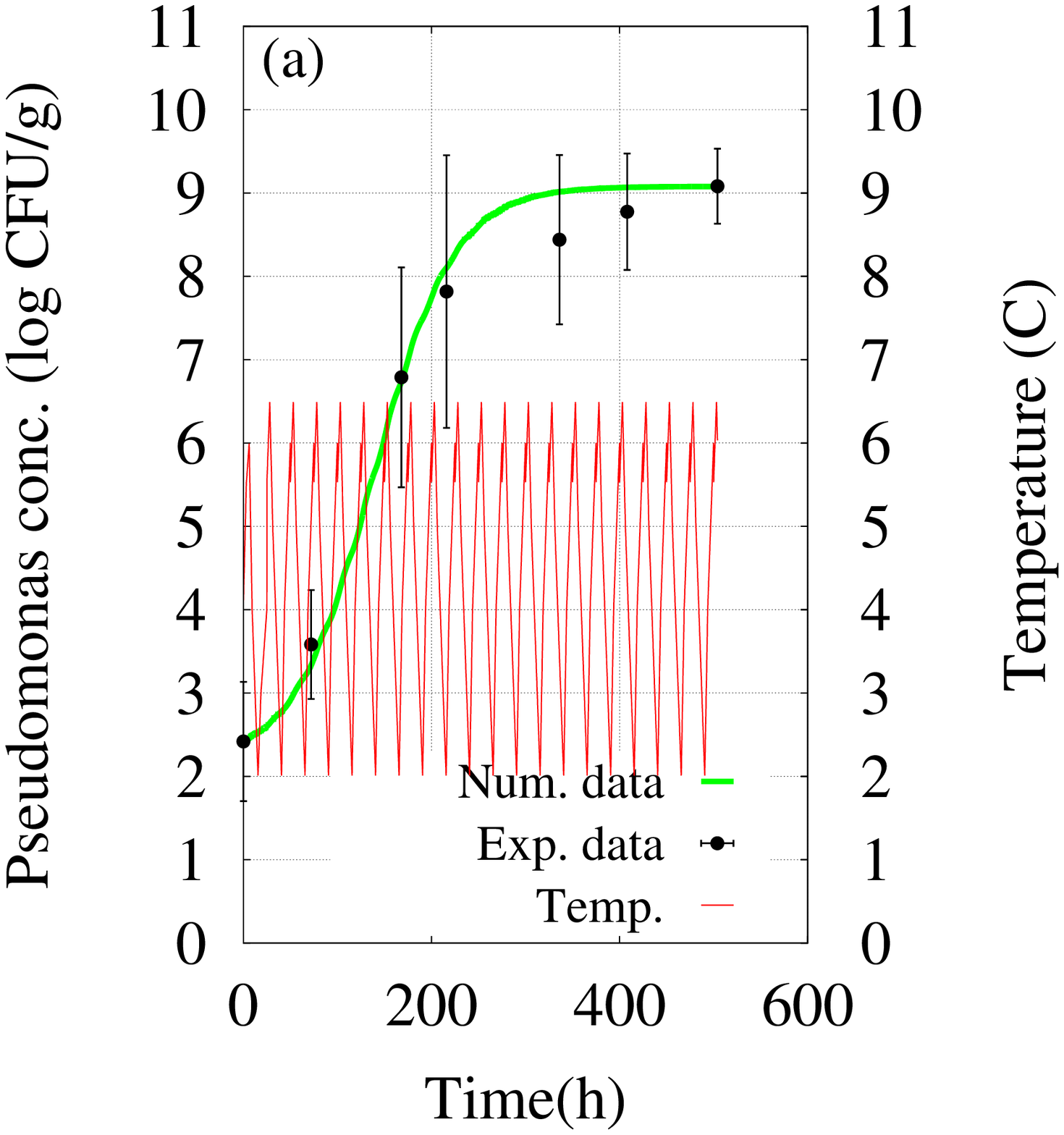}
\includegraphics[width=4.0cm]{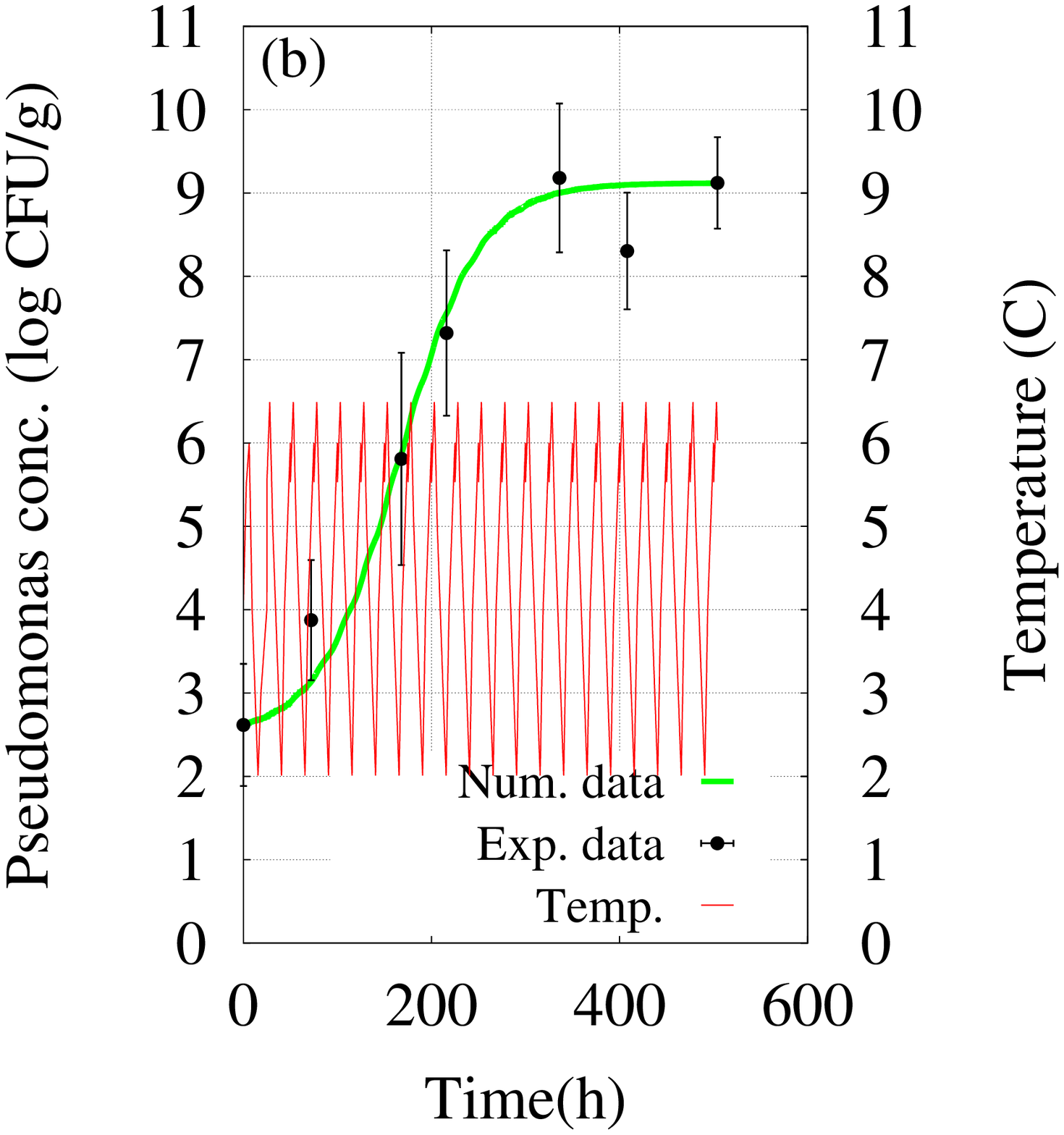}
\includegraphics[width=4.0cm]{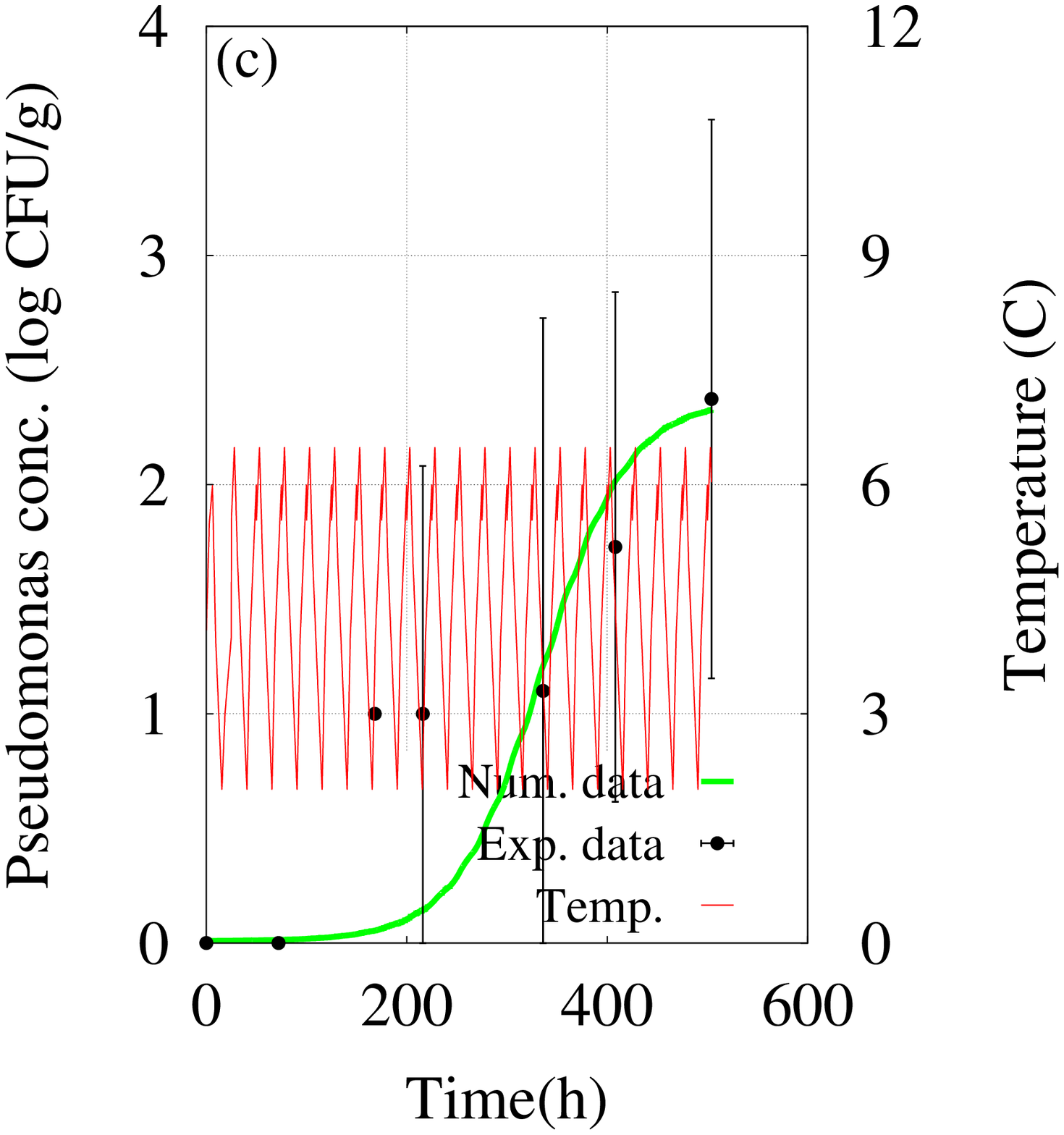}\\
\includegraphics[width=4.0cm]{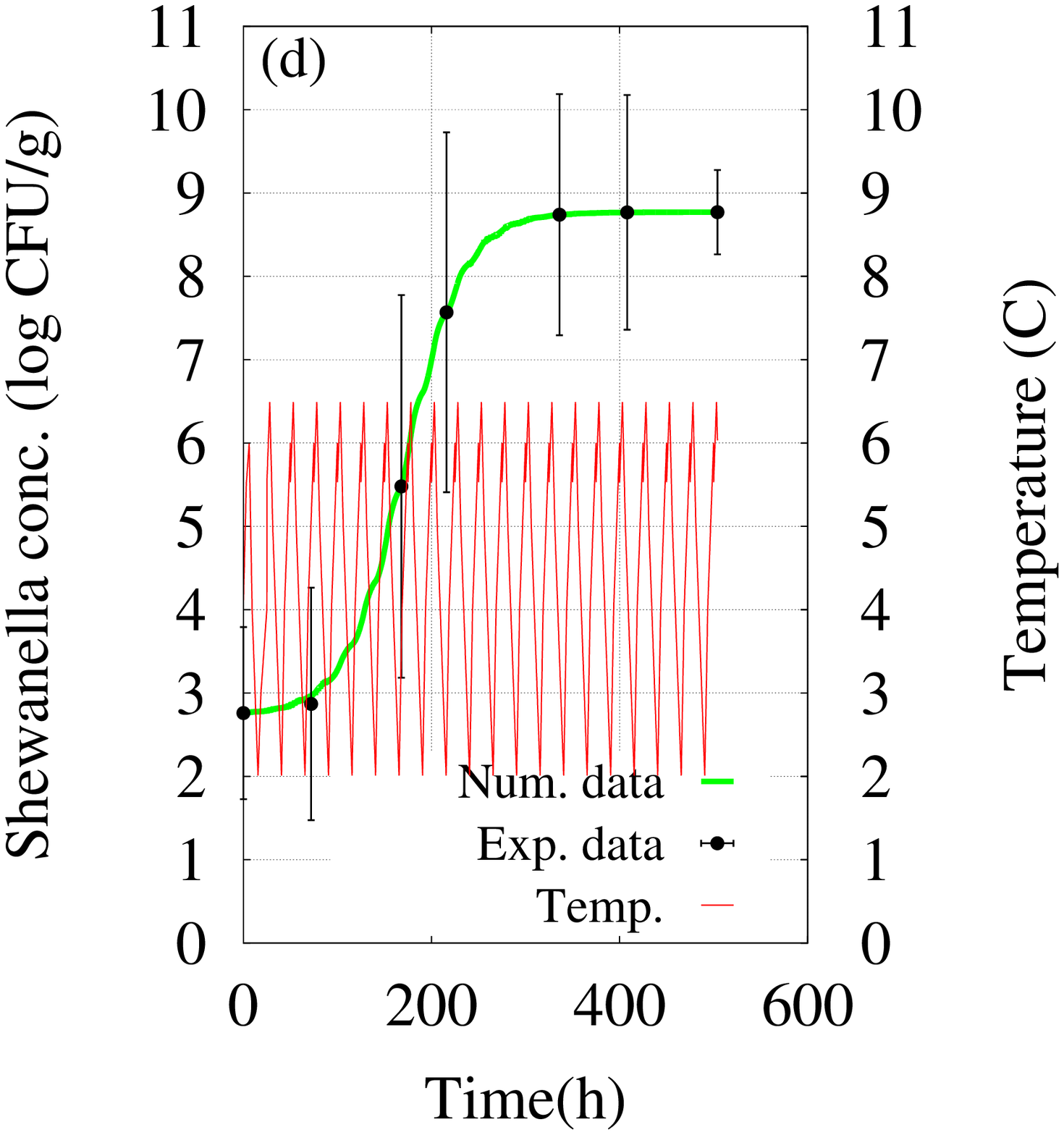}
\includegraphics[width=4.0cm]{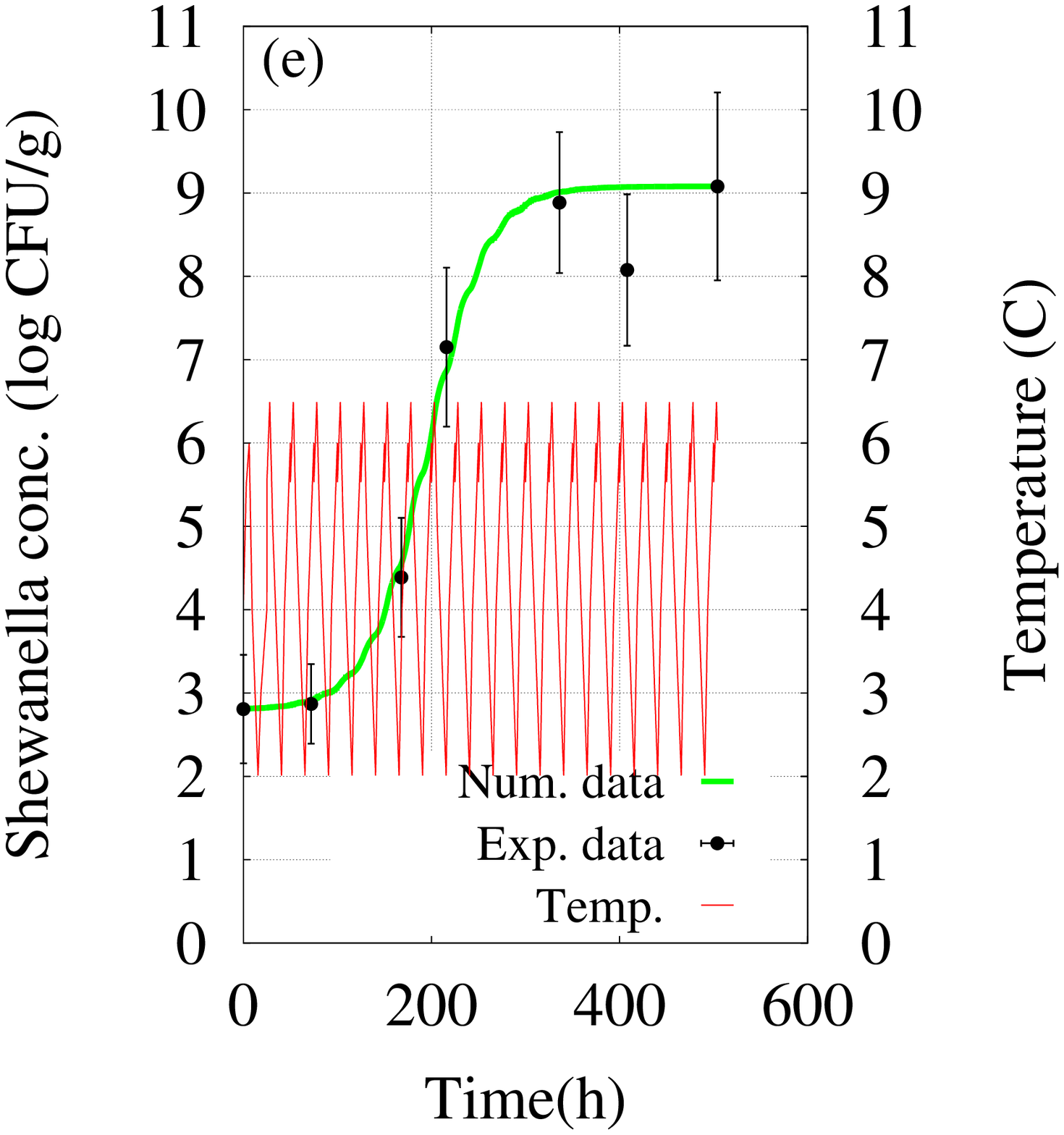}
\includegraphics[width=4.0cm]{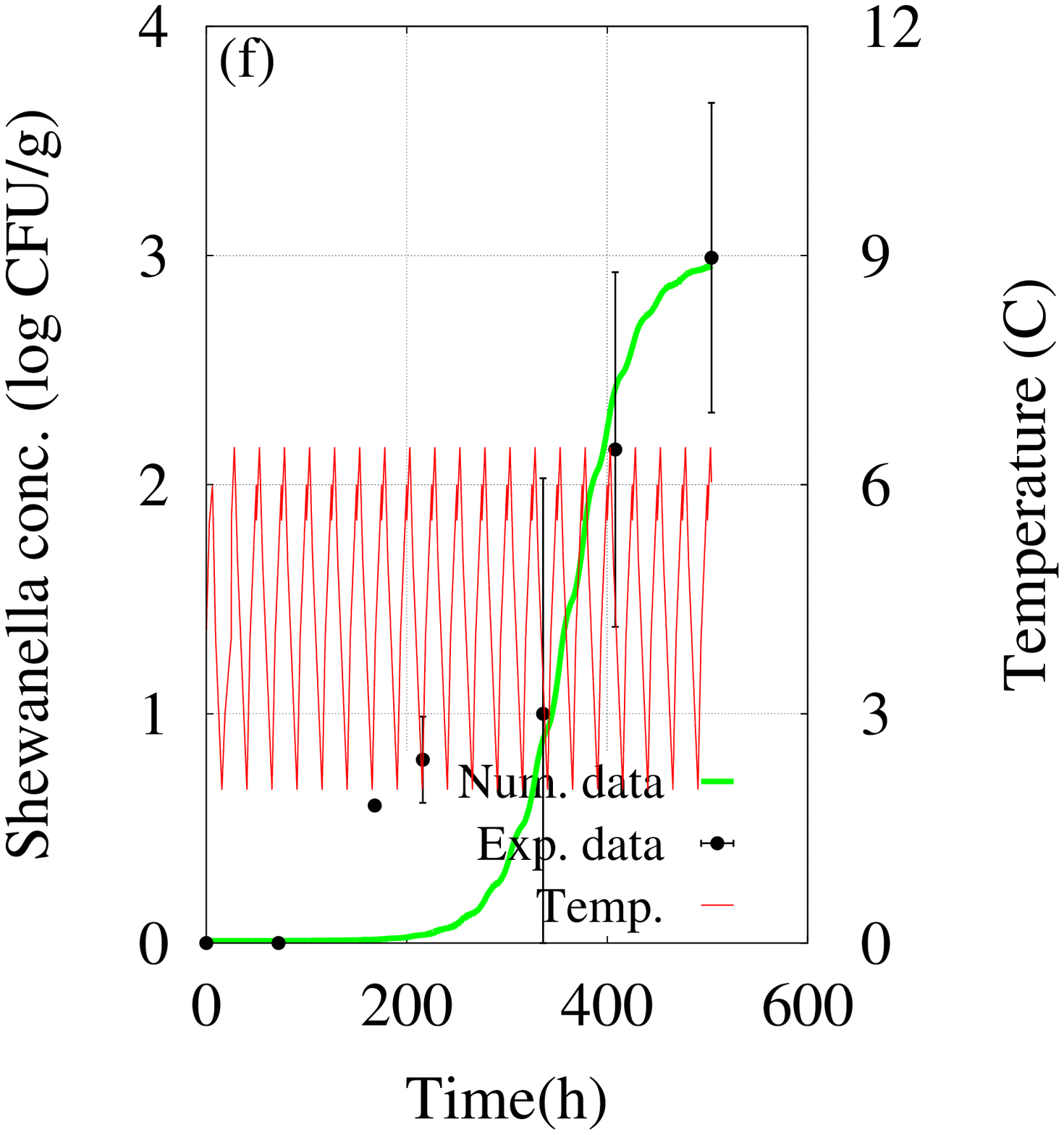}
\end{center}
\vspace{-0.8cm}\caption{\small \emph{Group~3. Comparison between
experimental data (black dots) and theoretical curves (green line)
of bacterial growth: Pseudomonas (white colonies) in panels a
(skin), b (gills), c (flesh); Shewanella (black colonies) in panels
d (skin), e (gills), f (flesh). Vertical bars indicate experimental
errors. Red curves represent the temperature profiles.}}
\vspace{-0.2cm} \label{Bacteria_Group3}
\end{figure}

\begin{figure}[h!]
\begin{center}
\includegraphics[width=4.0cm]{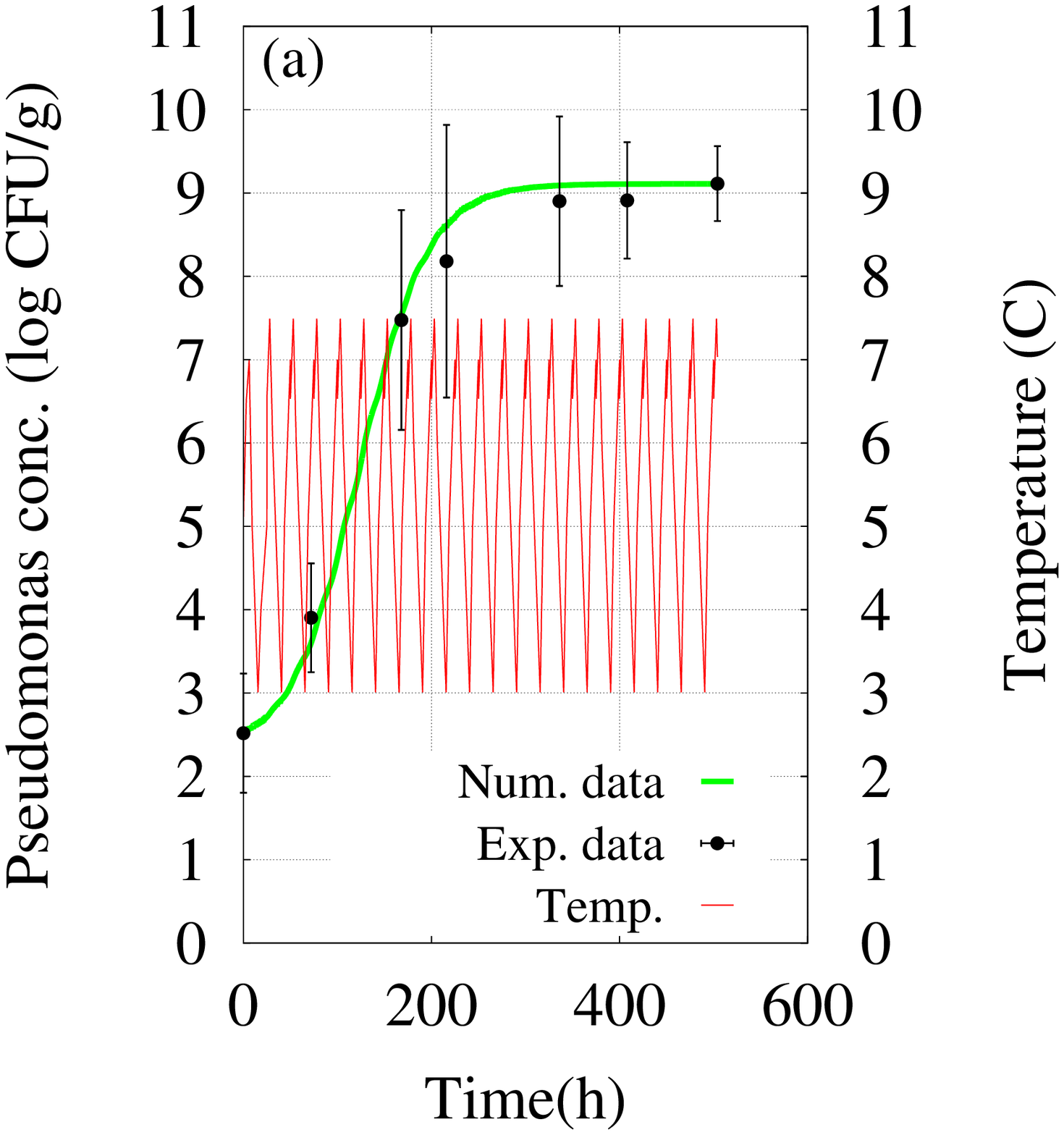}
\includegraphics[width=4.0cm]{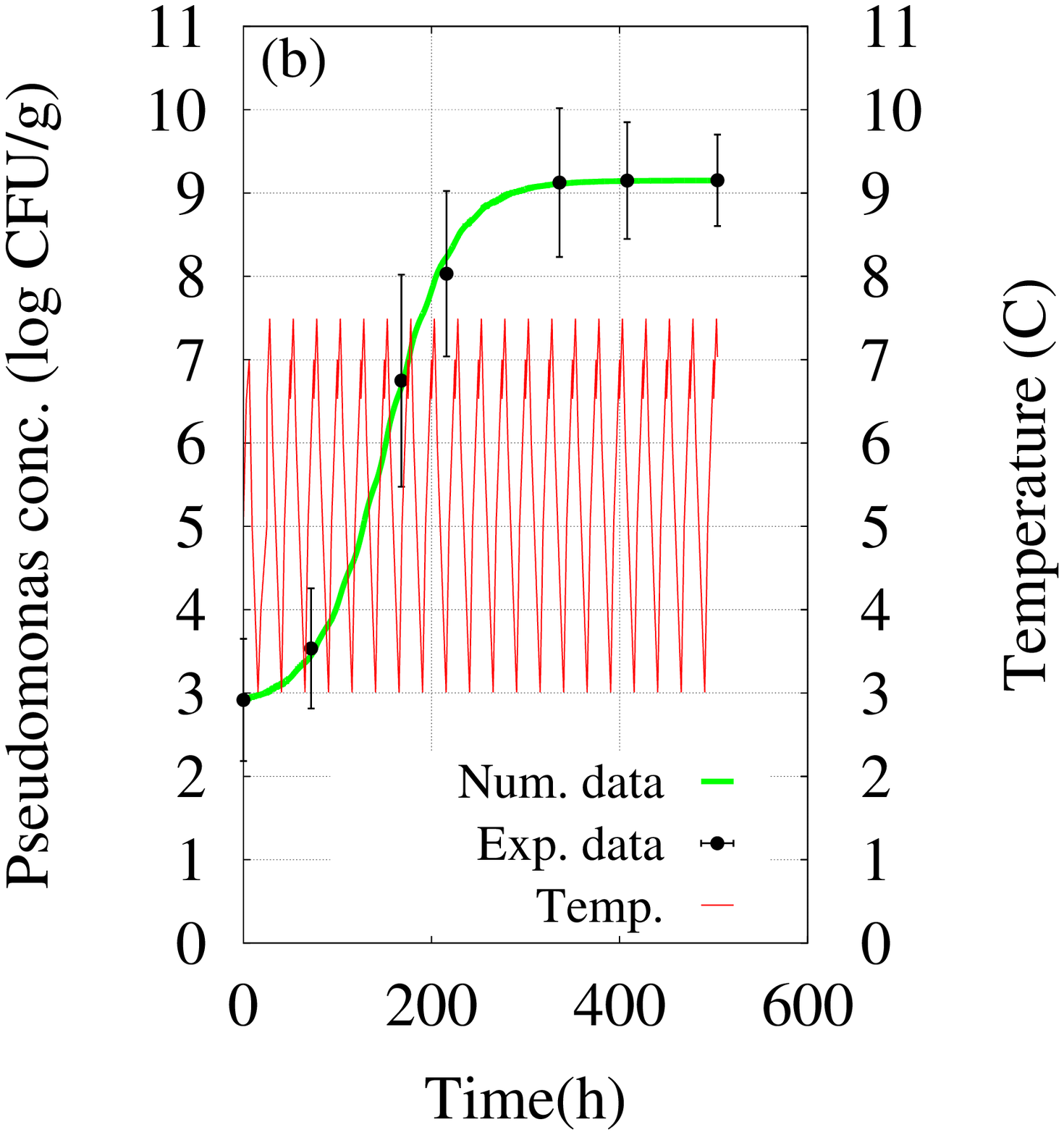}
\includegraphics[width=4.0cm]{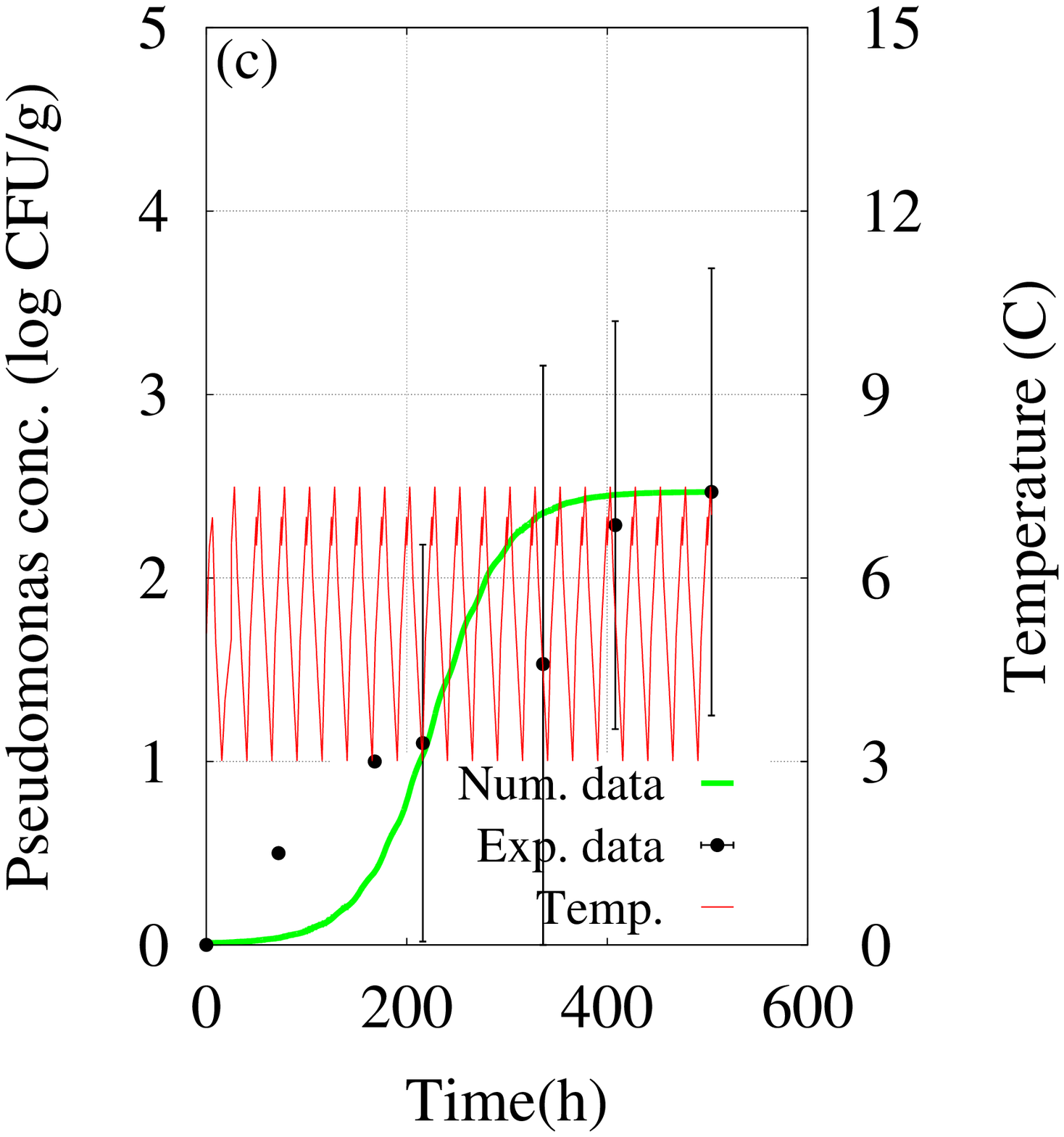}\\
\includegraphics[width=4.0cm]{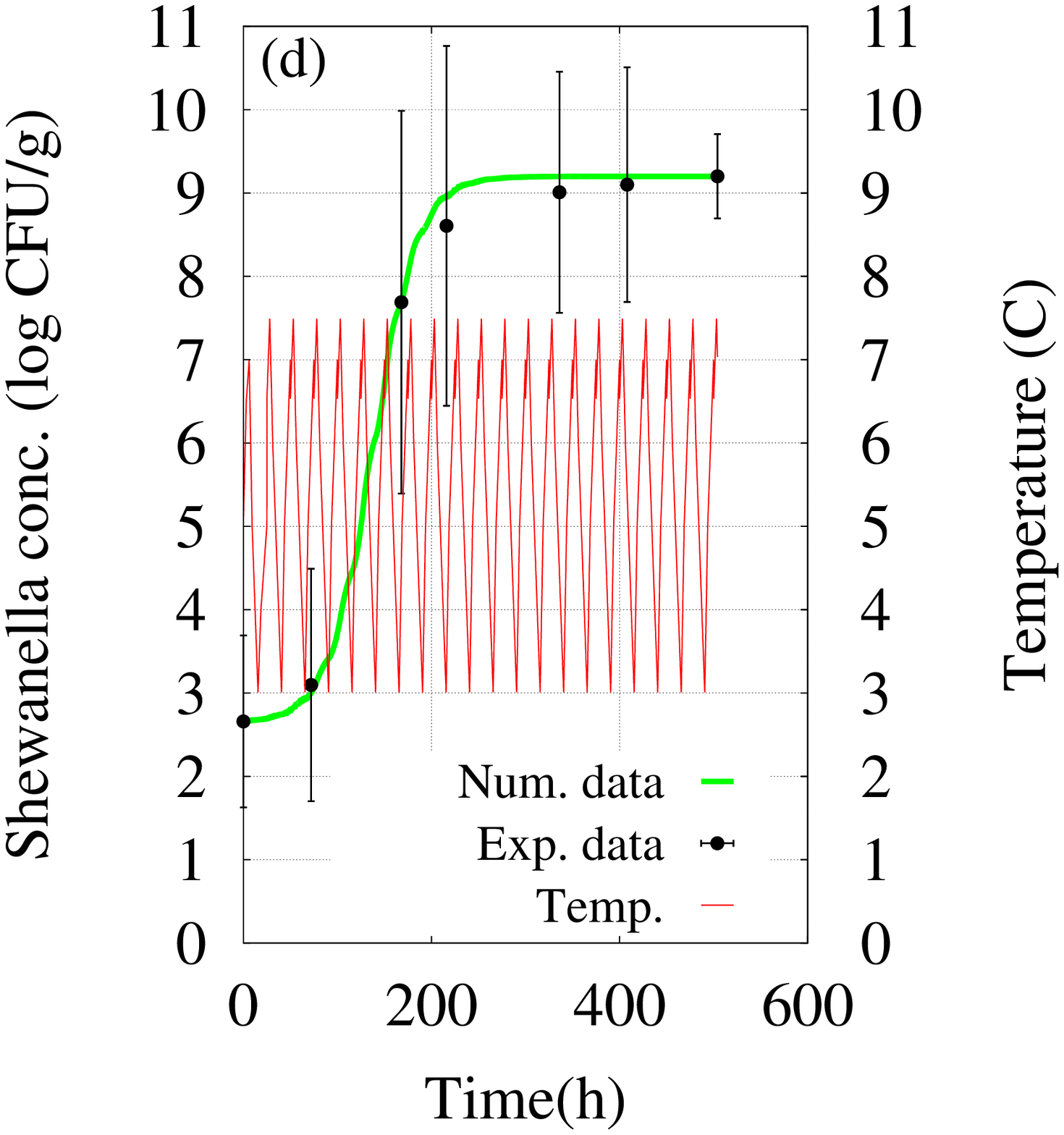}
\includegraphics[width=4.0cm]{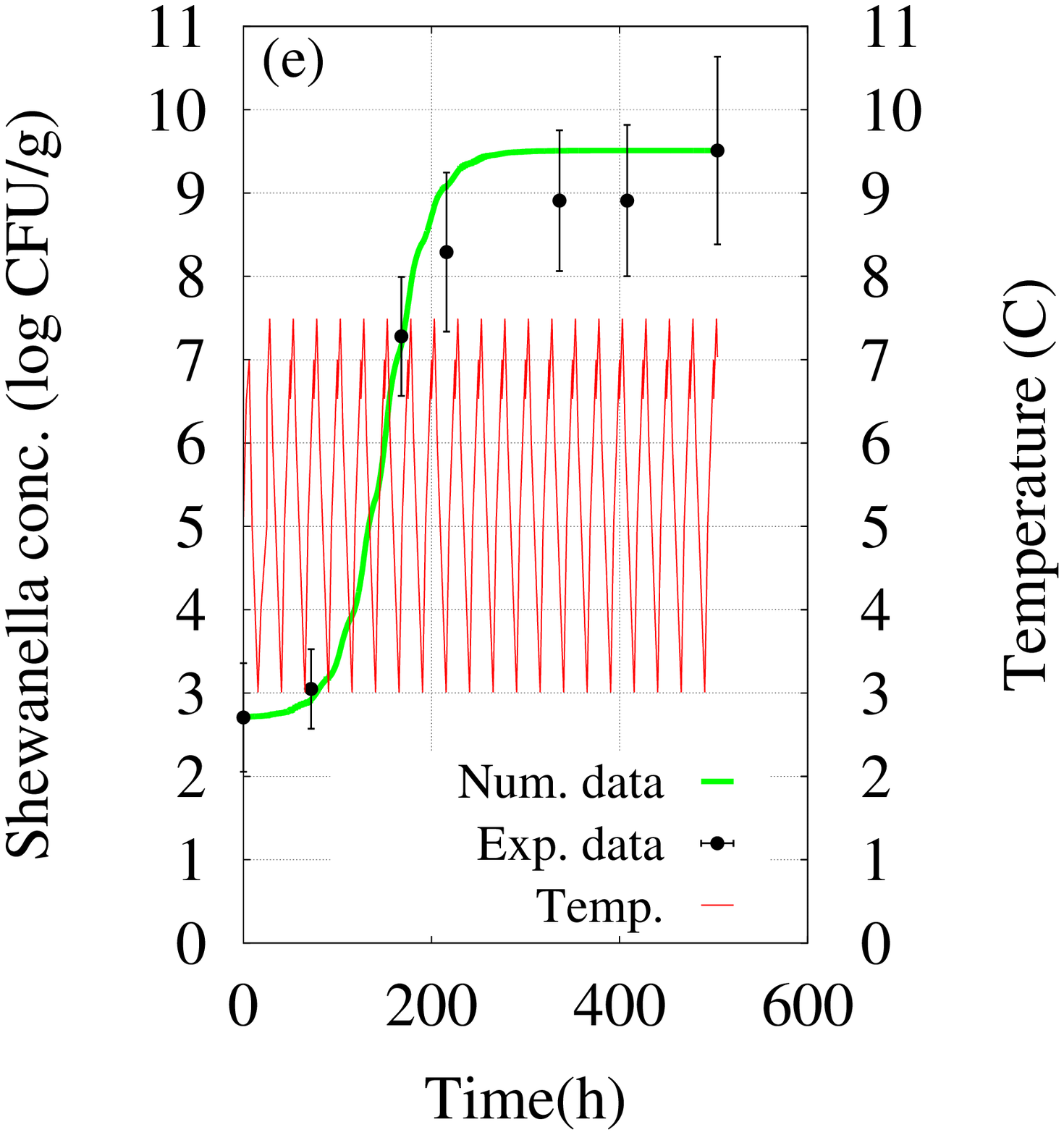}
\includegraphics[width=4.0cm]{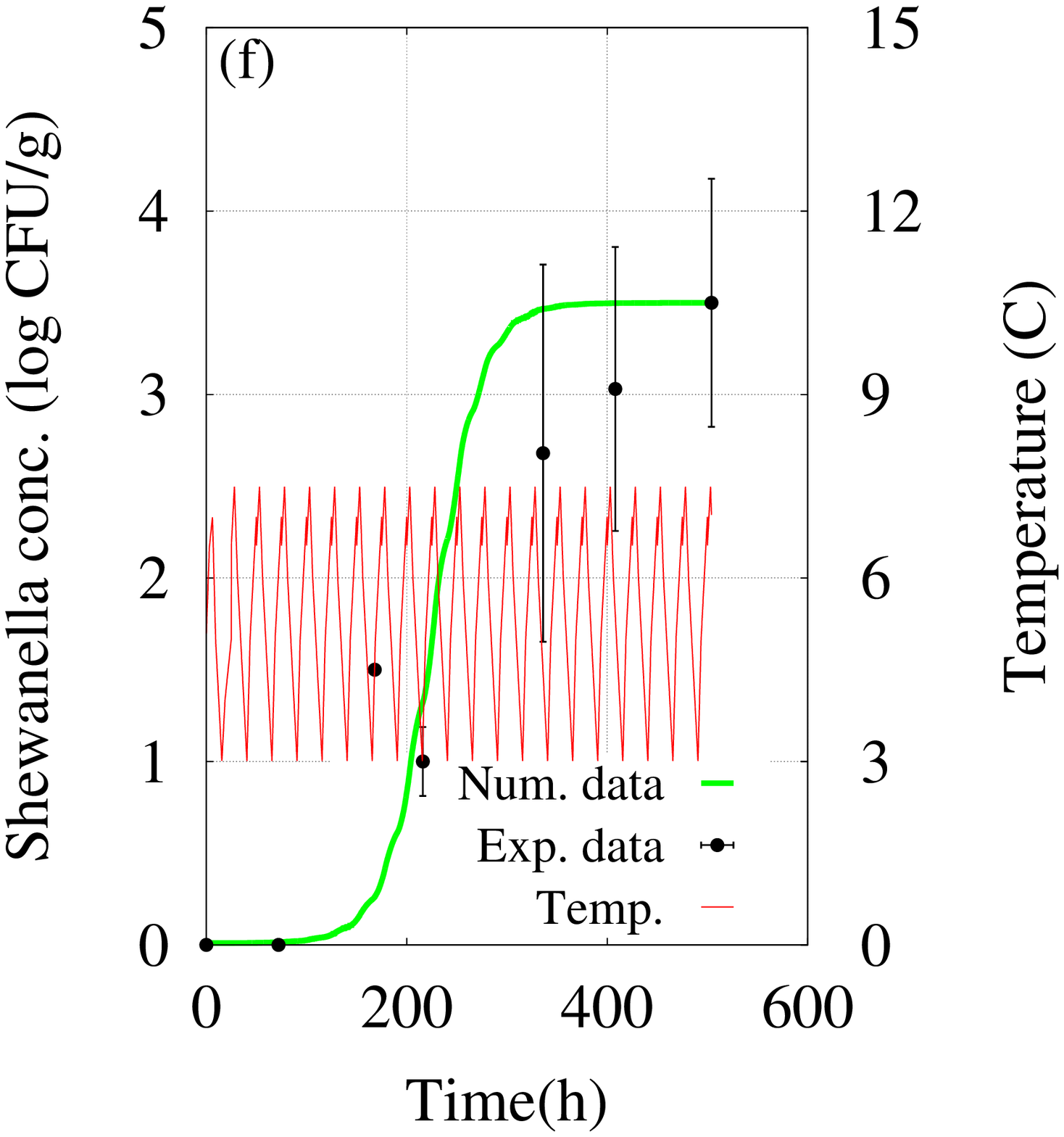}
\end{center}
\vspace{-0.8cm}\caption{\small \emph{Group~4. Comparison between
experimental data (black dots) and theoretical curves (green line)
of bacterial growth: Pseudomonas (white colonies) in panels a
(skin), b (gills), c (flesh); Shewanella (black colonies) in panels
d (skin), e (gills), f (flesh). Vertical bars indicate experimental
errors. Red curves represent the temperature profiles.}}
\vspace{-0.2cm} \label{Bacteria_Group4}
\end{figure}

As a first step, we solved Eqs.~(\ref{Nwi})-(\ref{Qbi}) by using: i)
secondary models given by Eqs.~(\ref{mu_max_w}),~(\ref{mu_max_b}),
where $pH$ is set at a constant value ($pH$=7.0); ii) initial
values, $N^0_{wi}$ and $N^0_{bi}$, and theoretical maximum
population concentrations, $N^{max}_{wi}$ and $N^{max}_{bi}$,
obtained by experimental data. Moreover, to fix $Q^0_{wi}$ and
$Q^0_{bi}$ (initial values of the physiological states) we used a
fitting procedure based on the minimization of the distance between
experimental and theoretical curves for bacterial concentrations,
i.e. the minimum of the root mean square error (RMSE). By this way,
we obtained the theoretical curves for the two bacterial
concentrations in the three different sites (skin, gills,
flesh).\\
We note that this procedure was performed by using separately
bacterial concentration data from Group~1, Group~2, Group~3, and
Group~4. As a result, for each site and each population, we obtained
one experimental growth curve as an average over the replicated
trials carried out in each group, and correspondingly one
theoretical curve (green line), which are shown in
Figs.~\ref{Bacteria_Group1}-\ref{Bacteria_Group4}, together with the
profile of temperature (red line) expressed in $^\circ C$.

Here we observe that, in general, the time behaviours of Pseudomonas
(white colonies) and Shewanella (black colonies) are comparable
for all groups.\\
More precisely, the initial concentrations of white and black
colonies, on skin and gills, were rather low ($<$Log 3 CFU
$g^{-1}$), but the flesh maintained a bacterial load even smaller
($<$Log~1~CFU~$g^{-1}$) until the $72^{nd}$ hour.\\
Moreover, the growth of both bacterial populations on skin and gills
was similar in all groups. In particular, in Groups~2,~3~and~4, the
growth was faster, and determined after 168–-216~h a concentration
$<$Log~8~CFU~$g^{-1}$ . Conversely, the bacterial load in the flesh
was characterized by values $<$~Log~4.5~CFU~$g^{-1}$. These results
confirm that skin and gills are the sites where the bacterial growth
occurs more quickly~\cite{Hus95}.

\subsection{Prediction of QI values}\label{SS:4.1}

\begin{figure}
\begin{center}
\includegraphics[width=6.0cm]{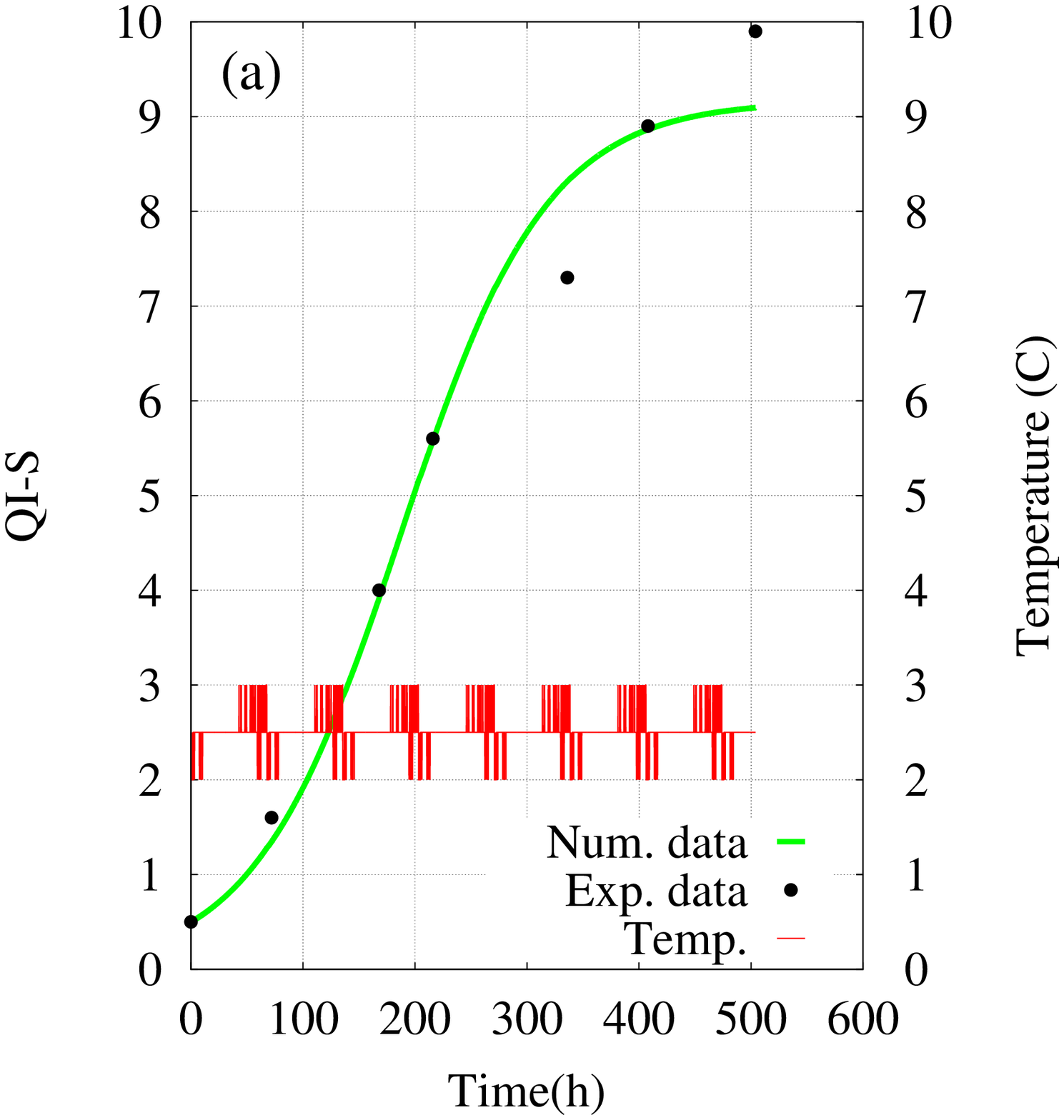}
\includegraphics[width=6.0cm]{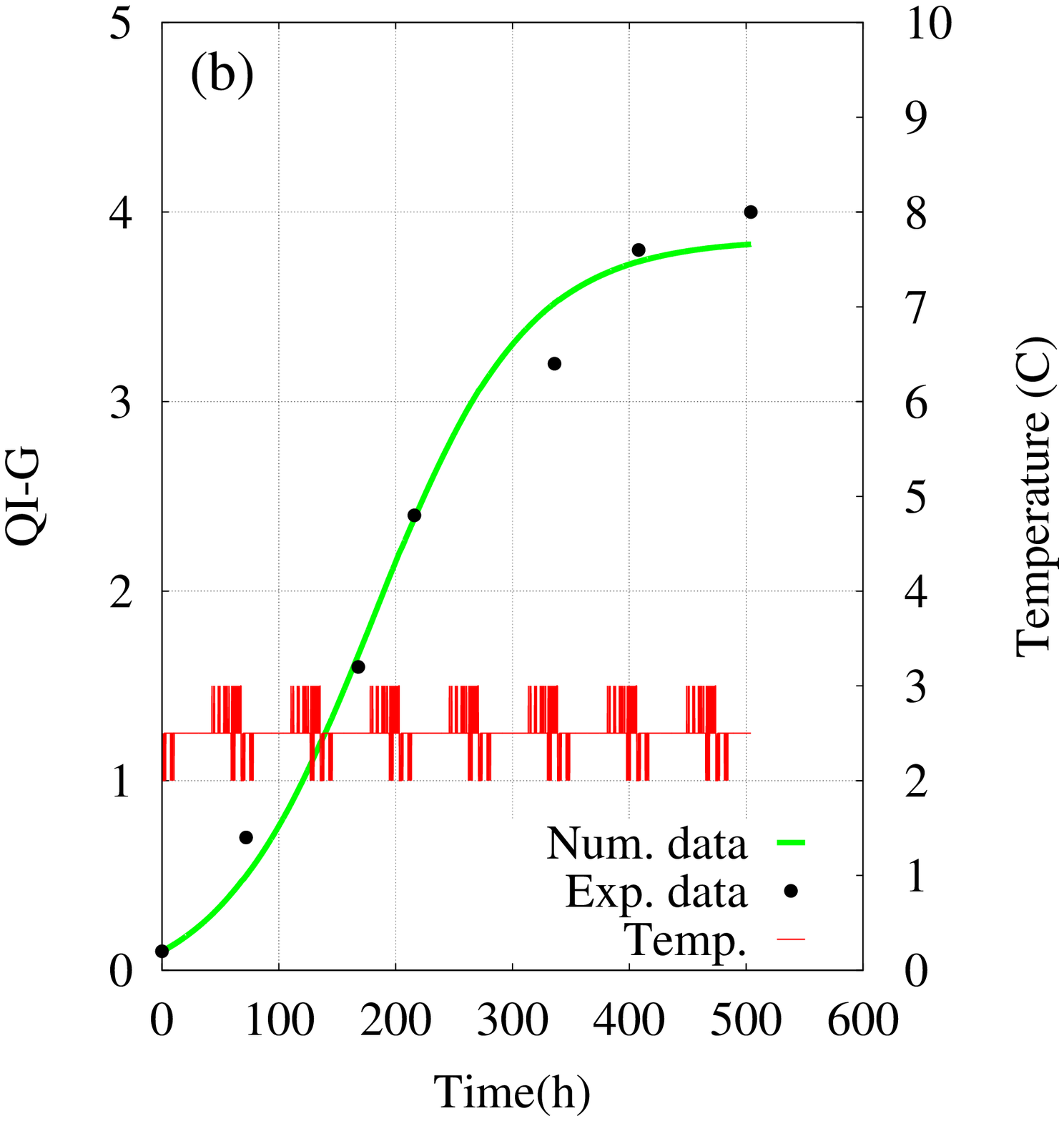}\\
\includegraphics[width=6.0cm]{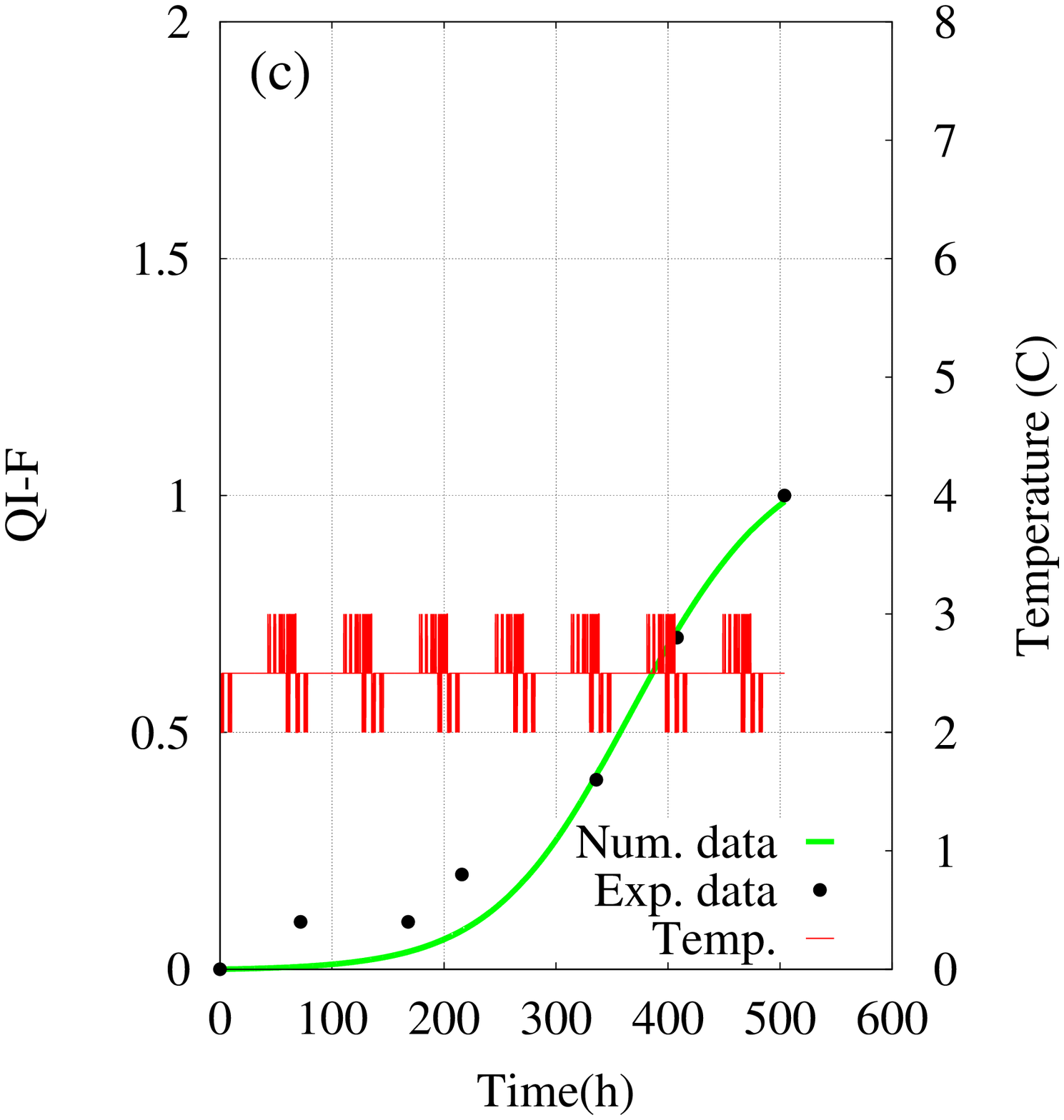}
\includegraphics[width=6.0cm]{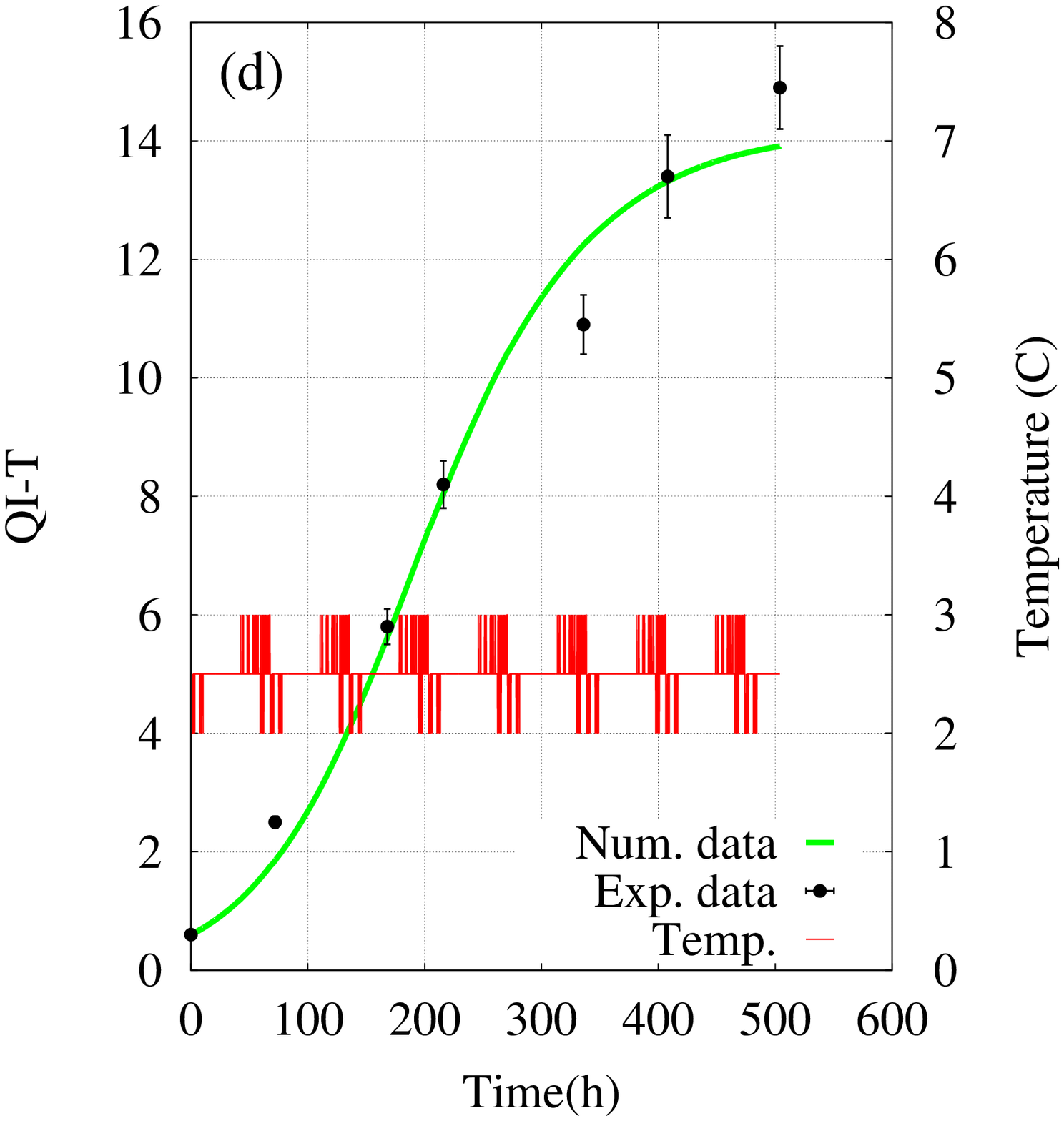}
\end{center}
\vspace{-0.8cm}\caption{\small \emph{Group~1. Comparison between
observed (black dots) and predicted (green line) quality indexes: a)
skin; b) gills; c) flesh. Panel d shows the total QI (summation of
the scores obtained in the three sites). Vertical bars indicate
experimental errors. Theoretical curves were calculated by
Eqs.~(\ref{QI_S})-(\ref{QI_F}), setting
$\beta_{1S}$~=~0.01$~\pm~0.01$, $\beta_{2S}$~=~1.39$~\pm~0.01$,
$\beta_{1G}$~=~0.01$~\pm~0.01$, $\beta_{2G}$~=~0.70$~\pm~0.01$,
$\beta_{1F}$~=~0.33$~\pm~0.01$, $\beta_{2F}$~=~0.16$~\pm~0.01$.
These values were determined by a fitting procedure (minimization of
the RMSE between sensory data and corresponding theoretical values).
Red curves represent the temperature profiles.}} \vspace{-0.2cm}
\label{QIs_Group1}
\end{figure}

As a second step, we focused on data from Group~1, by using
Eqs.~(\ref{QI_S})-(\ref{QI_F}) to convert theoretical bacterial
concentrations of this group (see green curves in
Fig.~\ref{Bacteria_Group1}) into QI values. As initial conditions,
$QI^0_S$, $QI^0_G$, and $QI^0_F$, we used those obtained by
sensory analysis.\\
Moreover, analogously to the procedure adopted to set $Q^0_{wi}$ and
$Q^0_{bi}$, we fixed the values of conversion coefficients
($\beta_{1S}$~=~0.01$~\pm~0.01$, $\beta_{2S}$~=~1.39$~\pm~0.01$,
$\beta_{1G}$~=~0.01$~\pm~0.01$, $\beta_{2G}$~=~0.70$~\pm~0.01$,
$\beta_{1F}$~=~0.33$~\pm~0.01$, $\beta_{2F}$~=~0.16$~\pm~0.01$) by
using a fitting procedure, based on the minimization of the RMSE
between experimental data and theoretical curves of $QI_S(t)$,
$QI_G(t)$, and $QI_F(t)$. By this way, we obtained the theoretical
values of the quality index in the three different sites (skin,
gills, flesh). Fig.~\ref{QIs_Group1} shows curves of  $QI_S(t)$,
$QI_G(t)$, and $QI_F(t)$ obtained by the model (green line),
corresponding experimental values (black dots), and temperature
profiles (red line). Specifically, panels a, b, and c display QIs
for the three different sites, while panel d refers to the total QI
score, defined as summation of the three partial quality indexes:
$QI(t)$=$QI_S(t)$+$QI_G(t)$+$QI_F(t)$.

We recall that this procedure was performed by using uniquely data
from Group~1. Accordingly, the values of $\beta_{1i}$, $\beta_{2i}$
($i=S,G,F$) were obtained by performing the best fitting between
predicted and observed QIs scores only for Group~1. We note also
that the observed QIs values used here were obtained as averages
over sensory evaluations performed by three expert persons.
\begin{figure}
\begin{center}
\includegraphics[width=6.0cm]{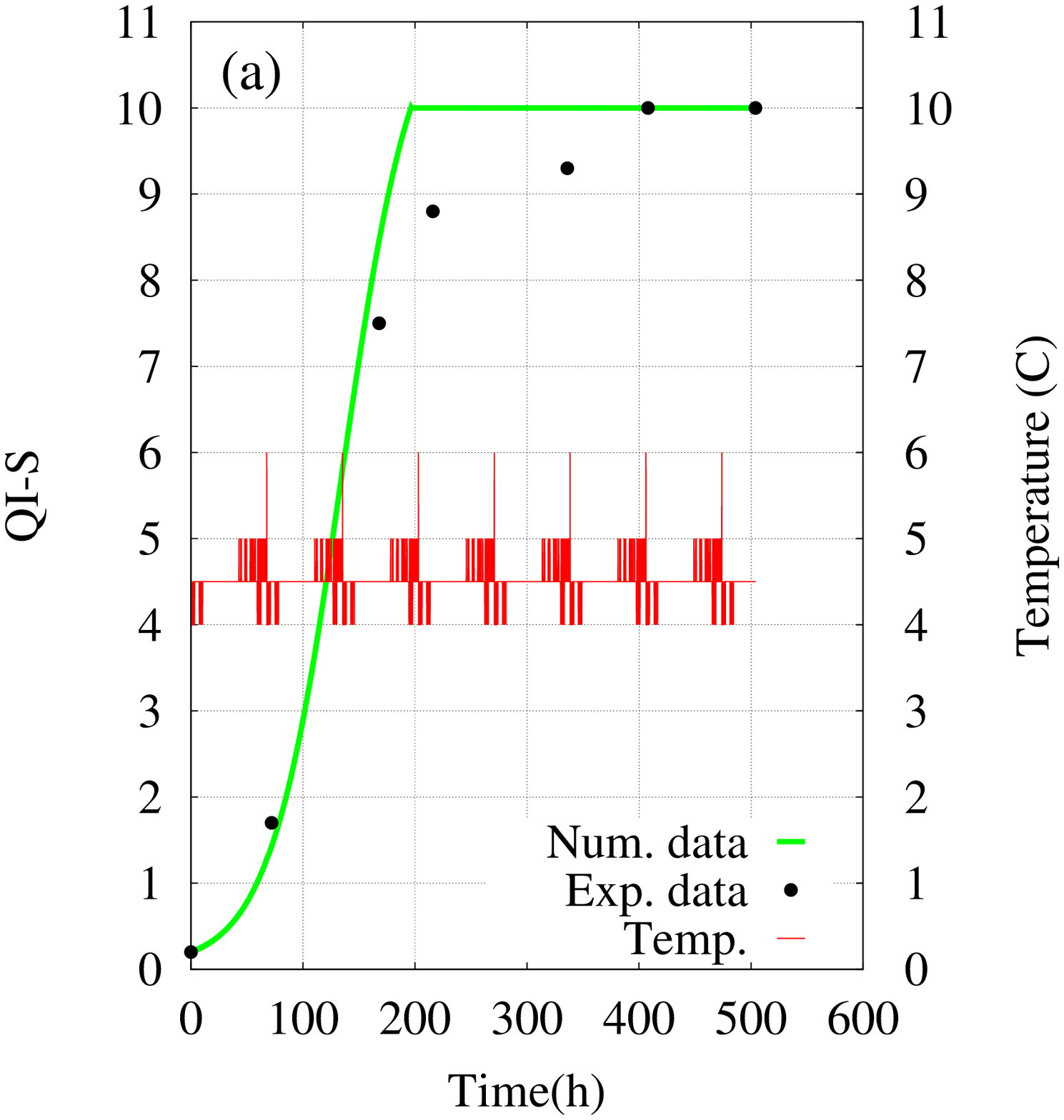}
\includegraphics[width=6.0cm]{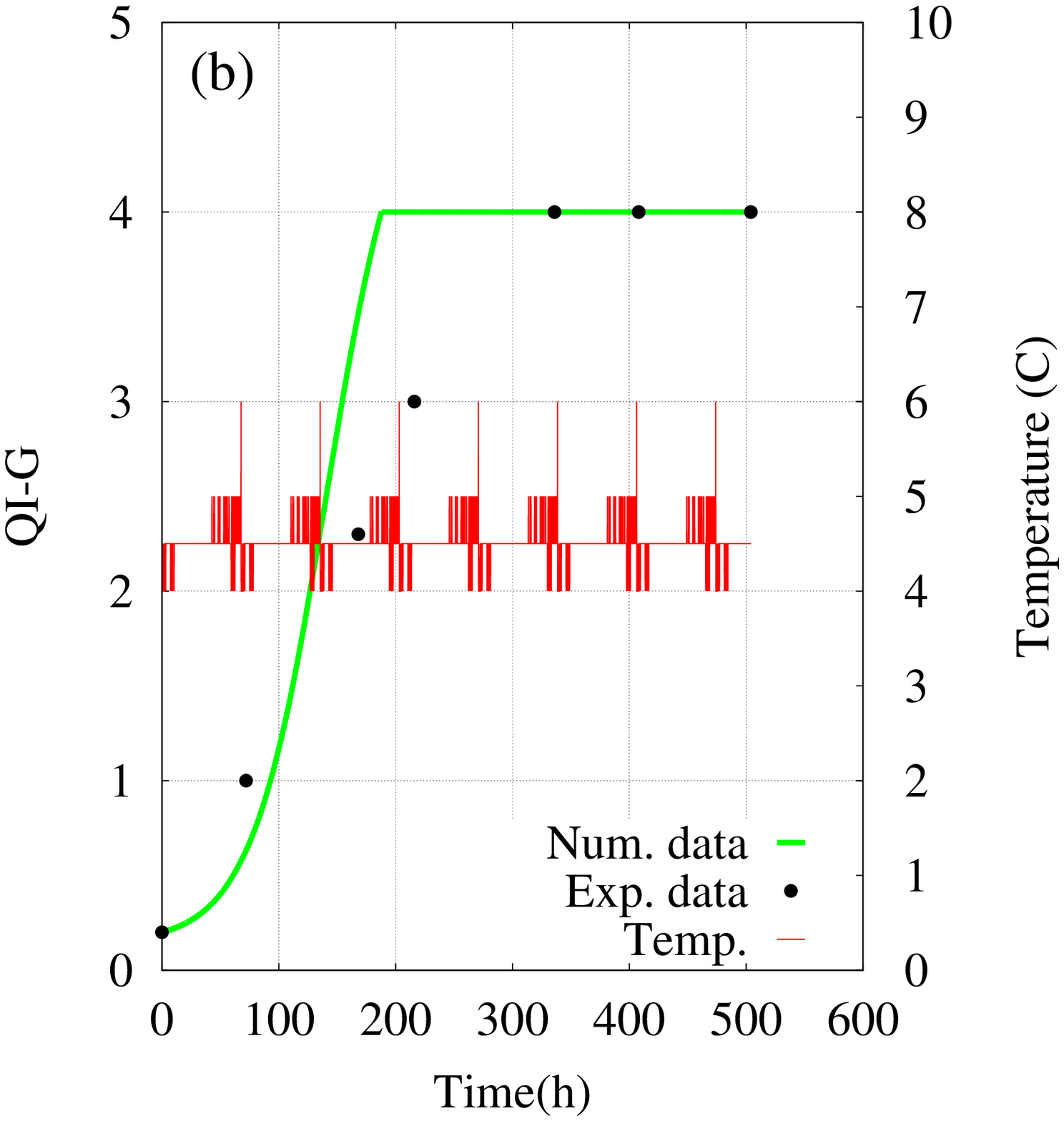}\\
\includegraphics[width=6.0cm]{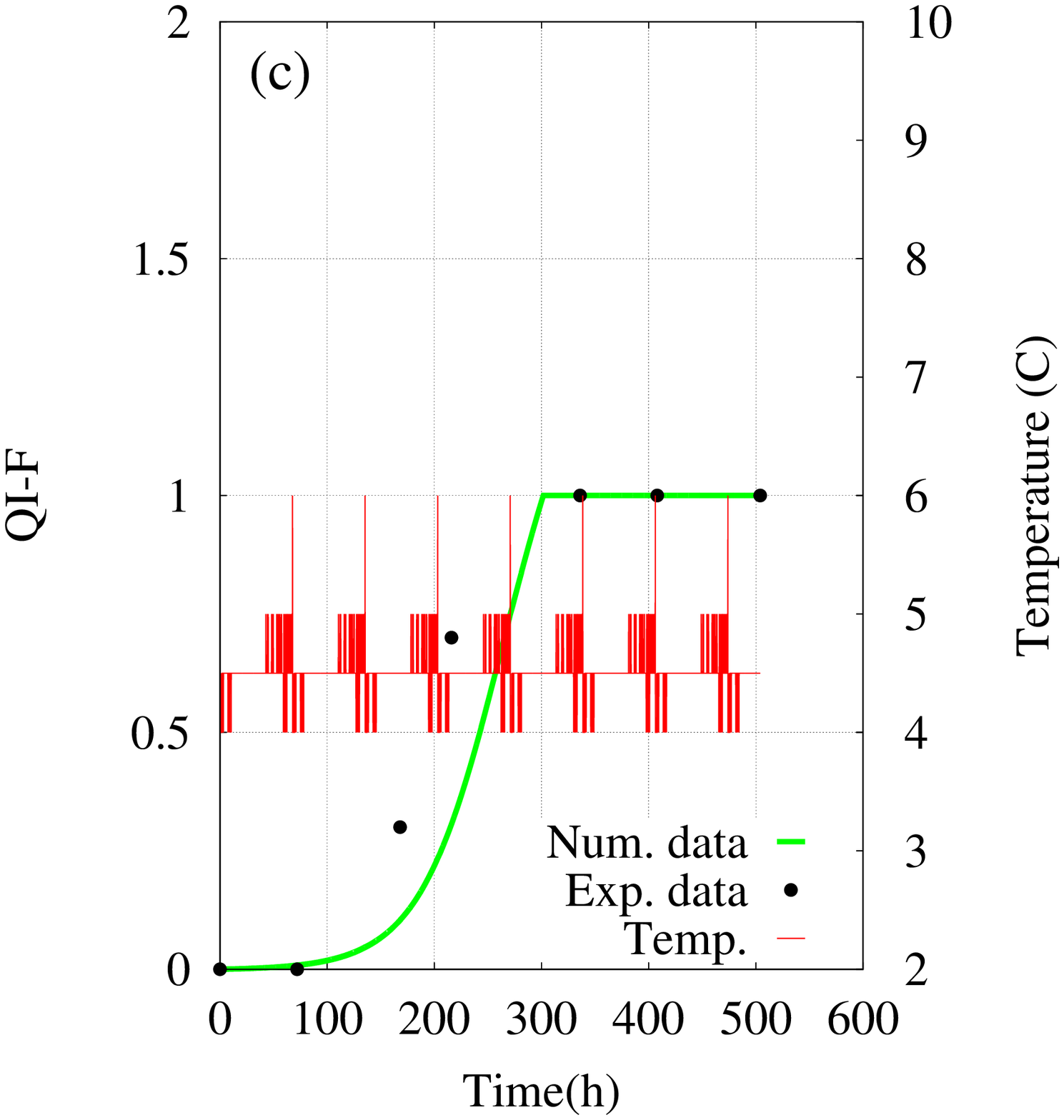}
\includegraphics[width=6.0cm]{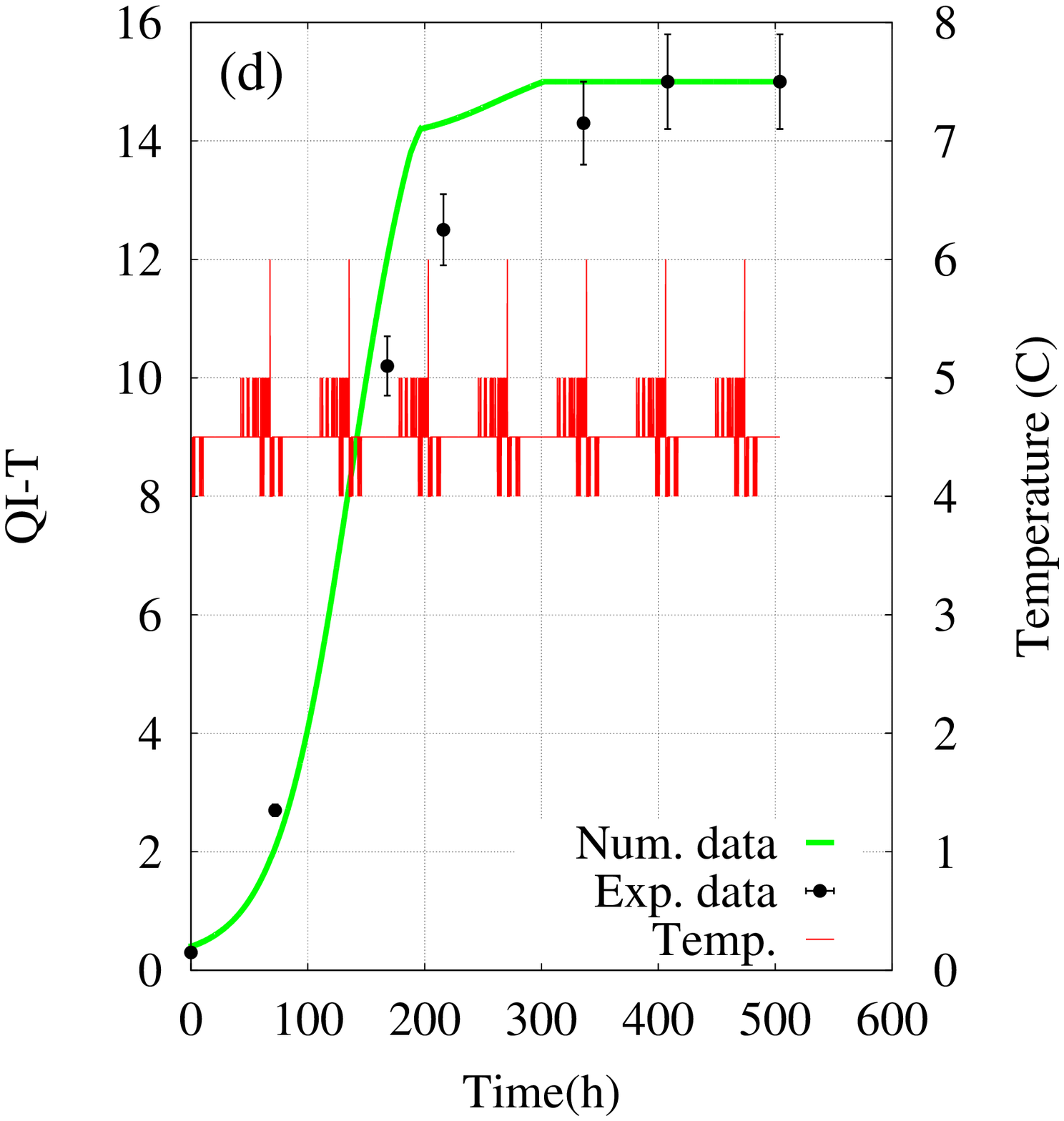}
\end{center}
\vspace{-0.8cm}\caption{\small \emph{Group~2. Comparison between
observed (black dots) and predicted (green line) quality indexes
for: a) skin; b) gills; c) flesh. Panel d shows the total QI
(summation of the scores obtained in the three sites). Vertical bars
indicate experimental errors. Theoretical curves were calculated
from Eqs.~(\ref{QI_S})-(\ref{QI_F}), by using the same values of
$\beta_{1i}$ and $\beta_{2i}$ ($i=S,G,F$) as in
Fig.~\ref{QIs_Group1}. Red curves represent the temperature
profiles.}} \vspace{-0.2cm} \label{QIs_Group2}
\end{figure}
\begin{figure}
\begin{center}
\includegraphics[width=6.0cm]{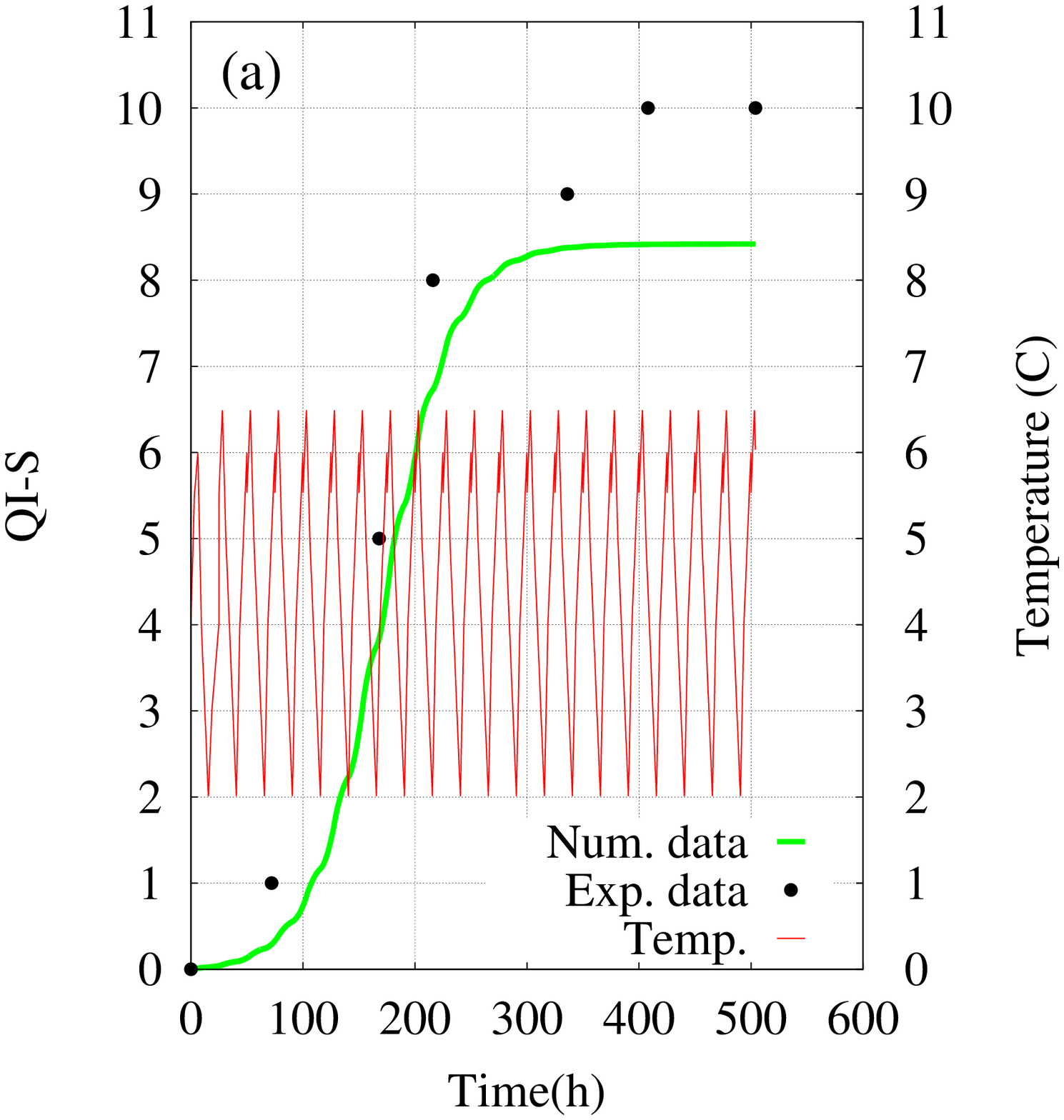}
\includegraphics[width=6.0cm]{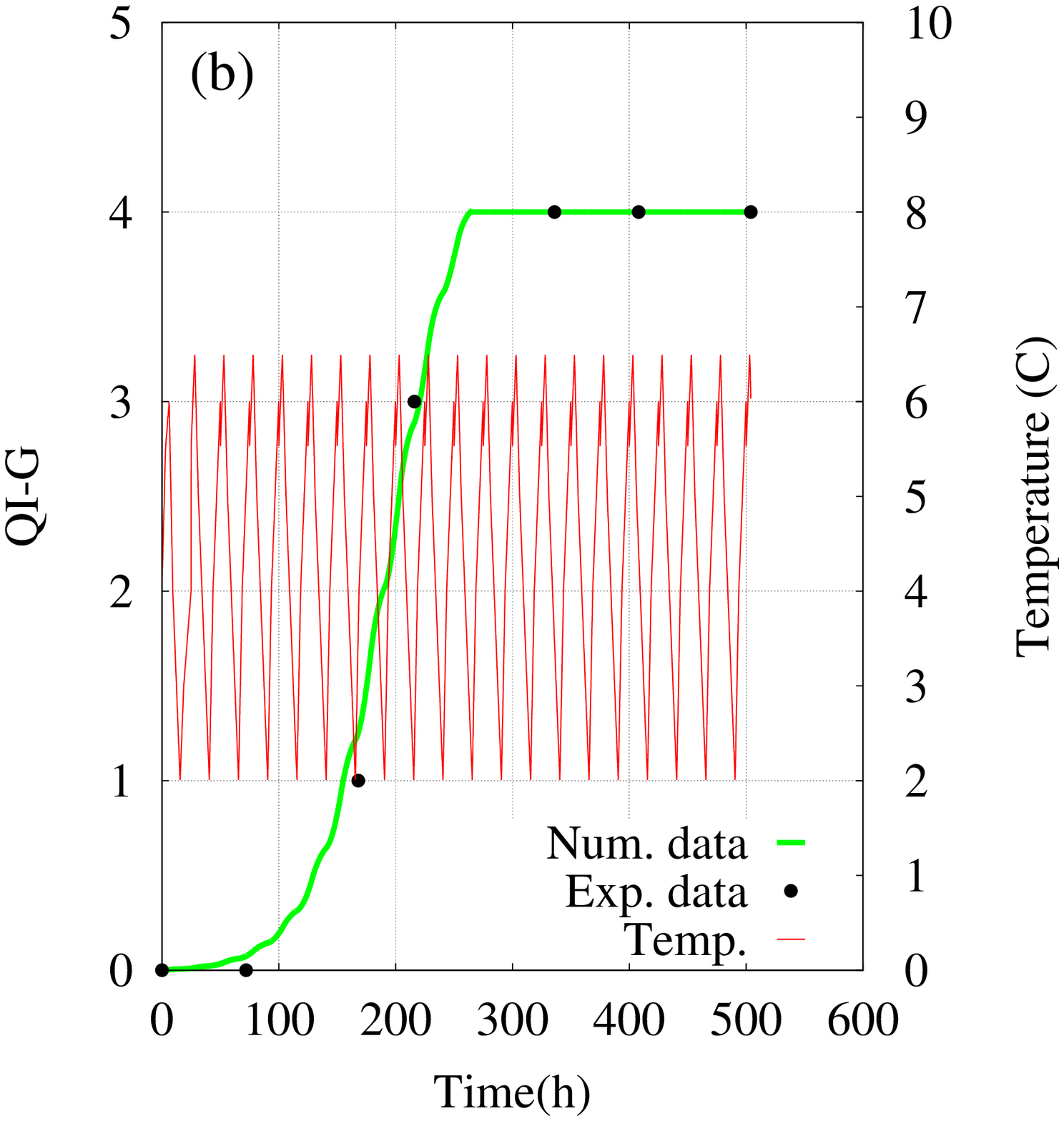}\\
\includegraphics[width=6.0cm]{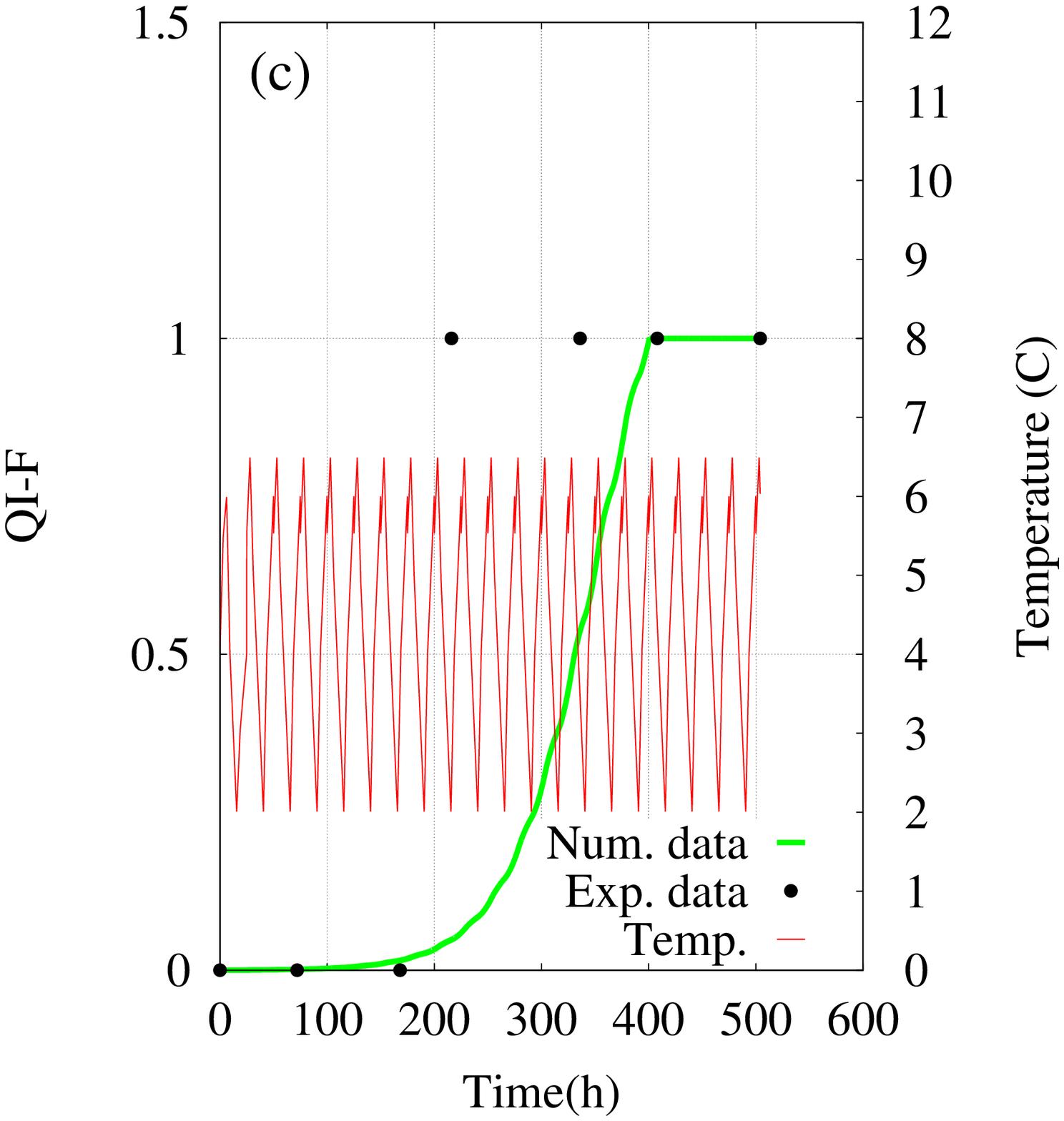}
\includegraphics[width=6.0cm]{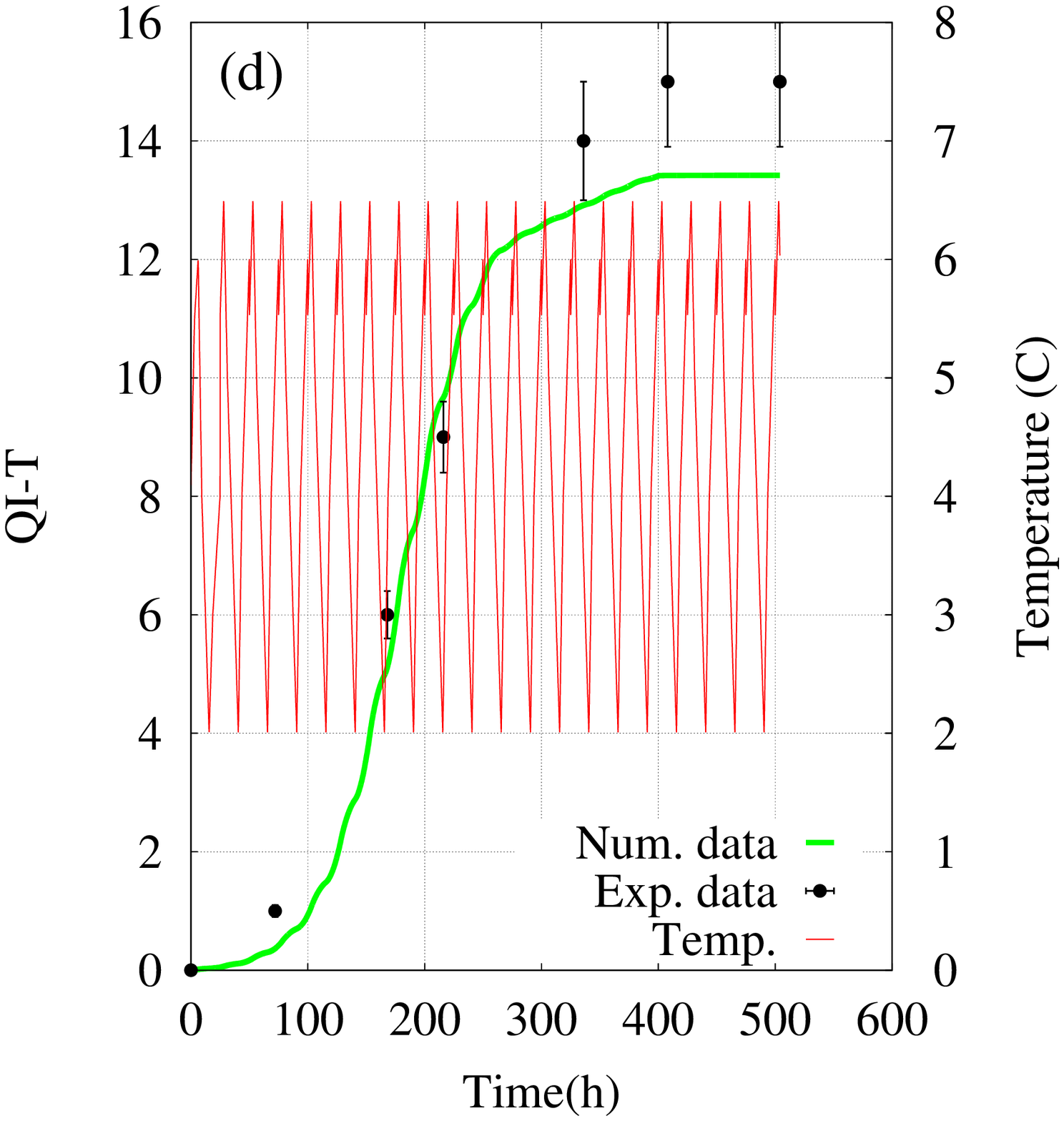}
\end{center}
\vspace{-0.8cm}\caption{\small \emph{Group~3. Comparison between
observed (black dots) and predicted (green line) quality indexes
for: a) skin; b) gills; c) flesh. Panel d shows the total QI
(summation of the scores obtained in the three sites). Vertical bars
indicate experimental errors. Theoretical curves were calculated
from Eqs.~(\ref{QI_S})-(\ref{QI_F}), by using the same values of
$\beta_{1i}$ and $\beta_{2i}$ ($i=S,G,F$) as in
Fig.~\ref{QIs_Group1}. Red curves represent the temperature
profiles.}} \vspace{-0.2cm} \label{QIs_Group3}
\end{figure}
\begin{figure}
\begin{center}
\includegraphics[width=6.0cm]{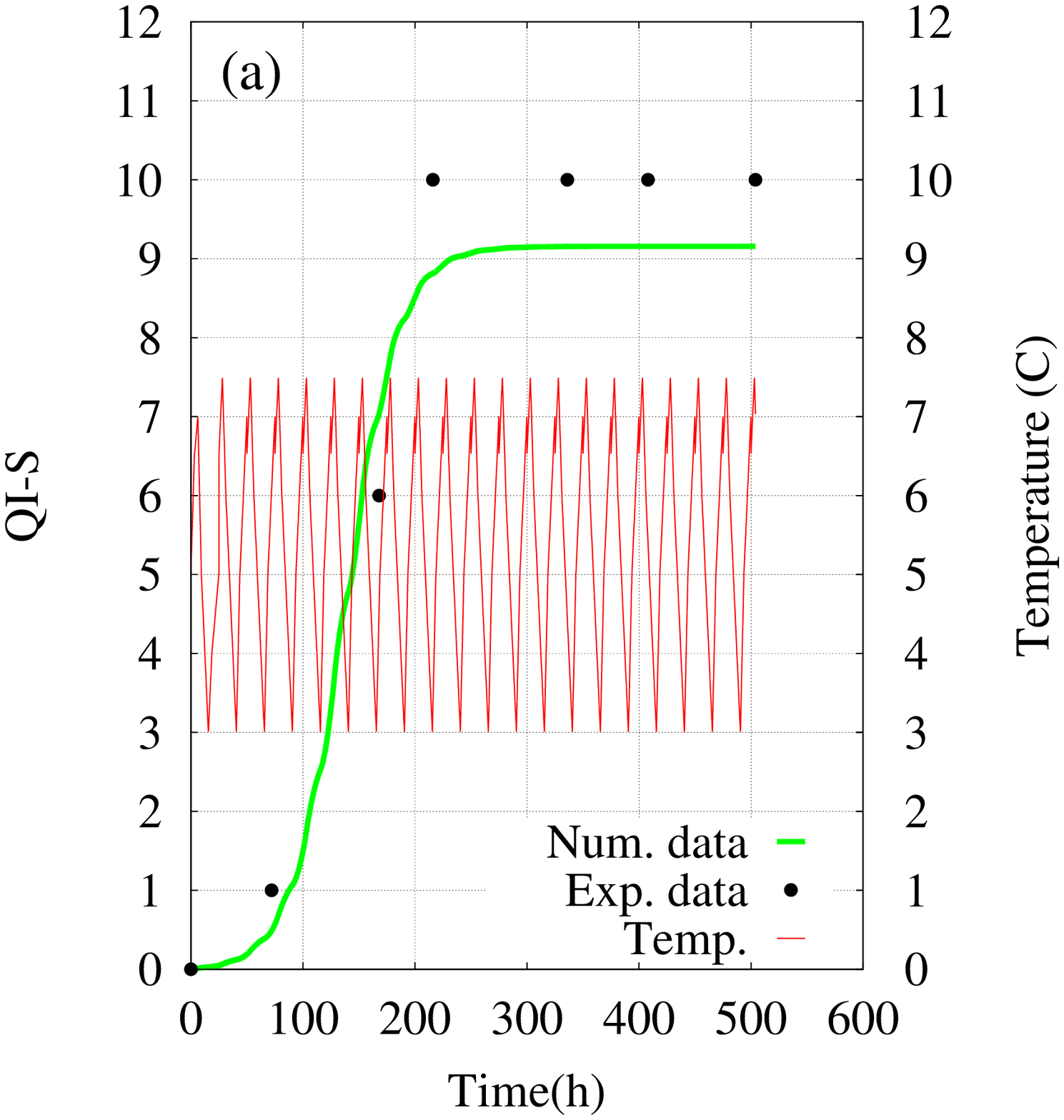}
\includegraphics[width=6.0cm]{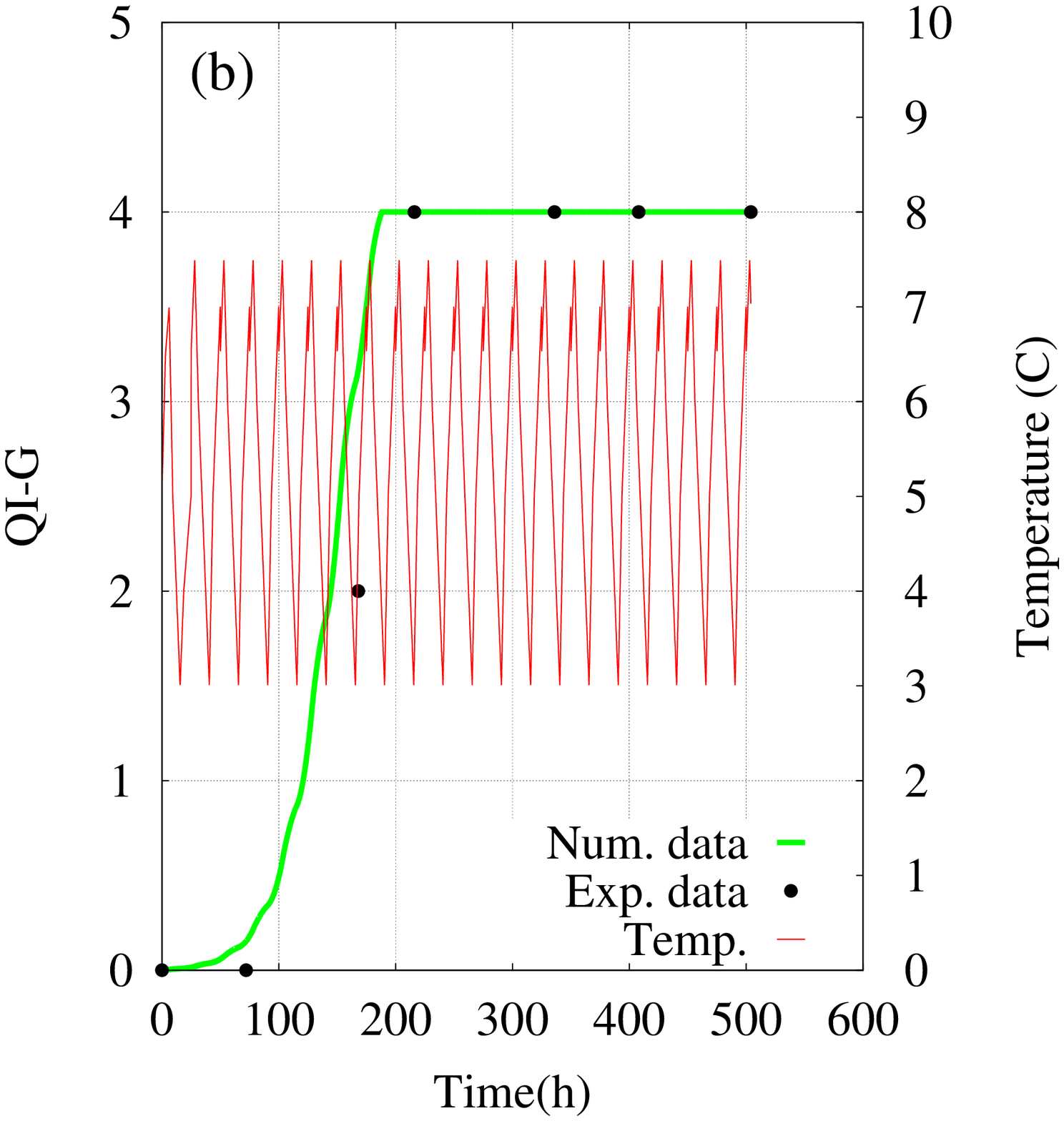}\\
\includegraphics[width=6.0cm]{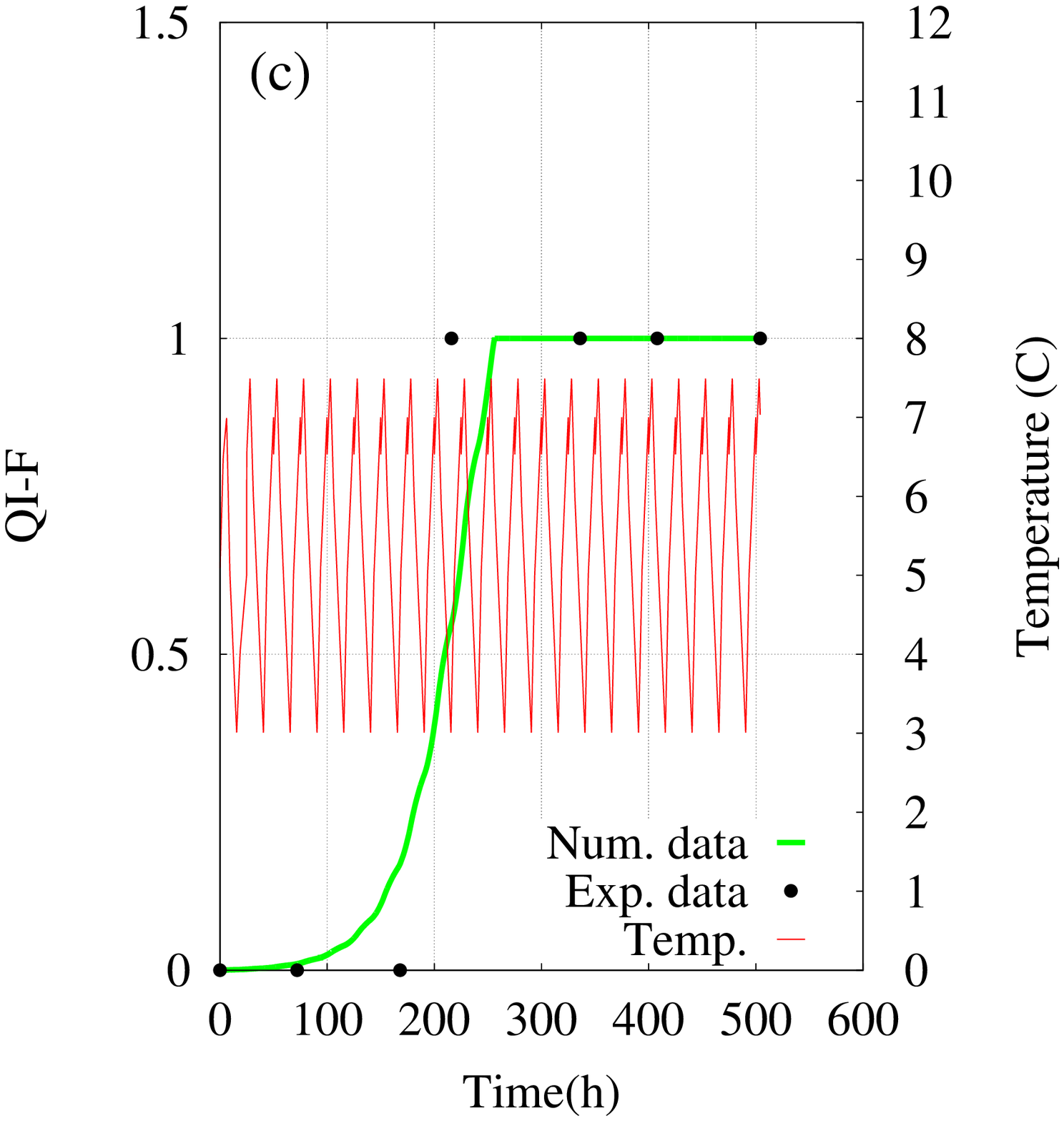}
\includegraphics[width=6.0cm]{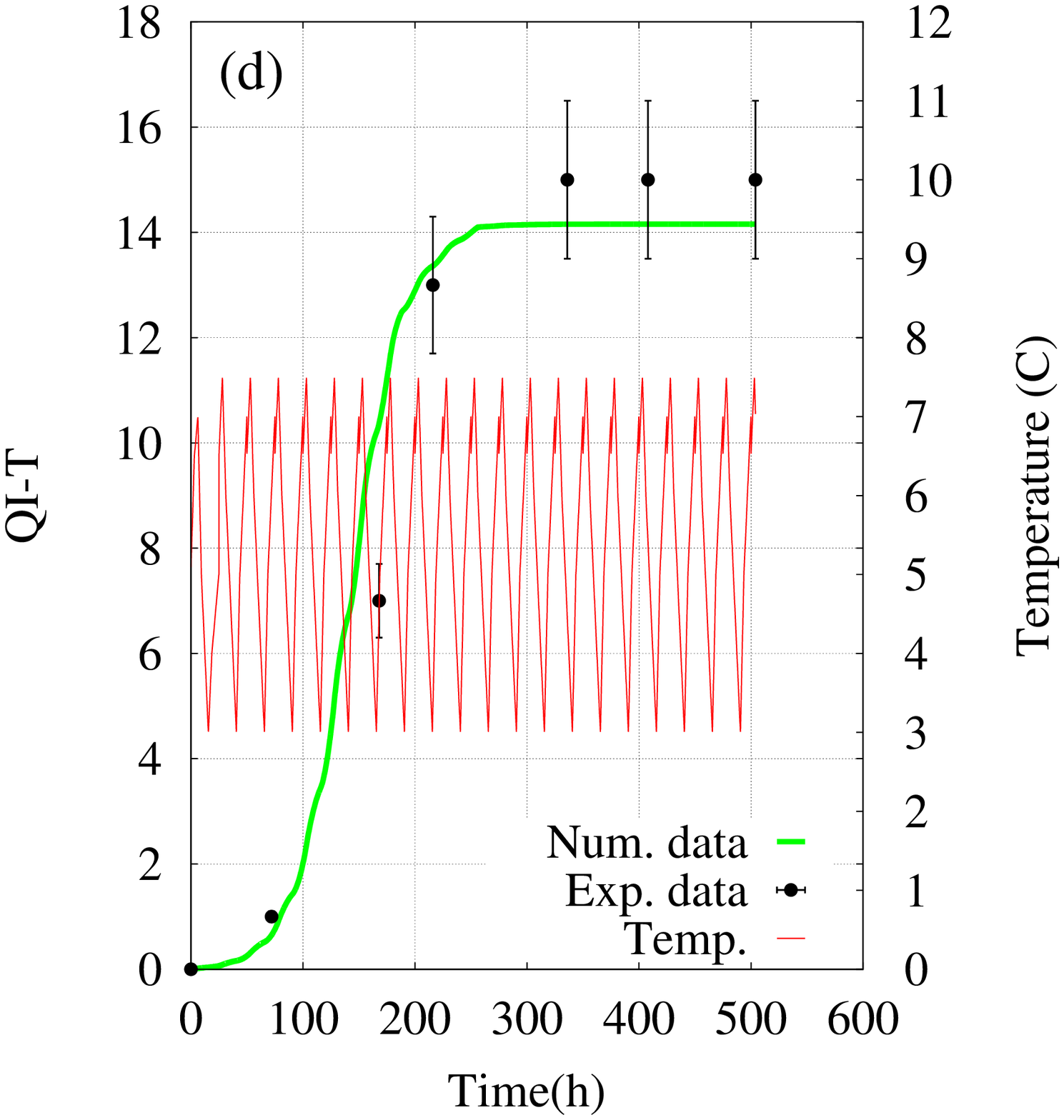}
\end{center}
\vspace{-0.8cm}\caption{\small \emph{Group~4. Comparison between
observed (black dots) and predicted (green line) quality indexes
for: a) skin; b) gills; c) flesh. Panel d shows the total QI
(summation of the scores obtained in the three sites). Vertical bars
indicate experimental errors. Theoretical curves were calculated
from Eqs.~(\ref{QI_S})-(\ref{QI_F}), by using the same values of
$\beta_{1i}$ and $\beta_{2i}$ ($i=S,G,F$) as in
Fig.~\ref{QIs_Group1}. Red curves represent the temperature
profiles.}} \vspace{-0.2cm} \label{QIs_Group4}
\end{figure}
As a further step, we intend to show that the conversion
coefficients $\beta_{1i}$ and $\beta_{2i}$, obtained by using
uniquely data from Group 1, can be used also for the other three
groups. As a consequence, one should observe a general
correspondence, mathematically expressed by
Eqs.~(\ref{QI_S})-(\ref{QI_F}), between bacterial concentrations and
predicted QIs. This correspondence should indicate that the
"conversion" depends mostly on the bacterial species (and eventually
strains), without being influenced by the specific dataset. At this
aim, in the following we use the values of $\beta_{1i}$,
$\beta_{2i}$ ($i=S,G,F$) calculated from Group 1 to predict QIs
scores also for Groups~2,~3,~4. Results are given in
Figs.~\ref{QIs_Group2}-\ref{QIs_Group4}, where sensory-analysis data
(black dots), corresponding predicted values
(green line), and temperature profiles (red line) are shown.\\
As one can expect, a good agreement between predicted and observed
QIs scores of Group~1 (see Fig.~\ref{QIs_Group1}) is found.
Moreover, the model is also able to reproduce the time behaviour of
quality indexes for Groups~2,~3,~4, even if the agreement between
predicted and observed QIs scores is less good (see
Figs.~\ref{QIs_Group2}-\ref{QIs_Group4}). This can be explained by
noting that in all groups the theoretical bacterial growth was
analyzed by assuming the presence of two populations, i.e.
Shewanella and Pseudomonas. However, discrepancies in the
composition of bacterial flora (both at level of species and
strains) among the four groups can be present, and determine for
Groups~2,~3,~4 experimental QI curves different from those
predicted by using the "conversion" coefficients obtained from the data of Group~1.\\
To conclude this section, we recall that this procedure was adopted
to verify whether, at least for a given fish species, a unique
correspondence, through the conversion coefficients $\beta_{1i}$,
$\beta_{2i}$, exists between spoilage bacteria concentrations and
QIs scores. The results obtained seem to confirm this hypothesis,
while indicating the presence of a behaviour qualitatively similar
between predicted values and observed QIs scores also for
Groups~2,~3,~and~4.\\

\vspace{-.7cm}
\section{Stochastic model}\label{S:5}

\vspace{-.2cm} In this paragraph we analyze how random perturbations
can affect the sensory evaluation performed by a panel of expert
persons. As a starting point we consider the discrepancies between
predicted values and observed QIs scores of Groups~2,~3,~and~4 (see
Figs.~\ref{QIs_Group2}-\ref{QIs_Group4}), and interpret them as a
consequence of the uncertainty and variability, which can be present
in sensory analysis. In fact, the results of a sensory evaluation
depend also on the abilities of the expert persons that perform the
analysis. Their skills can vary, in an unpredictable way, among
different persons. Moreover, the same expert can evaluate
differently the same fish sample, depending on the momentaneous
conditions of his/her sensory abilities such as precision and
reliability in evaluating surface and eyes appearance, odour,
elasticity of muscle and gills. To take into account these "sources"
of uncertainty and variability we modify our model by inserting
terms of multiplicative noise in Eqs.~(\ref{QI_S})-(\ref{QI_F}). As
a result, we obtain the following stochastic differential equations
\begin{eqnarray}
\frac{dQI_S}{dt}&=&\frac{dN_{wS}}{dt}\beta_{1S}+\frac{dN_{bS}}{dt}\beta_{2S}+QI_S\thinspace\xi_S
\label{QI_S_stoch}\\
\frac{dQI_G}{dt}&=&\frac{dN_{wG}}{dt}\beta_{1G}+\frac{dN_{bG}}{dt}\beta_{2G}+QI_G\thinspace\xi_G
\label{QI_G_stoch}\\
\frac{dQI_F}{dt}&=&\frac{dN_{wF}}{dt}\beta_{1F}+\frac{dN_{bF}}{dt}\beta_{2F}+QI_F\thinspace\xi_F,
\label{QI_F_stoch}
\end{eqnarray}
where $\xi_i(t)$ are statistically independent Gaussian white noises
with zero mean and correlation function $\langle
\xi_i(t)\xi_j(t')\rangle = \sigma_i \delta(t-t')\delta_{ij}$
($i,j=S,G,F$).

\vspace{-.2cm}
\subsection{Prediction of QIs values based on stochastic model}\label{SS:5.1}

According to the deterministic study, the bacterial growth curves
previously obtained (green lines in
Figs.~\ref{Bacteria_Group1}-\ref{Bacteria_Group4}) are used in
Eqs.~(\ref{QI_S_stoch})-(\ref{QI_F_stoch}), which are integrated
numerically.\\
It is important to note that the use of a random function, i.e.
noise source, to simulate the time behaviour of the system, makes
the single realization unpredictable and unique, and therefore
nonrepresentative of the real dynamics. As a consequence, one
possible choice to describe correctly the time evolution of the
system is to calculate the average of several realizations. This
procedure, indeed, allows to take into account different
''trajectories'' obtained by the integration of the stochastic
equations, without focusing on a specific realization~\cite{Den13a}.
\begin{table}
\begin{center}
\includegraphics[width=7.6cm]{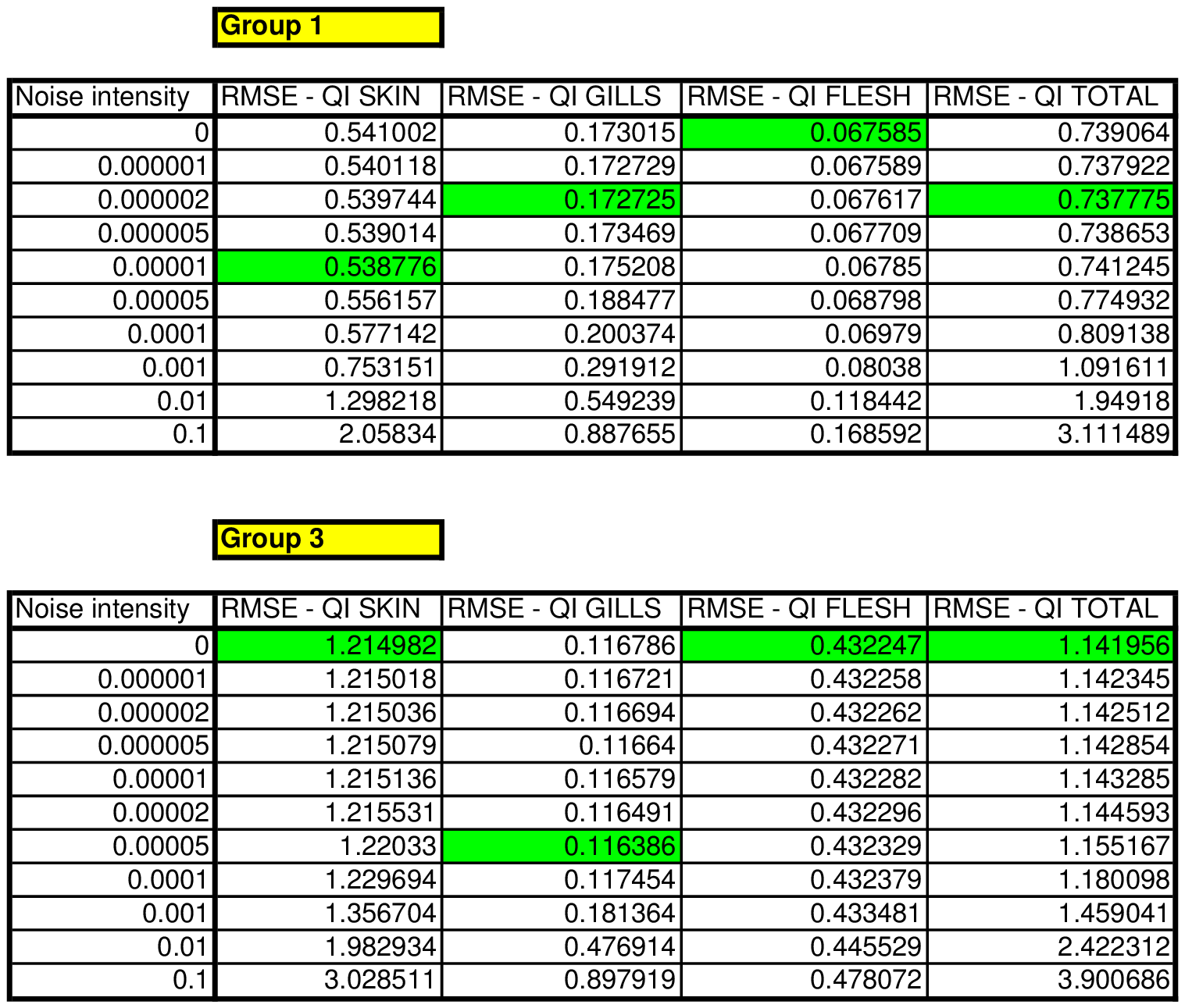}
\includegraphics[width=8.0cm]{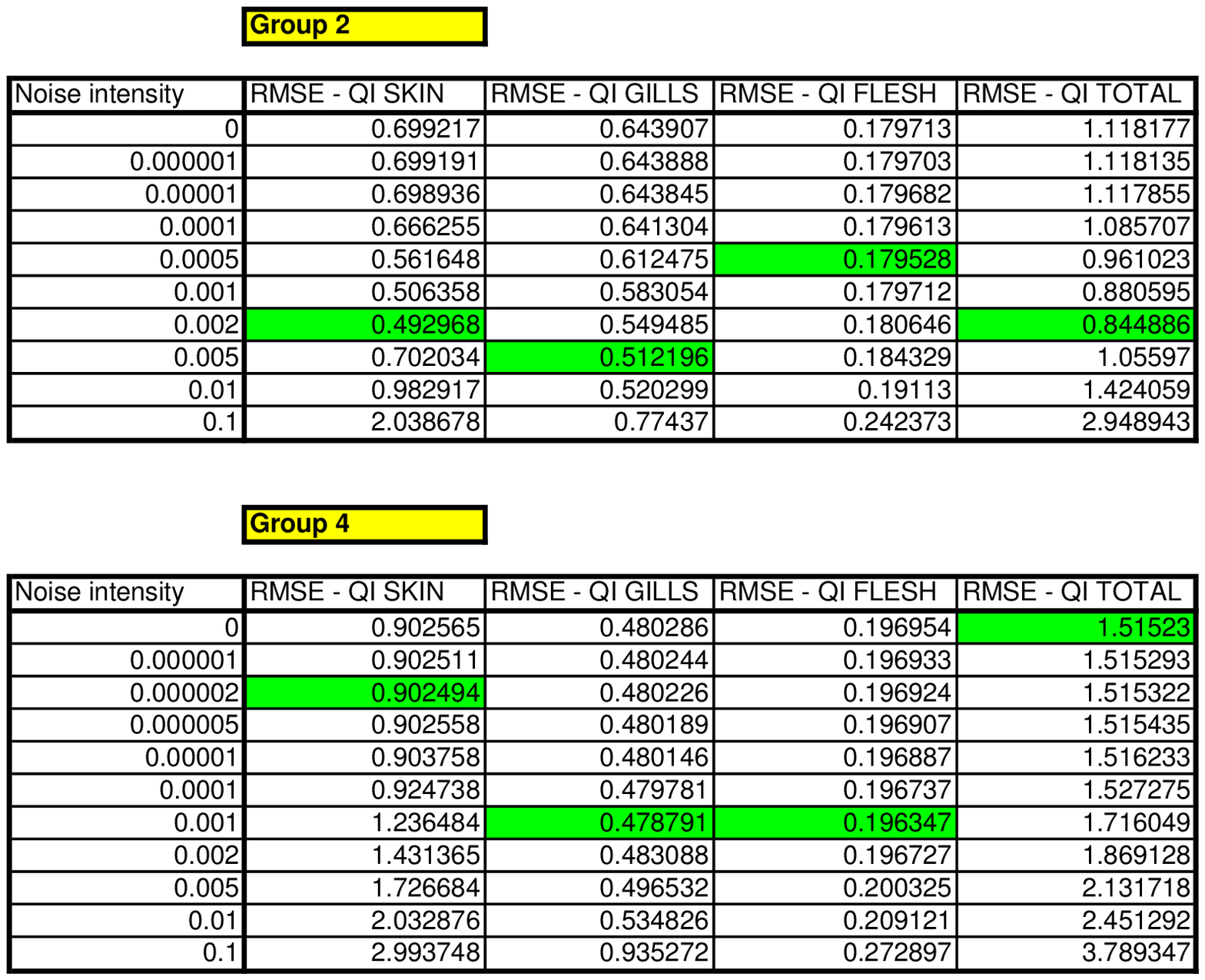}
\label{table1}
\end{center}
\vspace{-0.8cm}\caption{\small \emph{Root mean square error between
observed QIs scores and predicted values, calculated for different
noise intensities, $\sigma_S$, $\sigma_G$, $\sigma_F$.}}
\vspace{-0.2cm} \label{table1}
\end{table}
Thus, the time behaviours of QIs is reproduced for different values
of the three noise intensities ($\sigma_S$, $\sigma_G$, $\sigma_F$)
by averaging over $1000$ numerical
realizations~\cite{Den13b,Giu09a}. For each group, the values of QIs
predicted by the stochastic model are quantitatively compared with
the observed QIs scores by using a minimization procedure based on
the RMSE. Here we intend to focus on the predicted QIs characterized
by a minimum of RMSE (best agreement with observed QIs). Therefore,
as a preliminary analysis, in each site we obtain separately the
theoretical QIs, while determining the noise intensity for which the
RMSE is minimum. On the other side, it is reasonable to assume that,
in each group, all three sites are subject to the same noise
intensity. Therefore the total QI, including that corresponding to
the minimum RMSE (best agreement), is calculated by setting the
noise intensity to the same value in all three sites
($\sigma_S$~=~$\sigma_G$~=~$\sigma_F$). The results are given in
Table~\ref{table1}. Here, for all groups, the noise intensity for
which RMSE takes on the lowest value in each site is highlighted in
green. The statistical analysis performed for Groups~1,~2,~and~4
indicates that in most cases the agreement between predicted and
observed partial QIs is better for
noise intensities different from zero, that is when the stochastic approach is used.\\
\begin{figure}
\begin{center}
\includegraphics[width=6.0cm]{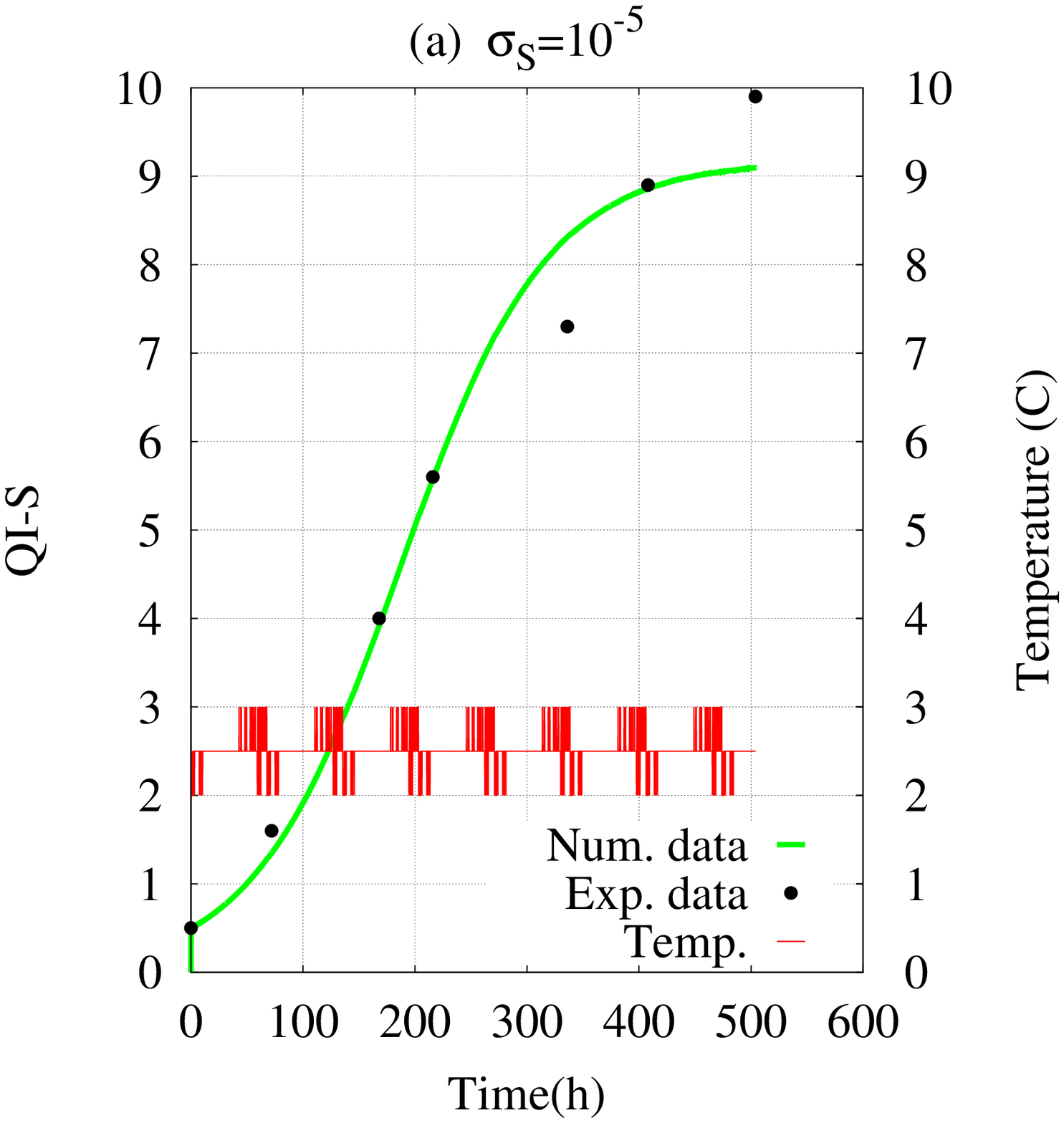}
\includegraphics[width=6.0cm]{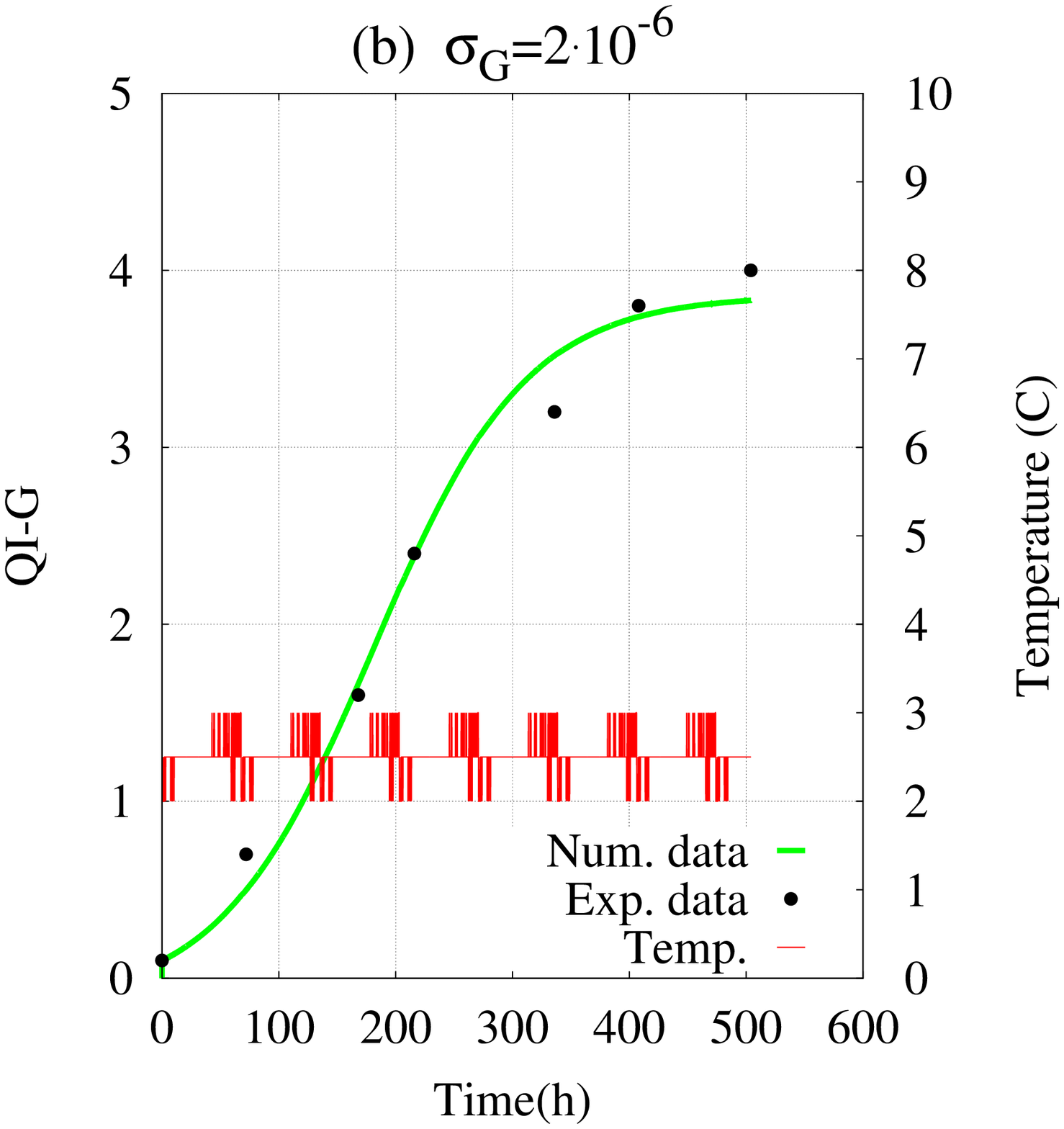}\\
\includegraphics[width=6.0cm]{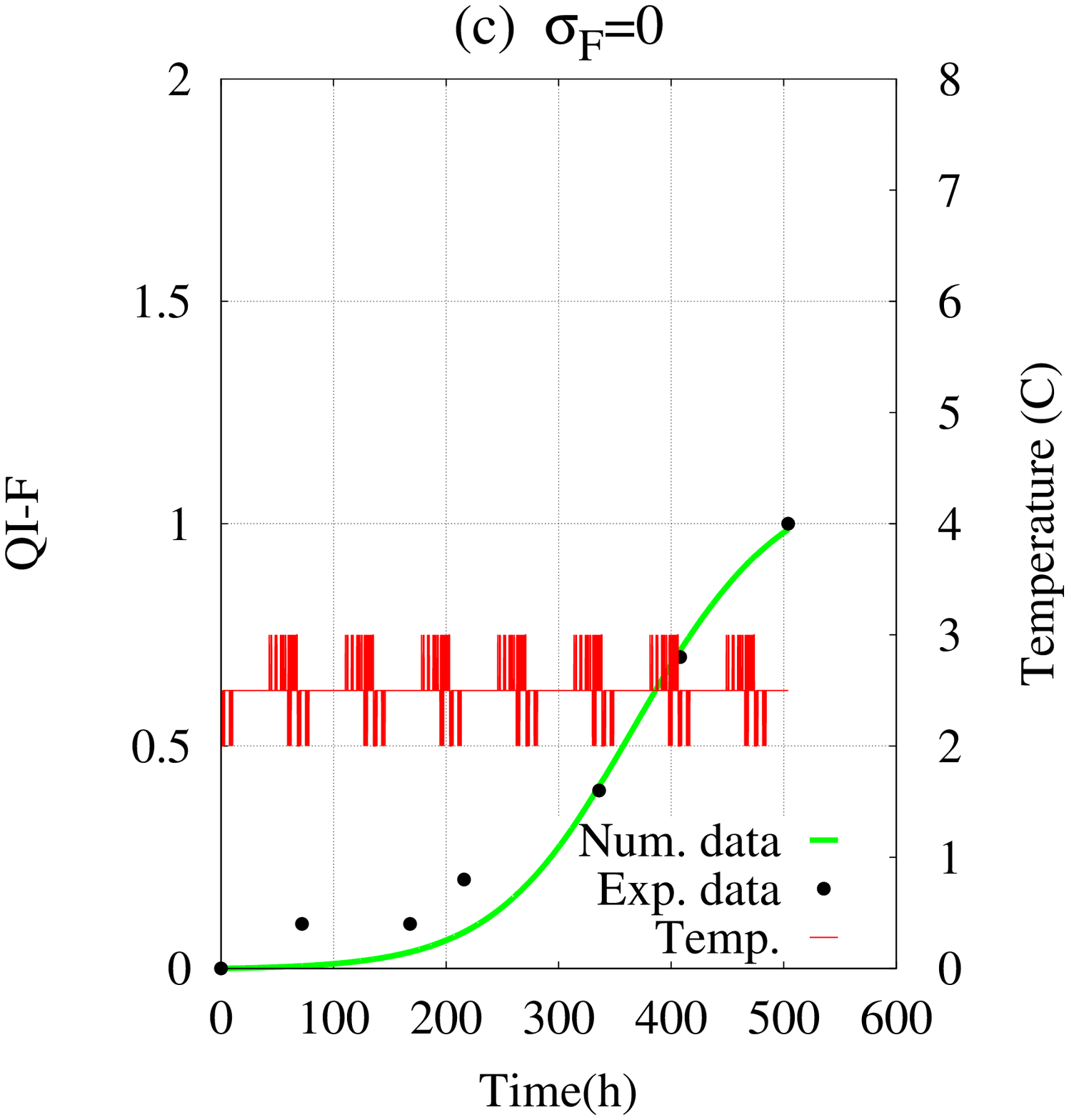}
\includegraphics[width=6.0cm]{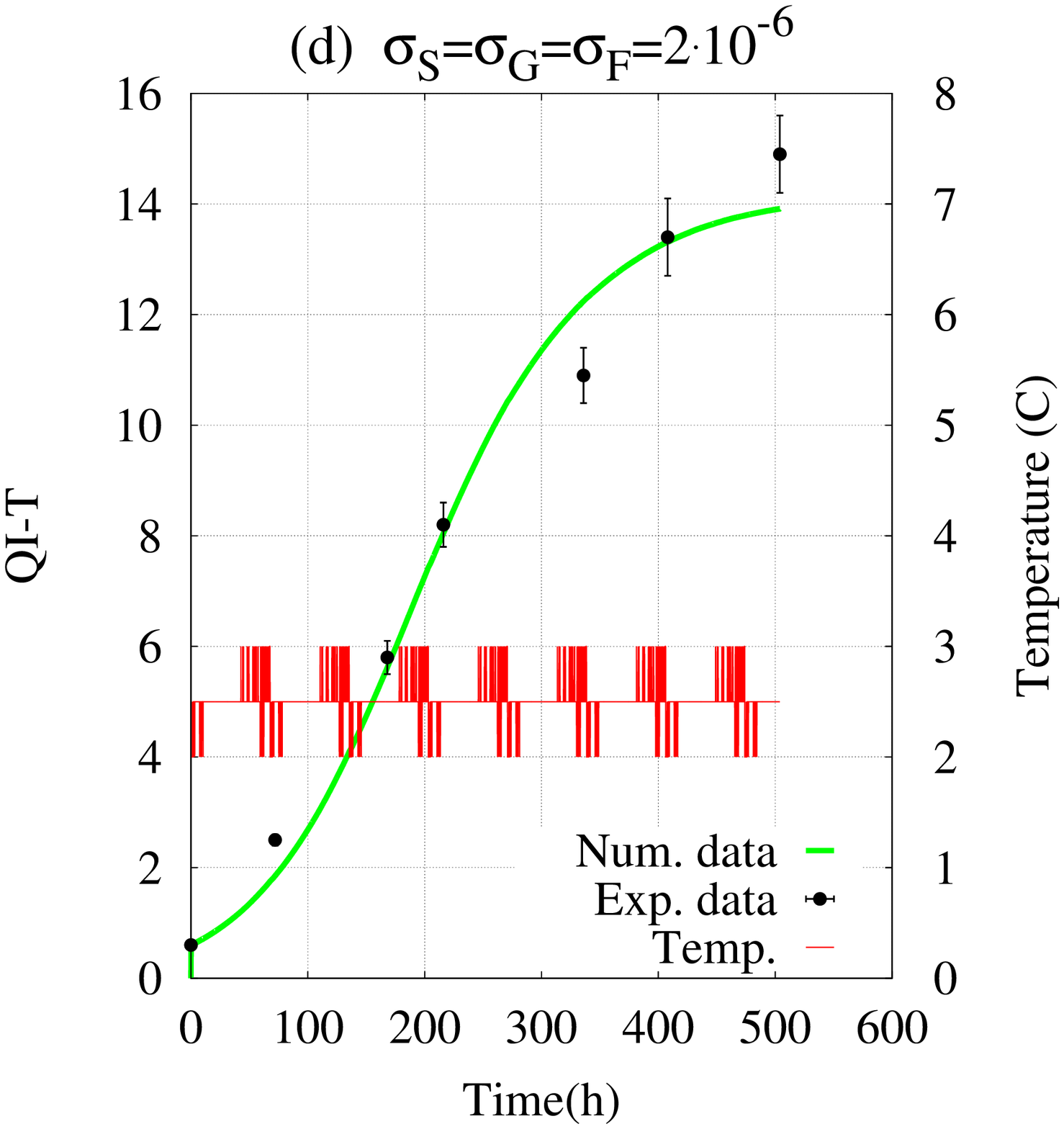}
\end{center}
\vspace{-0.8cm}\caption{\small \emph{Group~1. Comparison between
observed (black dots) and predicted by stochastic model (green line)
quality indexes: a) skin ($\sigma_S=10^{-5}$); b) gills
($\sigma_G=2\cdot 10^{-6}$); c) flesh ($\sigma_F=0$). Panel d shows
the total QI ($\sigma_S=\sigma_G=\sigma_F=2\cdot 10^{-6}$). Vertical
bars indicate experimental errors. Theoretical curves were
calculated from Eqs.~(\ref{QI_S_stoch})-(\ref{QI_F_stoch}), by using
the same values of $\beta_{1i}$ and $\beta_{2i}$ ($i=S,G,F$) as in
Fig.~\ref{QIs_Group1}. Red curves represent the temperature
profiles.}} \vspace{-0.2cm} \label{QIs_Stoch_Group1}
\end{figure}
\begin{figure}
\begin{center}
\includegraphics[width=6.0cm]{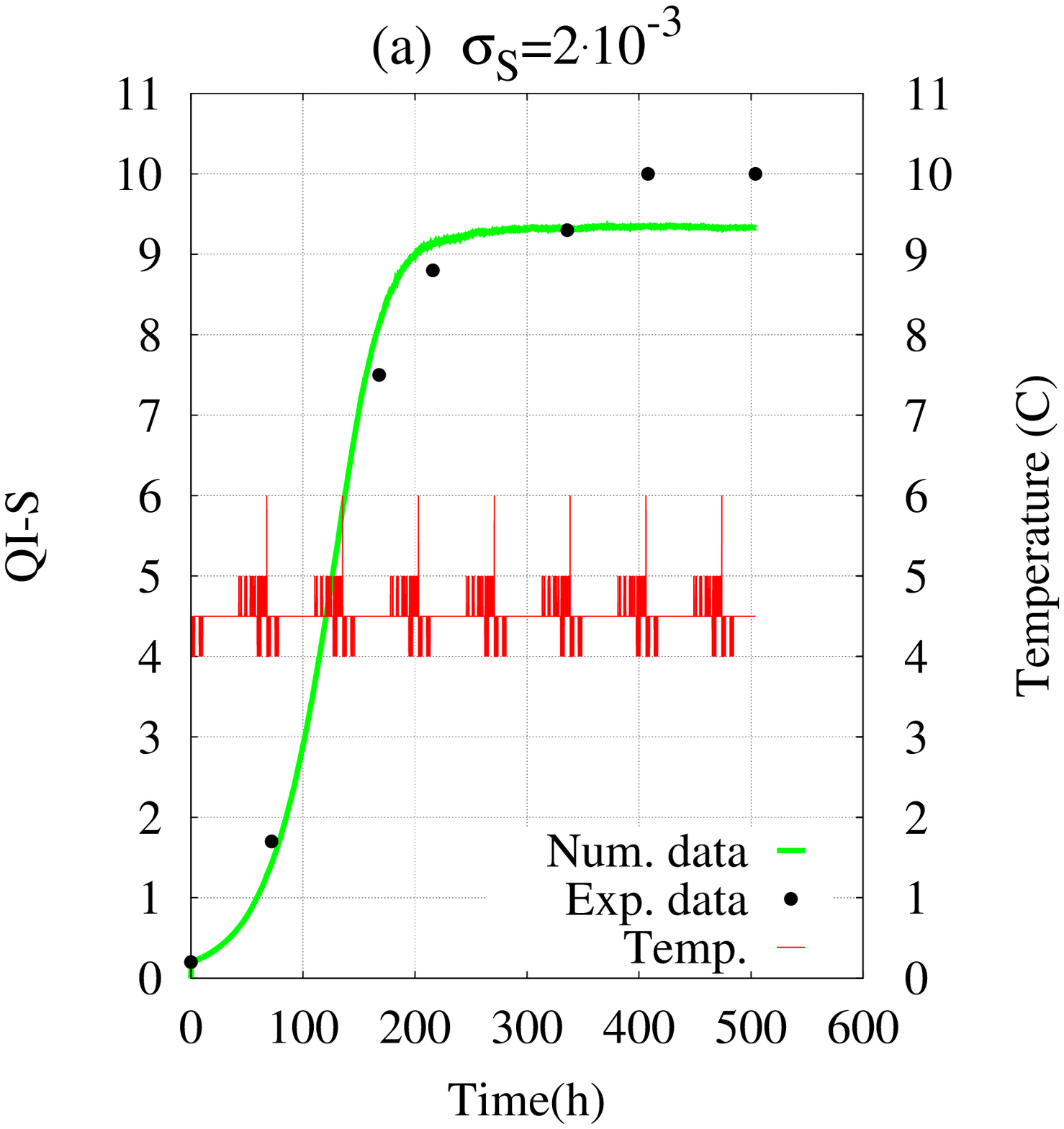}
\includegraphics[width=6.0cm]{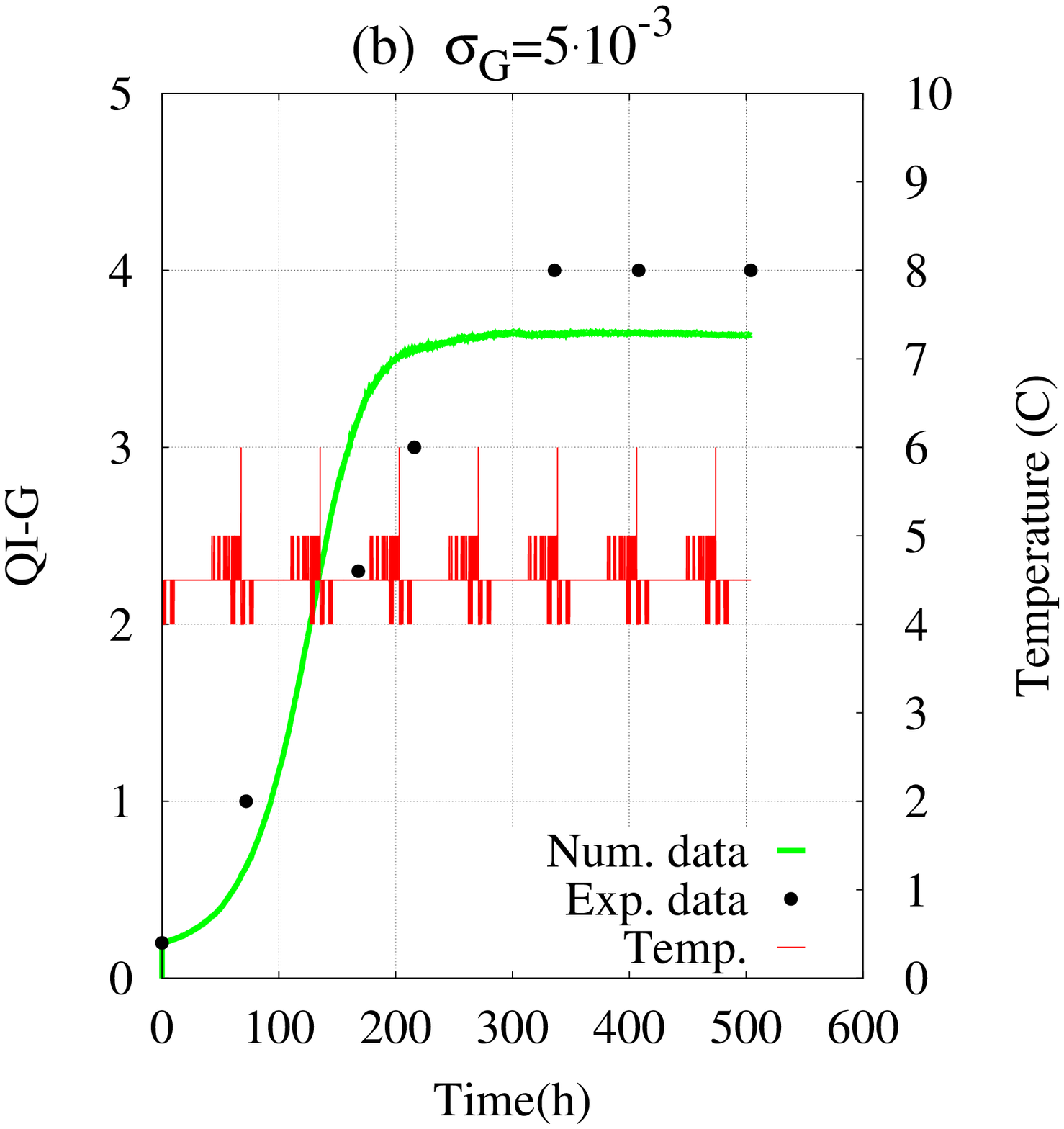}\\
\includegraphics[width=6.0cm]{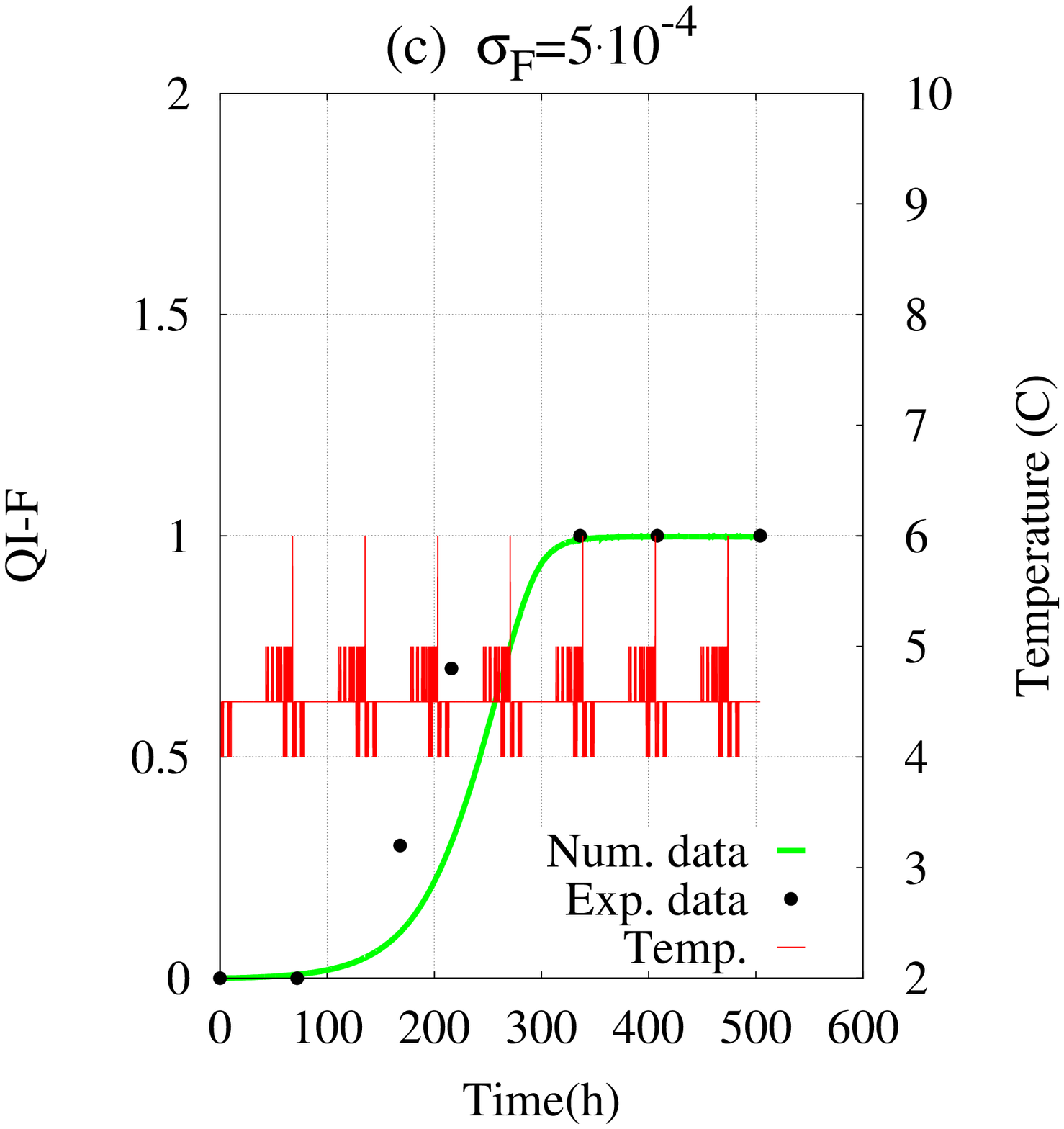}
\includegraphics[width=6.0cm]{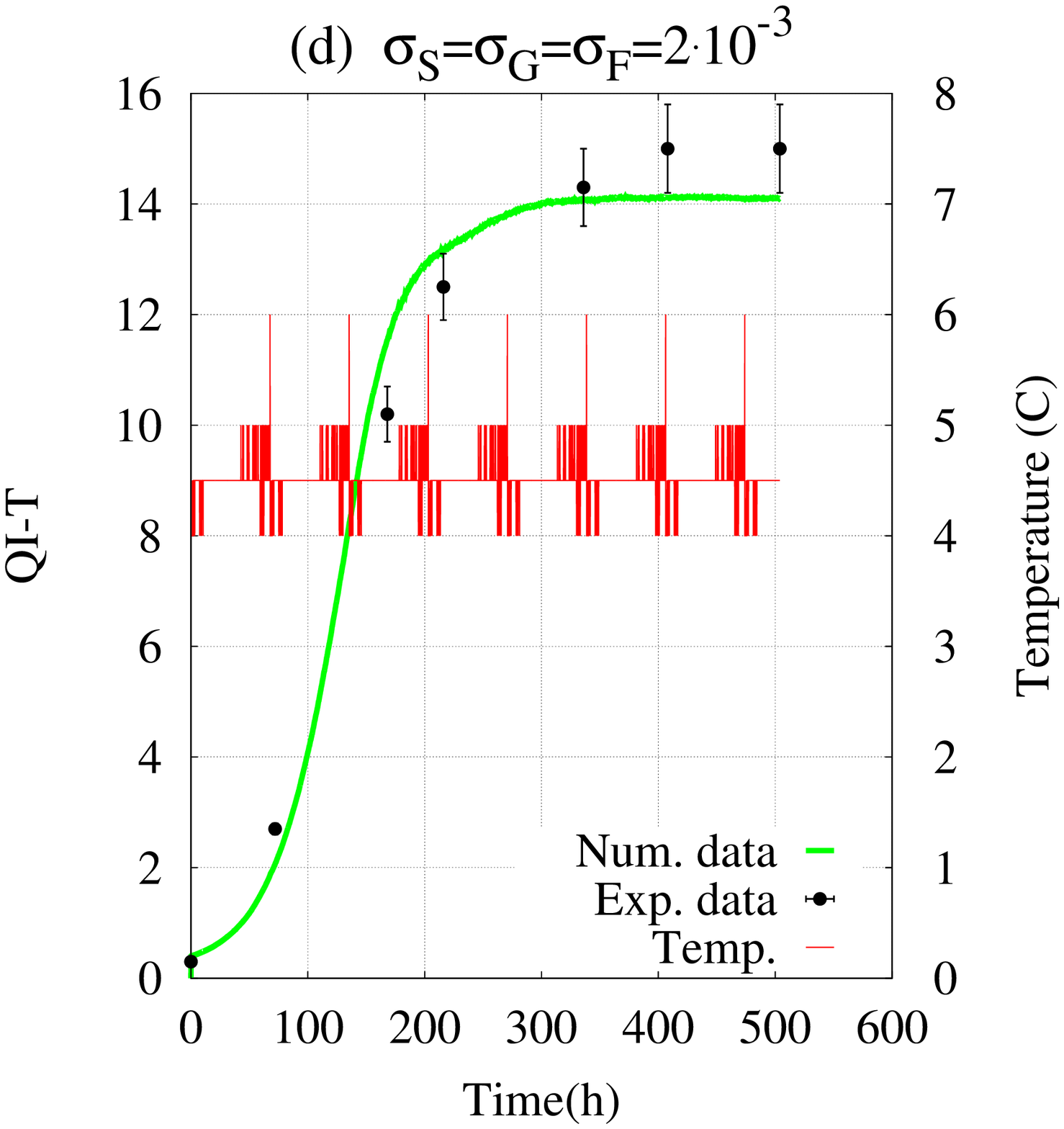}
\end{center}
\vspace{-0.8cm}\caption{\small \emph{Group~2. Comparison between
observed (black dots) and predicted by stochastic model (green line)
quality indexes: a) skin ($\sigma_S=2\cdot 10^{-3}$); b) gills
($\sigma_G=5\cdot 10^{-3}$); c) flesh ($\sigma_F=5\cdot 10^{-4}$).
Panel d shows the total QI ($\sigma_S=\sigma_G=\sigma_F=2\cdot
10^{-3}$). Vertical bars indicate experimental errors. Theoretical
curves were calculated from
Eqs.~(\ref{QI_S_stoch})-(\ref{QI_F_stoch}), by using the same values
of $\beta_{1i}$ and $\beta_{2i}$ ($i=S,G,F$) as in
Fig.~\ref{QIs_Group1}. Red curves represent the temperature
profiles.}} \vspace{-0.2cm} \label{QIs_Stoch_Group2}
\end{figure}
\begin{figure}
\begin{center}
\includegraphics[width=6.0cm]{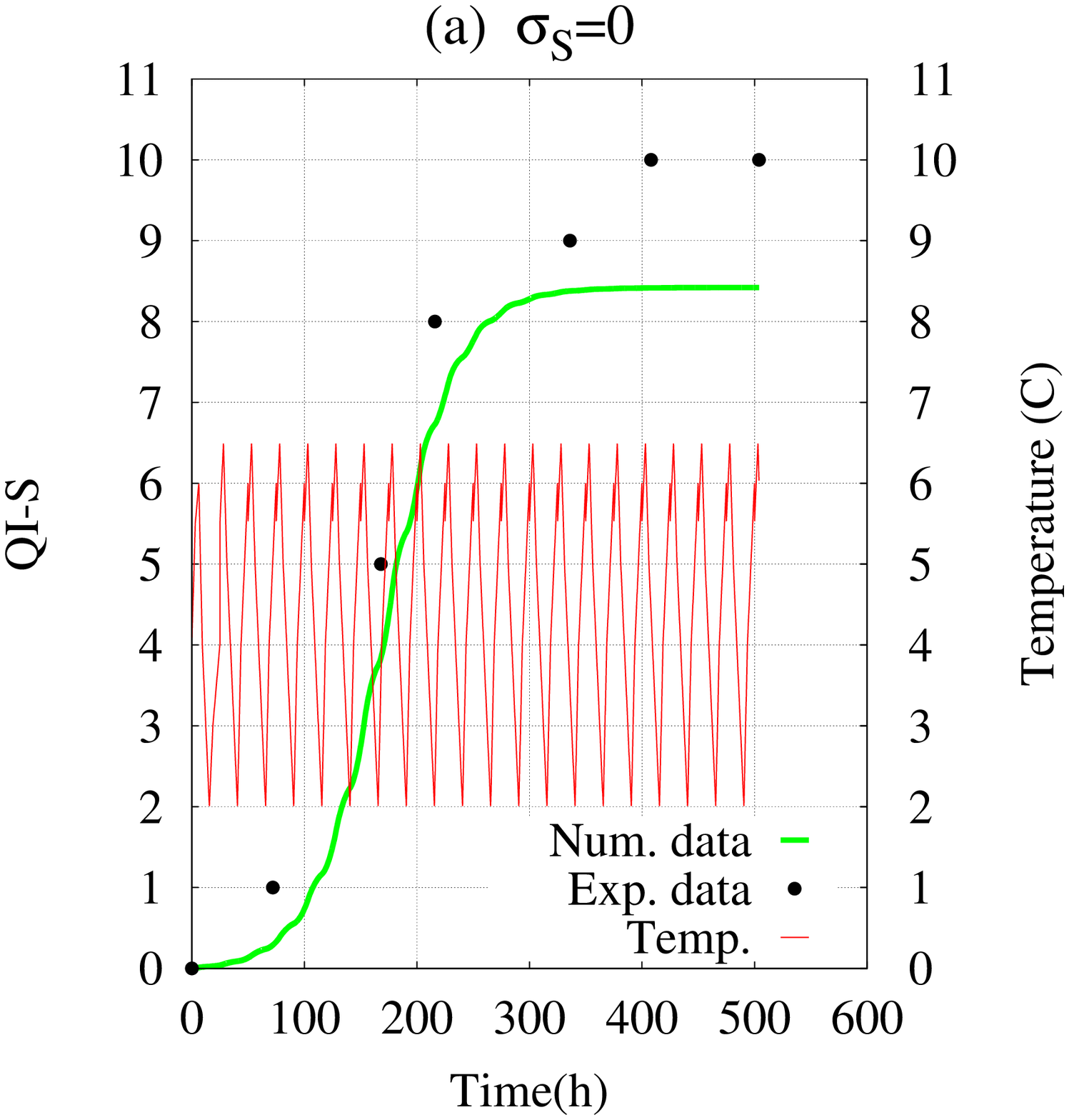}
\includegraphics[width=6.0cm]{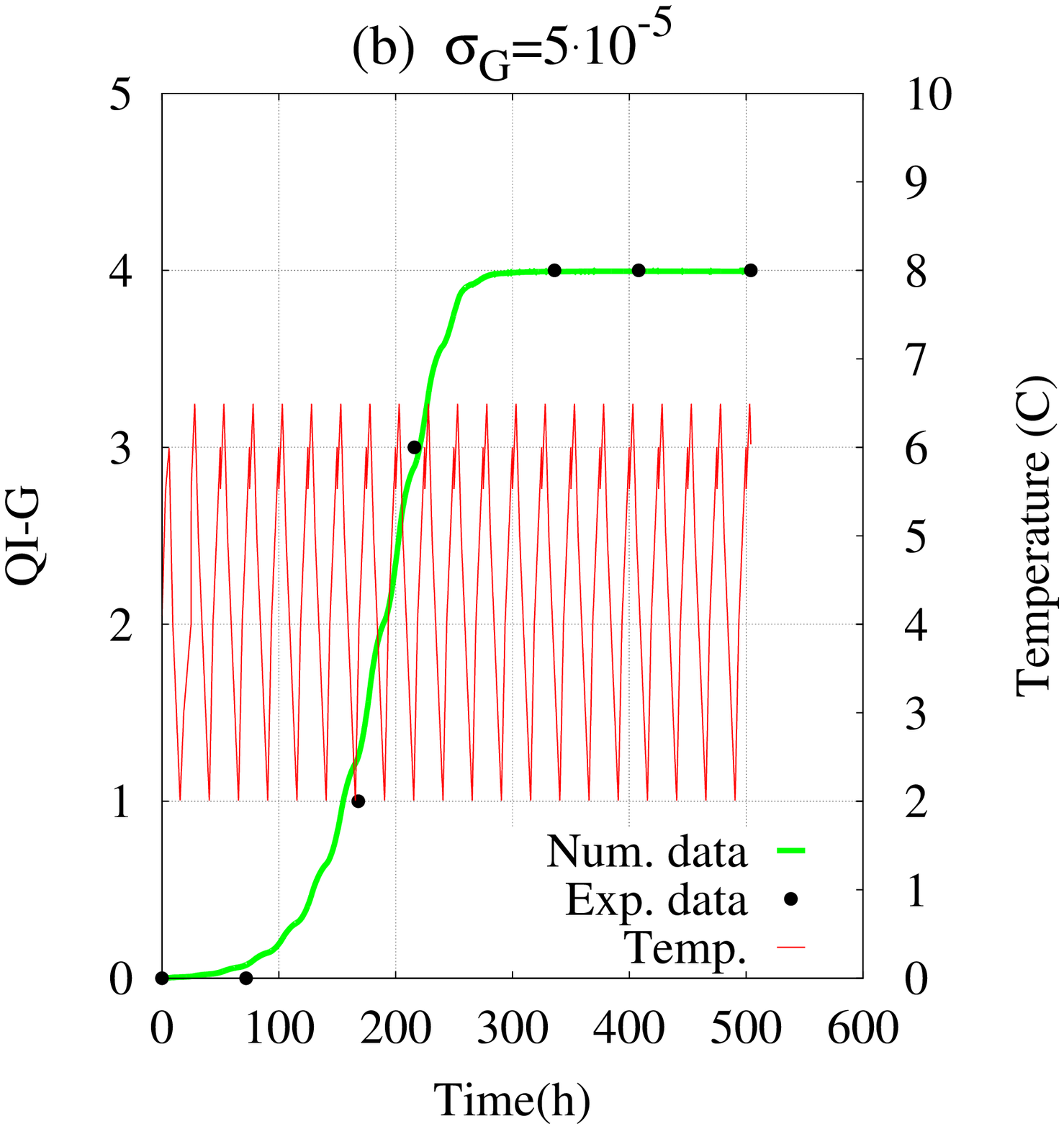}\\
\includegraphics[width=6.0cm]{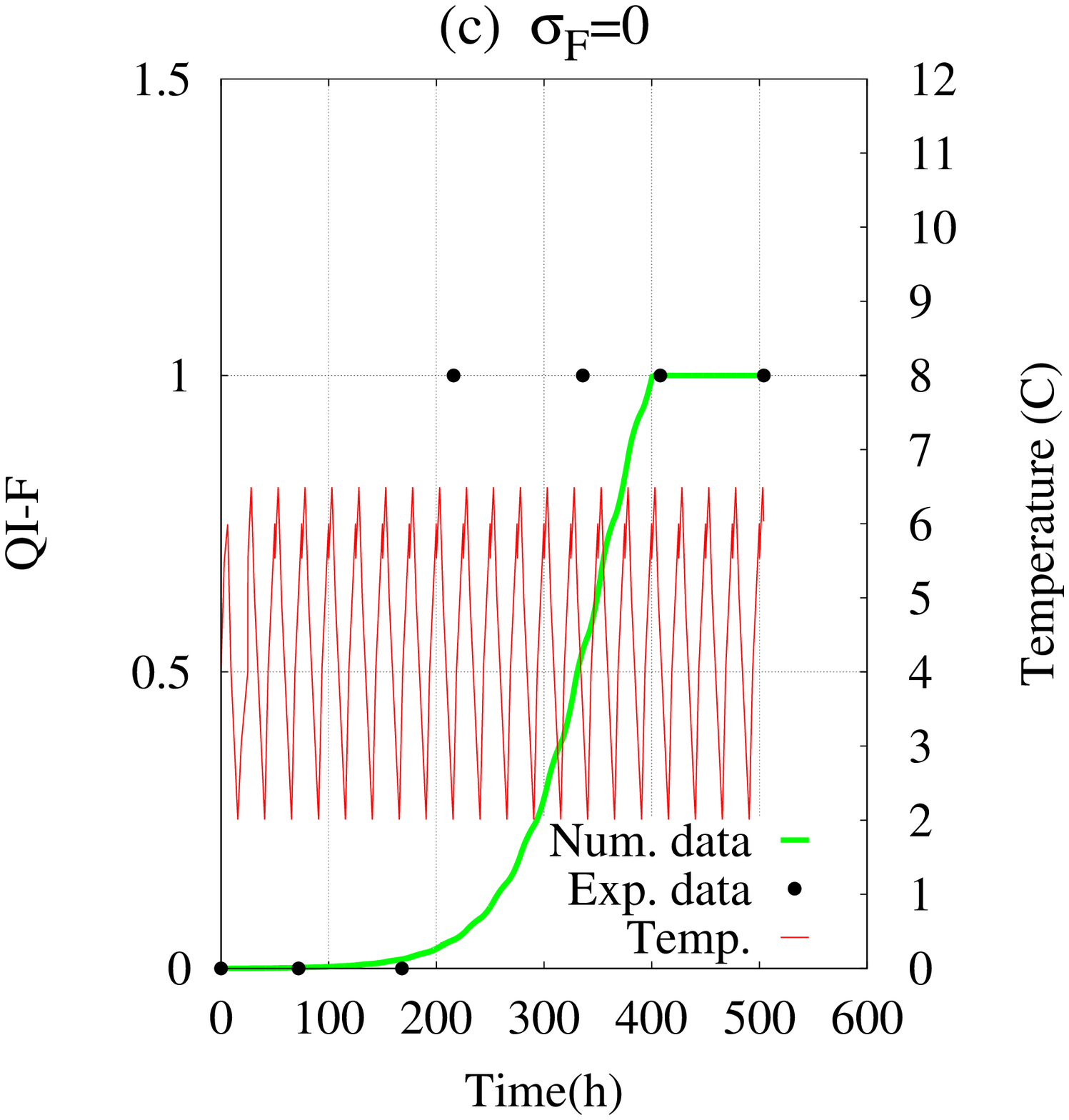}
\includegraphics[width=6.0cm]{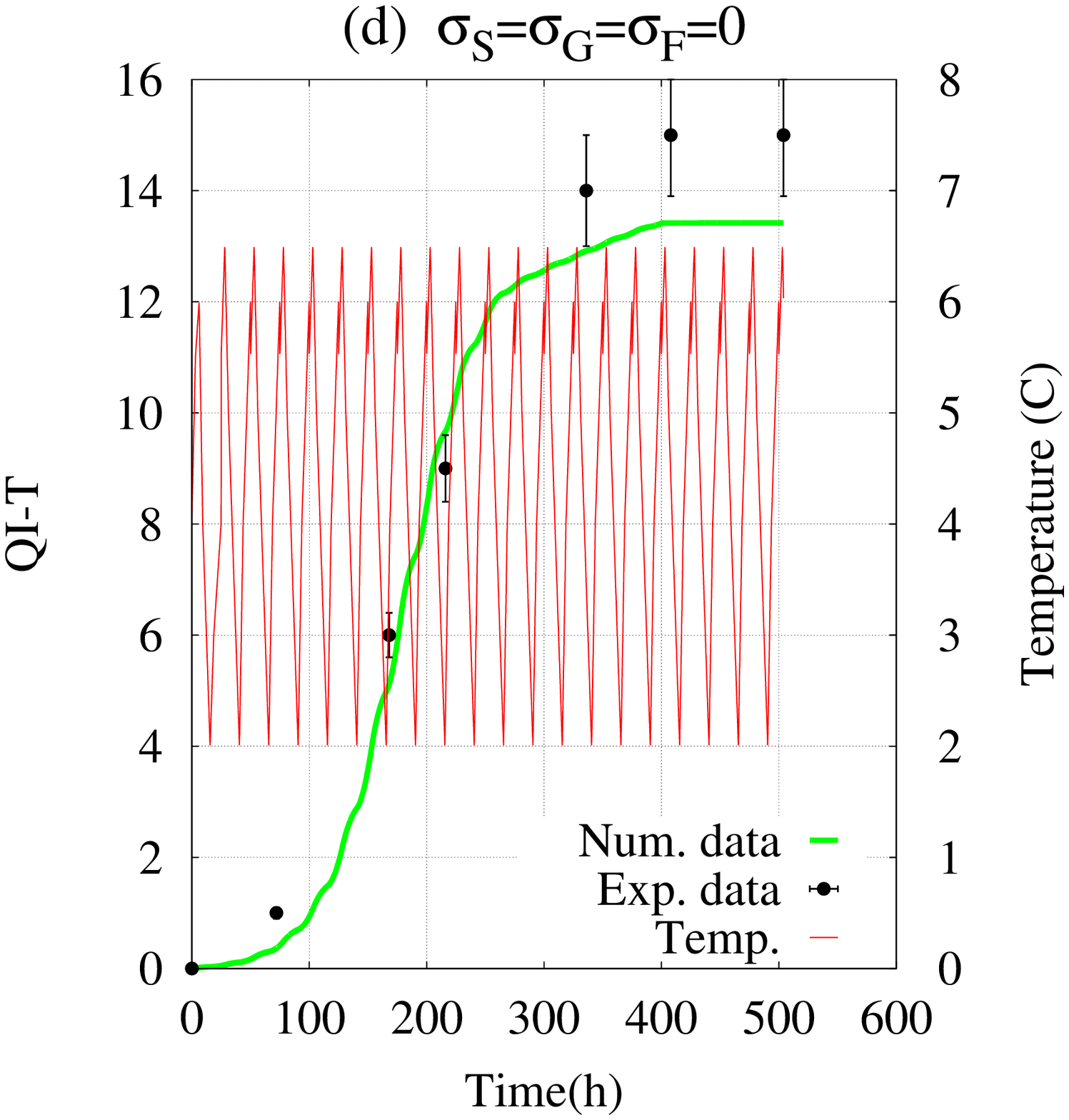}
\end{center}
\vspace{-0.8cm}\caption{\small\emph{Group~3. Comparison between
observed (black dots) and predicted by stochastic model (green line)
quality indexes: a) skin ($\sigma_S=0$); b) gills ($\sigma_G=5\cdot
10^{-5}$); c) flesh ($\sigma_F=0$). Panel d shows the total QI.
($\sigma_S=\sigma_G=\sigma_F=0$). Vertical bars indicate
experimental errors.} Theoretical curves were calculated from
Eqs.~(\ref{QI_S_stoch})-(\ref{QI_F_stoch}), by using the same values
of $\beta_{1i}$ and $\beta_{2i}$ ($i=S,G,F$) as in
Fig.~\ref{QIs_Group1}. Red curves represent the temperature
profiles.} \vspace{-0.2cm} \label{QIs_Stoch_Group3}
\end{figure}
\begin{figure}
\begin{center}
\includegraphics[width=6.0cm]{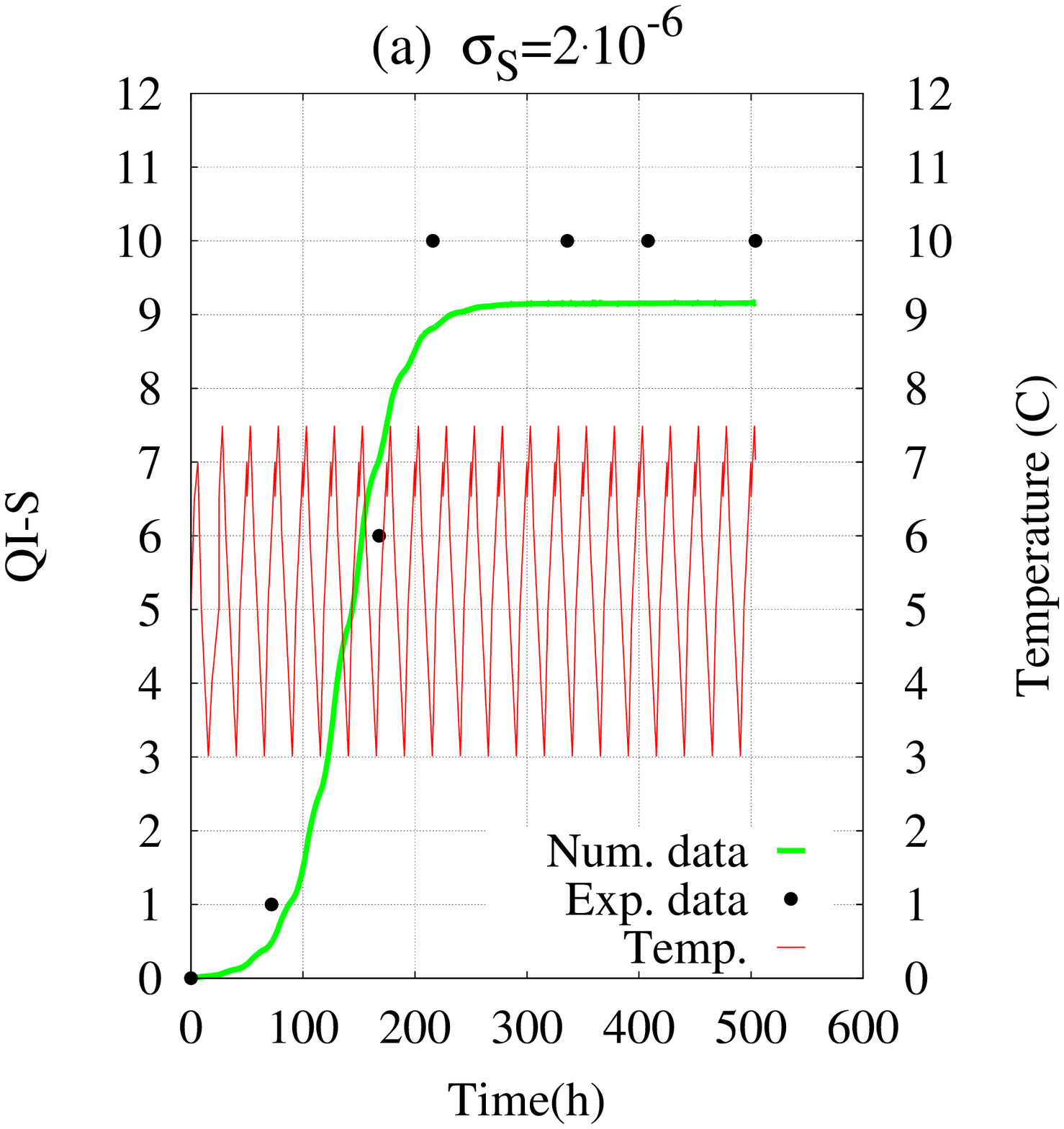}
\includegraphics[width=6.0cm]{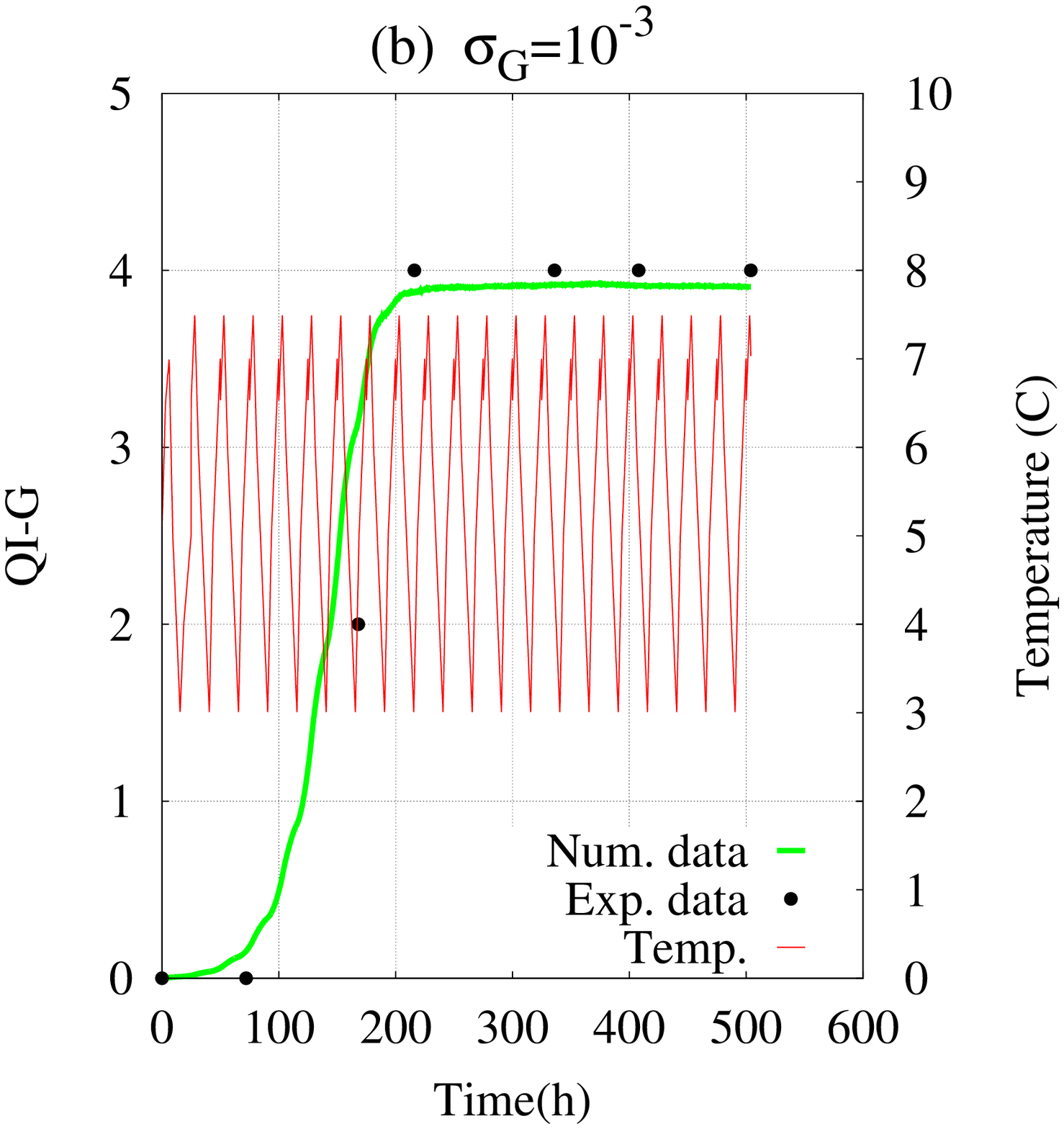}\\
\includegraphics[width=6.0cm]{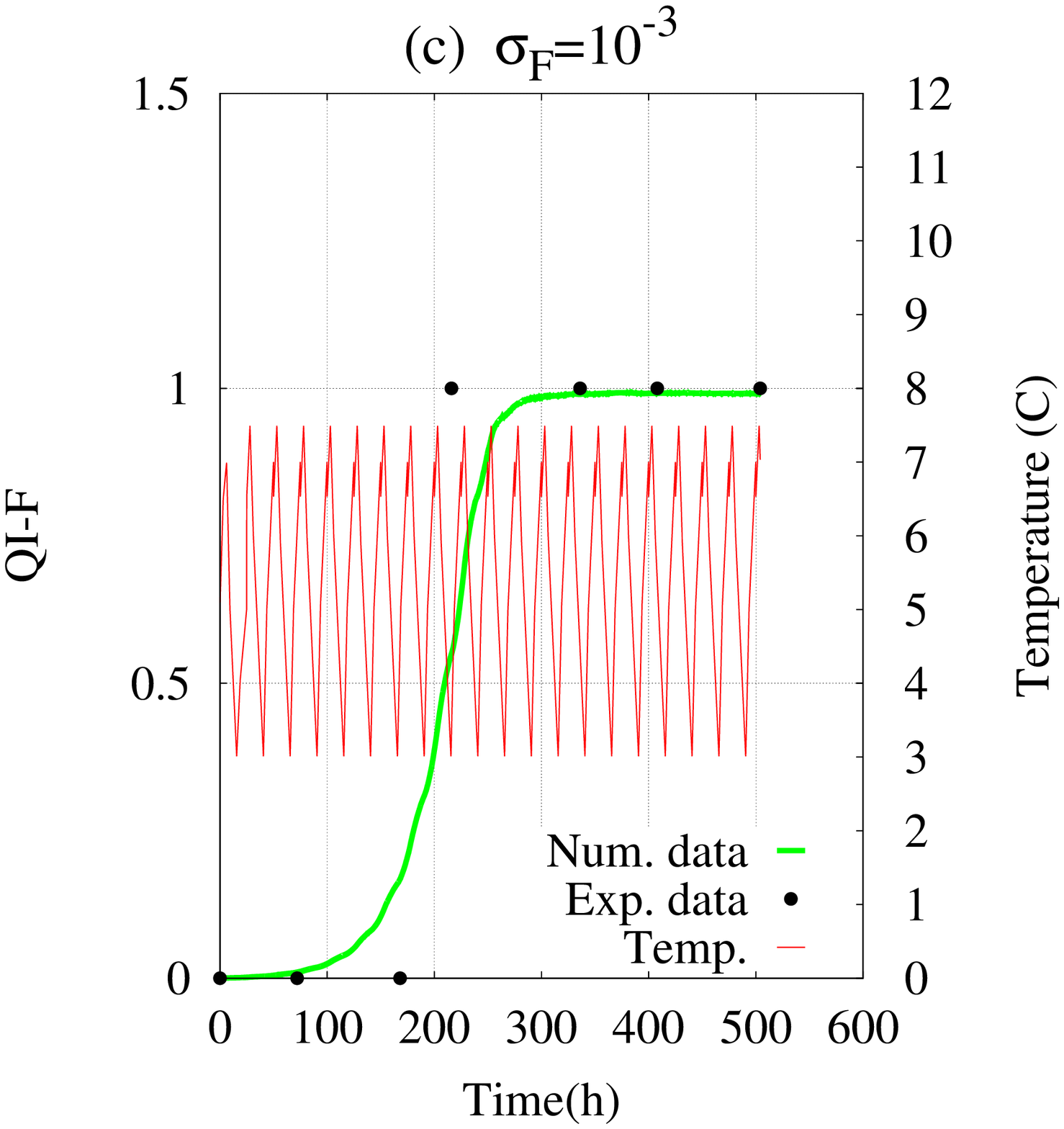}
\includegraphics[width=6.0cm]{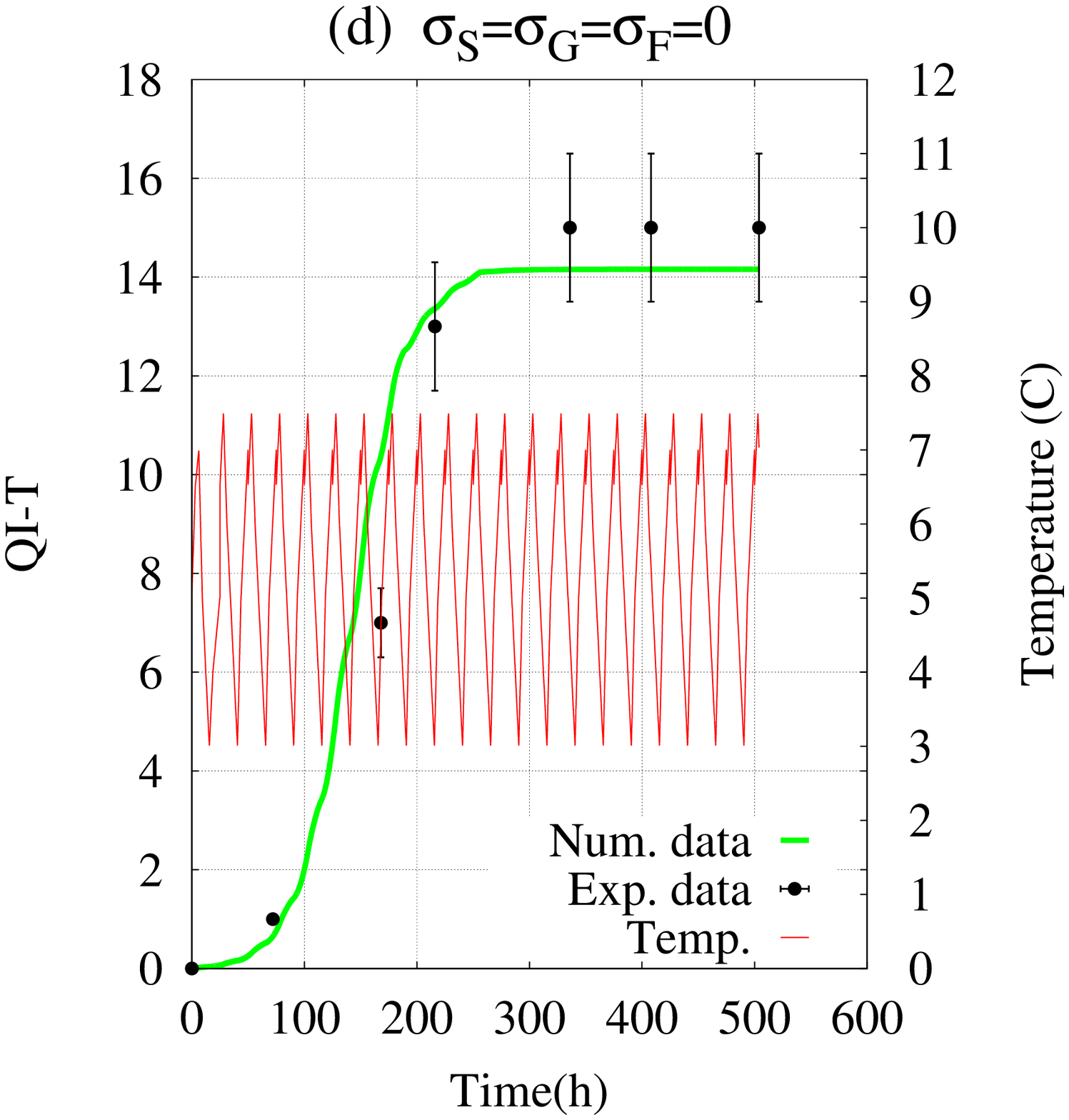}
\end{center}
\vspace{-0.8cm}\caption{\small \emph{Group~4. Comparison between
observed (black dots)and predicted by stochastic model (green line)
quality indexes: a) skin ($\sigma_S=2\cdot 10^{-6}$); b) gills
($\sigma_G=10^{-3}$); c) flesh ($\sigma_F=10^{-3}$). Panel d shows
the total QI ($\sigma_S=\sigma_G=\sigma_F=0$). Vertical bars
indicate experimental errors. Theoretical curves were calculated
from Eqs.~(\ref{QI_S_stoch})-(\ref{QI_F_stoch}), by using the same
values of $\beta_{1i}$ and $\beta_{2i}$ ($i=S,G,F$) as in
Fig.~\ref{QIs_Group1}. Red curves represent the temperature
profiles.}} \vspace{-0.2cm} \label{QIs_Stoch_Group4}
\end{figure}
Conversely, for Group~3 the RMSE becomes minimum for zero noise
intensity on skin and flesh, and for $\sigma_G=5\cdot10^{-5}$ in
gills.\\
Finally we note that for Groups 3 and 4 the minimum RMSE for the
total QI is obtained when all three noise intensities are zero, that
is in deterministic regime. This indicates that in the system
analyzed the fluctuations connected with the uncertainty
and variability of the sensory analysis affect weakly the QI evaluation.\\
The results of this section confirm that random perturbations
influence in most case the QIs scores, and play therefore a
non-negligible role in sensory analysis. These fluctuations can be
interpreted as a consequence of uncertainty and variability in
sensory
analysis.\\
Figs.~\ref{QIs_Stoch_Group1}-\ref{QIs_Stoch_Group4} show, for each
group and site, the partial and total predicted QIs (green curves)
for which the RMSE takes on the minimum value. As usual, black dots
and red lines indicate observed QIs and temperature profiles,
respectively. Note that curves obtained for zero noise intensity are
those calculated by the deterministic model.

\section{Conclusions}\label{S:6}

In this paper we studied a model which allows to reproduce the
dynamics of two bacteria populations, Pseudomonas and Shewanella,
responsible for food spoilage. Specifically, we studied the dynamics
of the two populations in specimens of Gilthead seabream
(\emph{Sparus aurata}), subdivided in four groups, by using a model
based on two logistic equations, and analyzing separately the
bacterial growths on skin, gills, and flesh.

By a fitting procedure we obtained theoretical growth curves for
bacterial concentrations in a very good agreement with experimental
data. As known, spoilage bacteria are responsible for loss of
quality and freshness in fish products. Therefore we used
theoretical growth curves for bacterial concentrations to predict
the time behaviour of some sensory characteristics of the fish food
analyzed. At this aim, we took into account the Quality Index Method
(QIM), a scoring system for freshness and quality sensory estimation
of fishery products, initially developed by the Tasmanian Food
Research Unit~\cite{Bre85}. The QIM scoring allows to assign demerit
points to each sensory parameter considered, providing, by a
summation of the partial scores, an overall sensory score named
Quality Index (QI).\\
To analyze the connection between the sensory characteristics of
fresh fish specimens and the two bacterial concentrations, we
reproduced the QI scores observed in sensory analysis, by a set of
differential equations, which allow to "translate" the bacterial
concentrations into QI scores. The investigation was carried out
separately on skin, gills and flesh, obtaining a QI score for each
site. In particular, we compared observed and predicted QIs. As a
result, depending on the group of specimens considered and sites
analyzed (skin, gills, or flesh), we found: i) a good agreement in
all sites of Group~1, in flesh of Group~2, on gills of Group~3, in
flesh and gills of Group~4; ii) a less good agreement in the other cases.\\
Finally, we took into account the effects of random fluctuations.
Specifically, we considered uncertainty and variability in sensory
analysis, by modeling them as effects of random fluctuations.
Therefore, we modified the differential equations for QI scores by
adding terms of multiplicative white Gaussian noise. By solving the
equations of stochastic model for different values of noise
intensity, we obtained different QI curves. These were
quantitatively compared with the experimental findings, coming from
sensory analysis, by using the root mean square error test. We found
that, for some groups of specimens considered and some sites
analyzed (skin, gills, or flesh), theoretical QI scores obtained in
the presence of suitable noise intensities are in a better agreement
with those observed experimentally with respect to those calculated
by the deterministic model.\\
Finally we note that our study could play a key role in view of
using microbial predictive models not only for a food risk
assessment, but also to develop a protocol which provides a
quantitative estimation of the organoleptic properties of food
products. Such a procedure, used together with experimental analyses
performed by "electronic noses", could permit to get a more
trustable evaluation of the quality index, while contributing to
make more precise the prediction of the changes occurring in the
organoleptic properties. This aspect agrees to the new European
approach to food quality assessment and management.

\vspace*{0.5cm}

\section*{Acknowledgements}
Authors acknowledge the financial support by Ministry of University,
Research and Education of Italian Government, Project "RITMARE
SP2\_WP1\_AZ3\_UO04 - Potenziamento  delle campagne scientifiche di
acquisizione di informazioni indipendenti dalla pesca sulle
risorse", and Project PON02\_00451\_3362121 "PESCATEC - Sviluppo di
una Pesca Siciliana Sostenibile e Competitiva attraverso
l'Innovazione Tecnologica".


\end{document}